\documentclass[11pt]{article}
\usepackage[usenames,dvipsnames,svgnames,table]{xcolor} 

\usepackage{enumitem} 
\usepackage{rotfloat} 
\usepackage{graphicx}
\usepackage{epsfig}
\usepackage{amssymb, amsmath}
\usepackage{verbatim}
\usepackage{natbib}
\usepackage{authblk}
\usepackage{kotex}
\usepackage{multirow}
\usepackage[colorlinks]{hyperref} 
\usepackage[colorinlistoftodos, textsize=scriptsize]{todonotes} 
\usepackage{subcaption} 

\newtheorem{theorem}{Theorem}

\newtheorem{proposition}{Proposition}
\newtheorem{corollary}{Corollary}

\newenvironment{proof}[1][Proof]{\begin{trivlist}
		\item[\hskip \labelsep {\bfseries #1}]}{\end{trivlist}}
\newenvironment{definition}[1][Definition]{\begin{trivlist}
		\item[\hskip \labelsep {\bfseries #1}]}{\end{trivlist}}

\newcommand{\bea}{\begin{eqnarray*}}
	\newcommand{\eea}{\end{eqnarray*}}
\newcommand{\bean}{\begin{eqnarray}}
\newcommand{\eean}{\end{eqnarray}}
\newcommand{\baln}{\begin{align}}
\newcommand{\ealn}{\end{align}}
\newcommand{\bdefi}{\begin{definition}}
	\newcommand{\edefi}{\end{definition}}
\newcommand{\benu}{\begin{enumerate}}
	\newcommand{\eenu}{\end{enumerate}}
\newcommand{\ben}{\begin{enumerate}}
	\newcommand{\een}{\end{enumerate}}

\newcommand{\E}{E}
\newcommand{\V}{{\rm var}}

\newcommand{\sg}{\Sigma}

\newcommand{\what}{\widehat}

\newcommand{\bbE}{E}

\newcommand{\bbP}{\mathbb{P}} 
\newcommand{\bbR}{\mathbb{R}}

\newcommand{\lra}{\longrightarrow}

\newcommand{\calC}{\mathcal{C}}

\begin{document}


\title{Maximum Pairwise Bayes Factors for Covariance Structure Testing}
\author[1]{Kyoungjae Lee}
\author[2]{Lizhen Lin}
\author[3]{David Dunson}
\affil[1]{Department of Statistics, Inha University}
\affil[2]{Department of Applied and Computational Mathematics and Statistics, The University of Notre Dame}
\affil[3]{Department of Statistical Science, Duke University}

\maketitle
\begin{abstract}
	Hypothesis testing of structure in covariance matrices is of significant importance, but faces great challenges in high-dimensional settings. Although consistent frequentist one-sample covariance tests have been proposed, there is a lack of simple, computationally scalable, and theoretically sound Bayesian testing methods for large covariance matrices. Motivated by this gap and by the need for tests that are powerful against sparse alternatives, we propose a novel testing framework based on the maximum pairwise Bayes factor. Our initial focus is on one-sample covariance testing; the proposed test can {\it optimally} distinguish null and alternative hypotheses in a frequentist asymptotic sense. We then propose diagonal tests and a scalable covariance graph selection procedure that are shown to be consistent. A simulation study evaluates the proposed approach relative to competitors. We illustrate advantages of our graph selection method on a gene expression data set.
\end{abstract}

Key words: Bayesian hypothesis test, covariance matrix, modularization

\section{Introduction}\label{sec:intro}

Consider a sample of observations from a high-dimensional normal model
\bean
X_1,\ldots,X_n \mid \sg_n &\overset{i.i.d.}{\sim}& N_p(0, \sg_n), \label{model}
\eean
where $\sg_n \in \bbR^{p\times p}$ is a covariance matrix. 
There is often interest in inferring the structure in $\sg_n$ and in comparing different alternative covariance structures.
This article focuses on this problem from a hypothesis testing perspective.
Let $X = (X_1,\ldots, X_n)^T \in \bbR^{n\times p}$ be the data matrix.
{\it A one-sample covariance test} can be reduced to the following simple form:
\bean\label{test}
H_0: \sg_n = I_p \quad \text{versus}\quad H_1: \sg_n \neq I_p ,
\eean
by noting that $H_0:\sg_n = I_p$ is equivalent to a null hypothesis $H_0: \sg_n = \sg_0$ for any given positive definite matrix $\sg_0$ by applying the linear transformation $X_i \mapsto \sg_0^{-1/2} X_i$.

Another important problem is testing diagonality
\bea
H_{0}: \sigma_{ij} = 0 \text{ for any }i\neq j  \quad\text{ versus }\quad H_{1}: \text{ not } H_0 ,
\eea
where $\sg_n = (\sigma_{ij})$.
Finally, we consider the problem of  {\it support recovery}, corresponding to estimating the nonzero elements of covariance matrices. 


We are interested in constructing  novel Bayesian procedures that are practically applicable with theoretical guarantees for the (i) one-sample covariance test,  (ii) diagonality test, and (iii) support recovery of the covariance matrix.
We consider the high-dimensional setting in which the number of variables $p$ can grow to infinity as the sample size $n$ gets larger and possibly be  much larger than $n$.
Although it is well known that assuming a restricted covariance class is necessary for consistent {\em estimation} of large covariance matrices \citep{johnstone2009consistency,lee2018optimal}, in a {\em testing} context we focus on alternative hypotheses $H_1$ that are unconstrained.  
One natural possibility is to assume a conjugate inverse-Wishart prior $IW_p(\nu_n,A_n)$ for $\Sigma_n$ under $H_1$.  
However, in order for the resulting posterior to be proper, it is necessary to choose the degrees of freedom $\nu_n > p-1$, suggesting an extremely informative prior in high-dimensional settings.  
The resulting test will certainly be highly sensitive to the choice of $A_n$, and hence is not very useful outside of narrow applications having substantial prior information.  
One could instead choose a non-conjugate prior for $\Sigma_n$ under $H_1$, but then substantial computational issues arise in attempting to estimate the Bayes factor.


From a frequentist perspective, 
\cite{chen2010tests} and \cite{cai2013optimal} suggested consistent one-sample covariance tests based on unbiased estimators of $\|\sg_n -I_p\|_F^2$, where $\|A\|_F = \big(\sum_{ij} a_{ij}^2 \big)^{1/2}$ is the Frobenius norm of a matrix $A=(a_{ij})$.
Under the null hypothesis, they showed that their test statistic is asymptotically normal. 
The test also has power tending to one as $n$ goes to infinity, but it requires the condition, $\|\sg_n - I_p\|_F^2 \, n/p \to \infty$ as $n\to\infty$.
This condition implies that they essentially adopted $H_1 = \{ \sg_n : \|\sg_n - I_p\|_F^2 \ge b_n  p/n \}$ for some $b_n \to \infty$ as $n\to\infty$ as the alternative class.
\cite{cai2013optimal} proved that if we consider an alternative class $H_1 = \{ \sg_n : \|\sg_n - I_p\|_F^2 \ge \epsilon_n \}$, say a dense alternative, the condition $\epsilon_n \ge b_n  p/n$ is inevitable for any level $\alpha$ test to have power tending to one.
This excludes  cases in which a finite number of the components of $\sg_n-I_p$ have a magnitude $(p/n)^{1/2}$, although $(p/n)^{1/2}$ can be a significant signal when $p \ge n$.

The above discussion motivates us to develop  hypothesis tests that are easy to implement in practice while possessing theory guarantees. In particular, we wish to construct tests that can perform well even when the condition $\|\sg_n - I_p\|_F^2 \, n/p \to \infty$ fails to hold.
We achieve this by proposing a novel Bayesian testing framework based on the maximum pairwise Bayes factor which will be introduced in Section \ref{subsec:mpbf}.
The basic strategy is to focus on the pairwise difference between $\sg_n$ and $I_p$ rather than the Frobenius norm or other matrix norms.
More precisely, instead of considering a usual Bayes factor based on a prior on the whole covariance matrix, we first consider the pairwise Bayes factors for each element of the matrix and combine them by taking a maximum over all possible pairs. 
This approach is analagous to  frequentist tests based on maximum-type statistics  \citep{jeng2013simultaneous,enikeeva2019high}.
Our construction enables us to consider a different alternative class, $H_1= \{\sg_n : \|\sg_n - I_p\|_{\max}^2 \ge C \log p/n \}$ for some constant $C>0$, say a sparse alternative, where $\|A\|_{\max} = \max_{i,j}|a_{ij}|$ for a matrix $A=(a_{ij})$. 
When the primary interest is not on  a collection of very weak signals, but on detecting at least one \emph{meaningful signal}, our test is much more effective than the frequentist methods mentioned above.


The proposed testing method is general, easily implementable and theoretically supported, being the first Bayesian test shown to be consistent in the high-dimensional setting for the one-sample or diagonal covariance testing problems.
Our procedure yields proven false discovery rate control and power improvement compared to existing methods.   
The proposed  one-sample test is rate-optimal in the sense that it can distinguish the  sparse alternative class $H_1= \{\sg_n : \|\sg_n - I_p\|_{\max}^2 \ge \epsilon_n\}$ from the null with the fastest rate of $\epsilon_n$, while guaranteeing consistency under the null.
We also propose a scalable graph selection method for high-dimensional covariance graph models using pairwise Bayes factors.
The proposed method consistently recovers the true covariance graph structure under a weaker or comparable condition to those in the existing frequentist literature.

Recently, \cite{leday2018fast}  suggested a fundamentally different pairwise approach to test marginal or conditional independence between two variables.
Their focus is on the joint distribution of  the $i$th and $j$th variables and an inverse-Wishart prior for $\sg_n$ was imposed.
For each $i\neq j$, the hypothesis testing problem $H_{0,ij}^M: \sigma_{ij}=0$ versus $H_{1,ij}^M: \sigma_{ij}\neq 0$ was considered. 
Since the resulting Bayes factors for the pairwise tests are not scale-invariant, they proposed {\it scaled} versions.
P-values under the {\it conditional null distribution} were obtained by shuffling or permuting labels of observations \citep{jiang2017bayesian}.
For  support recovery, they suggest using standard multiplicity correction procedures to control the false discovery rate, obtaining a frequentist procedure.
Selection consistency results were not provided.  

\verb|R| code for implementation of our empirical results are available at https://github.com/leekjstat/mxPBF.
Proofs of our main results are included in Supplementary Material.

\section{Preliminaries}\label{sec:prel}

\subsection{Notations}\label{subsec:notation}

For any real values $a$ and $b$, we denote $a\vee b$ as the maximum between $a$ and $b$.
For any positive sequences $a_n$ and $b_n$, we denote $a_n \ll b_n$ or $a_n = o(b_n)$ if $a_n / b_n \to 0$ as $n\to\infty$.
For any vector $x = (x_1,\ldots,x_p)^T \in \bbR^p$, we define the vector $\ell_1$- and $\ell_2$-norm as $\|x\|_1= \sum_{j=1}^p |x_i|$ and $\|x\|_2 = (\sum_{j=1}^p x_j^2 )^{1/2}$, respectively.
Let $\calC_p$ be the set of all $p\times p$ positive definite matrices.
We denote $\chi_k^2(\lambda)$ as the non-central chi-square distribution with degrees of freedom $k$ and non-centrality $\lambda \ge 0$, and let $\chi_k^2 = \chi_k^2(\lambda=0)$.
For positive real values $a$ and $b$, $IG(a,b)$ denotes the inverse gamma distribution with shape $a$ and scale $b$.

\subsection{Maximum Pairwise Bayes Factor}\label{subsec:mpbf}
In this subsection, we introduce our approach focusing on the one-sample covariance test.
As described before, the basic strategy is to concentrate on the \emph{pairwise difference} between $\sg_n$ and $I_p$.
Let $\tilde{X}_j\in \bbR^n$ be the $j$th column vector of $X$.
For any indices $i$ and $j$, based on the joint distribution \eqref{model}, the conditional distribution of $\tilde{X}_i$ given $\tilde{X}_j$ is
\bean\label{ij_reg_model}
\tilde{X}_i  \mid \tilde{X}_j  &\sim& N_n \Big( a_{ij} \tilde{X}_j ,\, \tau_{ij}^2 I_n  \Big), 
\eean
where $a_{ij} \in \bbR$ and $\tau_{ij}>0$. 
We can view  \eqref{ij_reg_model} as a linear regression model given a design matrix $\tilde{X}_j$.
For each paired conditional model \eqref{ij_reg_model}, we consider a testing problem
\bean\label{hypo_ij}
H_{0,ij}: a_{ij}=0 ,\, \tau_{ij}^2 =1 \quad\text{versus} \quad H_{1,ij}: \text{ not } H_{0,ij} .
\eean
If $H_{0,ij}$ is true, $\sigma_{ij} = 0 $ and $\sigma_{ii} =1$ because $a_{ij} = \sigma_{ij}/ \sigma_{jj}$ and $\tau_{ij}^2 = \sigma_{ii} (1- \rho_{ij}^2)$, where $\sg_n = (\sigma_{ij})$ and $R_n = (\rho_{ij})$ are covariance and correlation matrices, respectively.
We suggest the following prior distribution under the alternative hypothesis $H_{1,ij}$ in \eqref{hypo_ij},
\bean
\begin{split}\label{prior}
	a_{ij} \mid \tau_{ij}^2 \,\,\sim\,\,  N \Big( 0 , \,  \frac{\tau_{ij}^2}{\gamma} \| \tilde{X}_j \|_2^{-1} \Big)  , &\quad\, \tau_{ij}^2 \,\,\sim\,\, IG \big( a_0, \, b_{0,ij}  \big) ,
\end{split}
\eean
where $\gamma= (n \vee p)^{-\alpha}$ and $a_0, b_{0,ij}$ and $\alpha$ are positive constants.
The induced Bayes factor is 
\bea
B_{10}(\tilde{X}_i, \tilde{X}_j) &=& \frac{p(\tilde{X}_i\mid \tilde{X}_j , H_{1,ij}) }{p(\tilde{X}_i \mid \tilde{X}_j , H_{0,ij})} 
\\
&=& \frac{b_{0,ij}^{a_0}}{ \Gamma(a_0)} \Big(\frac{\gamma}{1+\gamma} \Big)^{1/2} \Gamma\Big(\frac{n}{2}+a_0 \Big)\, e^{ n \what{\tau}_i^2/2 }  \, \Big(\frac{n}{2}\what{\tau}_{ij, \gamma}^2 + b_{0,ij} \Big)^{- n/2 - a_0}  ,
\eea
where $n \what{\tau}_i^2 = \|\tilde{X}_i\|_2^2$,  $n \what{\tau}_{ij, \gamma}^2 = \tilde{X}_i^T \{ I_n - (1+\gamma)^{-1} H_j  \} \tilde{X}_i$ and $H_j = \tilde{X}_j (\tilde{X}_j^T \tilde{X}_j)^{-1} \tilde{X}_j^T$.
The choice of hyperparameters $a_0$ and $b_{0,ij}$ is discussed in Section \ref{subsec:sim_one}.

The null hypothesis in the one-sample covariance test, $H_0: \sg_n = I_p$, is true if $H_{0,ij}$ is true for all pairs $(i,j)$ such that $i \neq j$.
We  aggregate the information from each pairwise Bayes factor $B_{10}(\tilde{X}_i, \tilde{X}_j)$ via the {\it maximum pairwise Bayes factor},
\bean\label{map_one}
B_{\max, 10}(X) &=& \max_{ i \neq j} B_{10}(\tilde{X}_i, \tilde{X}_j) .
\eean
A large value for $B_{\max, 10}(X)$ provides evidence supporting the alternative hypothesis.
By taking a maximum, $B_{\max, 10}(X)$ supports the alternative hypothesis if at least one of the pairwise Bayes factors  supports the alternative.
A natural question is whether false positives increase as we take a maximum over more and more pairs.
Indeed, we find that this is not the case, either asymptotically based on our consistency results (Theorems \ref{thm:MBF} and \ref{thm:diag_BF}) or in finite samples based on simulations.

\section{Main Results}\label{sec:main}

\subsection{One-sample Covariance Test}\label{subsec:one-sample}

In this subsection, we show consistency of $B_{\max, 10}(X)$ defined in \eqref{map_one} for 
the one-sample covariance test \eqref{test}.
We first introduce assumptions for consistency under $H_1 : \sg_n \neq I_p$.
Let $\sg_0 = (\sigma_{0,ij}) \in \calC_p$ be the true covariance matrix, implying the conditional distribution of $\tilde{X}_i$ given $\tilde{X}_{j}$ is 
\bean\label{true_conditional}
\tilde{X}_i \mid \tilde{X}_{j}  &\sim&  N_n \big( a_{0,ij} \tilde{X}_{j} , \tau_{0,ij}^2 I_n \big) 
\eean
under $\bbP_0$, where $a_{0,ij} = \sigma_{0,ij}/ \sigma_{0,jj}$, $\tau_{0,ij}^2 = \sigma_{0,ii} \{ 1- \sigma_{0,ij}^2 / (\sigma_{0,ii} \sigma_{0,jj}  ) \}$, $\bbP_0$ is the probability measure corresponding to model \eqref{model} with $\sg_n =\sg_0$, and  $\tau_{0,ij}^2 = \sigma_{0,ii}$ if and only if $a_{0,ij}=0$.
Under the alternative $H_1: \sg_n \neq I_p$, we assume that $\sg_0$ satisfies {\it at least one} of the following conditions:
\begin{itemize}
	\item[(A1)] There exists a pair $(i,j)$ satisfying
	\bean\label{sigma_betamin2}
	\big| \sigma_{0,ii} - 1  \big| &\ge&  \Big[ 4 \sigma_{0,ii} {C_1}^{1/2} + C_2 + \frac{2b_{0,ij}}{ \{n \log (n \vee p)\}^{1/2} }  \Big] \left\{\frac{\log (n \vee p)}{n} \right\}^{1/2} 
	\eean
	for some constants $C_1 >0$ and $C_2 > 2(\alpha +2)^{1/2}$.
	
	\item[(A2)] 
	There exists a pair $(i,j)$ satisfying
	\bean\label{sigma_betamin}
	\big|\tau_{0,ij}^2 - 1 \big| &\ge&  \Big[ 4 \tau_{0,ij}^2 {C_1}^{1/2} + C_2 + \frac{2b_{0,ij} + \tau_{0,ij}^2}{ \{n \log (n \vee p)\}^{1/2} }  \Big] \left\{\frac{\log (n \vee p)}{n} \right\}^{1/2} 
	\eean
	
	\item[(A3)] 
	There exists a pair $(i,j)$ satisfying
	\bean\label{betamin}
	\sigma_{0,ij}^2 &\ge& \frac{\sigma_{0,jj} }{1- 2 {C_1}^{1/2}\epsilon_0} \left\{\frac{9C_1 \tau_{0,ij}^2}{(1-C_3)^2} \vee \frac{C_4 (\alpha+2)}{C_3} \right\}   \frac{\log (n \vee p)}{n} \quad\quad
	\eean
	for some constants $0<C_3<1$ and $C_4 >1$.
	
\end{itemize}
Throughout the paper, $C_1, C_2, C_3$ and $C_4$ are fixed global constants.
For a given small constant $\epsilon > 0$, they can be considered as $C_1 = \epsilon,  C_2 = 2(\alpha+2)^{1/2} +\epsilon, C_3 = 1 - \epsilon^{1/4}$ and $C_4 = 1 + \epsilon$.

Condition (A1) is required to detect a non-unit variance  $\sigma_{0,ii}$, and can be interpreted as a {\it beta-min condition} for $|\sigma_{0,ii} -1|$.
The beta-min condition gives a lower bound for nonzero parameters and is essential for model selection consistency \citep{castillo2015bayesian,martin2017empirical}.
Interestingly, the rate of lower bound in (A1) is given by $\{\log (n \vee p)/n \}^{1/2}$, which has been commonly used in the variable selection literature.
Condition (A2) is similar to condition (A1), which can be interpreted as a beta-min condition for $|\tau_{0,ij}^2-1|$.
Condition (A3) is a beta-min condition for off-diagonal elements of the covariance matrix.
In summary, conditions (A1)--(A3) imply  the sparse alternative
\bea
\sg_0 &\in& H_1 \,=\, \Big\{ \, \sg_n :  \| \sg_n - I_p \|_{\max}^2 \ge C \, \frac{\log p}{n} \,  \Big\}
\eea
for some constant $C>0$, which corresponds to the {\it meaningful} difference we mentioned earlier.
In fact,  the rate $\log p/n$ is {\it optimal}  for guaranteeing the consistency under both hypotheses (Theorem \ref{thm:LB}). 
Our method is not designed to detect dense alternatives in which all differences are very small, but requires at least one difference to be sufficiently large. 

Theorem \ref{thm:MBF} shows  consistency for the one-sample covariance test even in the high-dimensional setting as long as $\log p \le \epsilon_0^2 n$ for some small constant $\epsilon_0>0$.

\begin{theorem}\label{thm:MBF}
	Consider model \eqref{model} and the one-sample covariance testing problem \eqref{test}.
	Consider prior \eqref{prior} under $H_{1,ij}$ in \eqref{hypo_ij} with $\alpha> 8 (1 + {2}^{1/2} \epsilon_0)^2/ \{1- {2}^{3/2}\epsilon_0 (1 + {2}^{1/2}\epsilon_0)\}$ for some small constant $0<\epsilon_0< 3 \,(4 C_2)^{-1}$.
	Assume that $\log p \le \epsilon_0^2 \, n$ for all large $n$. 
	Then under $H_0: \sg_n =I_p$, for some constant $c>0$,
	\bea
	B_{\max, 10}(X) &=& O_p \big\{ (n\vee p)^{-c} \big\} .
	\eea
	If, under $H_1: \sg_n \neq I_p$, $\sg_0$ satisfies at least one of conditions (A1)--(A3), for some constant $c' >0$,
	\bea
	B_{\max, 10}(X)^{-1} &=& O_p \big\{ (n\vee p)^{-c'}  \big\}.
	\eea
\end{theorem}

We first prove that the pairwise Bayes factor $B_{10}(\tilde{X}_i, \tilde{X}_j)$ is consistent on a large event $E_{ij}$ such that $\bbP_0 (E_{ij}^c) \to 0$ as $n\to\infty$.
To show  consistency under $H_0$, it suffices to prove that $\sum_{i\neq j} \bbP_0 (E_{ij}^c) \to 0$ as $n\to\infty$, which means that the false discovery rate converges to zero.
The condition for $\alpha$ in Theorem \ref{thm:MBF} is closely related to this requirement.
It also has connections with the variable selection literature in regression  \citep{fernandez2001benchmark, narisetty2014bayesian,yang2016computational} where the prior dispersion needs to depend on $(n \vee p^2)$ or $p$ for consistency.
Our theory requires a larger dispersion of order roughly $(n \vee p)^8$ mainly due to the larger number of parameters compared to the regression setting. 

To show consistency under $H_1$, it suffices to show $ \bbP_0 (E_{ij}^c) \to 0$ as $n\to\infty$ for some index $(i,j)$ satisfying at least one of conditions (A1)--(A3).
Interestingly, the rate of convergence is similar under both hypotheses, unlike most Bayesian testing procedures with the notable exception of non-local prior based methods \citep{johnson2010use,johnson2012bayesian}.

The next theorem shows the optimality of the alternative class which is considered in Theorem \ref{thm:MBF} (Conditions (A1)--(A3)).
It says, when the alternative class is defined based on the element-wise maximum norm, the condition $\|\sg_0 - I_p\|_{\max}^2 \ge C \log p/n$ for some constant $C>0$ is necessary for any consistent test to have power tending to one.
Thus, conditions (A1)--(A3) are rate-optimal to guarantee the consistency under $H_0$ as well as $H_1$.

\begin{theorem}\label{thm:LB}
	Let $\bbE_{\sg}$ be the expectation corresponding to model \eqref{model}.
	For a given constant $C_\star>0$, define 
	$H_1(C_\star) = \Big\{ \sg \in \calC_p : \|\sg- I_p\|_{\max}^2 \ge C_\star^2 \log p/n    \Big\}.$
	If $C_\star^2 \le 2$, then for any consistent test $\phi$ such that $\bbE_{I_p} \phi \lra 0$ as $n\to\infty$,
	\bea
	\limsup_{n\to\infty} \inf_{\sg \in H_1(C_\star) } \bbE_{\sg} (\phi) \le \frac{1}{2} .
	\eea
\end{theorem}

\subsection{Testing Diagonality}\label{subsec:diag}

We now consider testing of diagonality of the covariance matrix: 
\bean\label{diag_test}
H_{0}: \sigma_{ij} = 0 \text{ for any }i\neq j  \quad\text{ versus }\quad H_{1}: \text{ not } H_0 ,
\eean
where $\sg_n = (\sigma_{ij})$.
The above hypothesis testing problem can be modularized into many pairwise independence tests
\bean\label{pair_test}
H_{0,ij}: \sigma_{ij} =0 \quad\text{ versus }\quad H_{1,ij}: \sigma_{ij} \neq 0
\eean
for all $1\le i< j \le p$.
We can adopt the maximum pairwise Bayes factor idea to aggregate the pairwise testing information from \eqref{pair_test} for all possible pairs $(i,j)$ such that $i \neq j$ to test \eqref{diag_test}.
Based on the conditional distribution \eqref{ij_reg_model}, the null hypothesis $H_{0,ij}$ in \eqref{pair_test} is equivalent to $H_{0,ij}': a_{ij}=0$.
We suggest  the prior $\pi(\tau_{ij}^2) \propto \tau_{ij}^{-2}$ under both $H_{0,ij}$ and $H_{1,ij}$, and the prior $\pi(a_{ij} \mid \tau_{ij}^2)$  defined in \eqref{prior} under $H_{1,ij}$, 
which leads to the pairwise Bayes factor
\bea
\tilde{B}_{10} (\tilde{X}_i, \tilde{X}_j) 
&=& \Big(\frac{\gamma}{1+\gamma} \Big)^{1/2} \left( \frac{ \what{\tau}_{ij, \gamma}^2 }{ \what{\tau}_{i}^2 } \right)^{- n/2 }.
\eea
The improper prior $\pi(\tau_{ij}^2) \propto \tau_{ij}^{-2}$ does not cause any problem because we use the same priors under $H_{0,ij}$ and $H_{1,ij}$.
We suggest using 
\bean\label{mxPBF_tilde_diag}
\tilde{B}_{\max, 10}(X) &=& \max_{ i < j} \tilde{B}_{10} (\tilde{X}_i, \tilde{X}_j) 
\eean
for the hypothesis testing problem \eqref{diag_test}.
Theorem \ref{thm:diag_BF} states the consistency of $\tilde{B}_{\max, 10}(X)$ for testing \eqref{diag_test} under regularity conditions.
For consistency under the alternative hypothesis, we assume the following condition: \\
(A4) There exists a pair $(i,j)$ satisfying
\bea
\sigma_{0,ij}^2 &\ge&  \frac{C_4 \sigma_{0,jj} }{1- 2\epsilon_0 {C_1}^{1/2}}   \left\{ \frac{9C_1 \tau_{0,ij}^2}{(1-C_3)^2} \vee \frac{ \alpha(1+\gamma)  (1+4\epsilon_0{C_1}^{1/2}) \sigma_{0,ii}  }{C_3 } \right\}  \frac{\log (n\vee p)}{n}  \quad\quad
\eea
for constants $C_1>0, 0<C_3<1$ and $C_4>1$ defined in Section \ref{subsec:one-sample}.

\begin{theorem}\label{thm:diag_BF}
	Consider model \eqref{model} and the diagonality testing problem \eqref{diag_test}. 
	For a given pair $(i,j)$ such that $i\neq j$, consider the prior $\pi(\tau_{ij}^2) \propto\tau_{ij}^{-2}$ under both $H_{0,ij}$ and $H_{1,ij}$, and the prior $\pi(a_{ij} \mid \tau_{ij}^2)$  defined in \eqref{prior} under $H_{1,ij}$ in \eqref{pair_test} with $\alpha > 4/(1- {2}^{1/2} 3 \epsilon_0)$ for some small constant $0<\epsilon_0< 1/({2}^{1/2} 3)$.
	Assume that $\log p \le \epsilon_0^2 \, n$ for all large $n$. 
	Then under $H_{0}: \sigma_{ij} = 0 \text{ for any }i\neq j$, for some constant $c>0$,
	\bea
	\tilde{B}_{\max, 10} (X) &=& O_p \big\{ (n\vee p)^{-c} \big\} .
	\eea
	If, under $H_{1}:\text{ not } H_0$, $\sg_0$ satisfies condition (A4), for some constant $c'>0$,
	\bea
	\tilde{B}_{\max, 10} (X)^{-1} &=&  O_p\big\{ (n\vee p)^{-c'} \big\} .
	\eea
\end{theorem}

Condition (A4) is the beta-min condition for off-diagonal elements of the true covariance matrix.
It indicates that if one of the off-diagonal elements satisfies the beta-min condition (A4), $\tilde{B}_{\max, 10}(X)$ consistently detects the true alternative hypothesis.
Similar to Theorem \ref{thm:MBF}, the condition for $\alpha$ is required to control the false discovery rate, and $\tilde{B}_{\max, 10}(X)$ has similar rates of convergence under both hypotheses.

Although the maximum pairwise Bayes factor idea is not limited to the test of diagonality, we introduce a few procedures that have been proposed for testing  diagonality in the literature.
\cite{yao2018testing} and \cite{leung2018testing} proposed $L_2$-type tests for  dependence in model-free settings. 
These tests are powerful against dense alternatives, while our focus is on the sparse setting.
\cite{han2017distribution} proposed two families of maximum-type rank tests of diagonality, which include Kendall's tau and Spearman's rho as special cases, respectively.

Although our procedure has a Bayesian motivation, one can use it as a frequentist test statistic.
In the following proposition, we derive the {\it limiting null distribution} of the maximum pairwise Bayes factor in \eqref{mxPBF_tilde_diag}.
This enables us to construct a test having size $\alpha$ asymptotically.
\begin{proposition}\label{prop:diag_limiting_null}
	Under the conditions of Theorem \ref{thm:diag_BF}, further assume that $p = p_n \to \infty$ as $n\to\infty$ and $\log p = o(n^{1/3})$.
	If $H_0: \sigma_{ij} = 0 \text{ for any }i\neq j $ is true, $2\log \tilde{B}_{\max, 10}(X) - C_{n,p} $ converges in distribution to a type I extreme value distribution with distribution function
	\bea
	F(z) &=& \exp \big\{  - (8\pi)^{-1/2} e^{- z/2}  \big\} , \quad z \in \bbR ,
	\eea
	as $n\to\infty$, where $C_{n,p} = 0.5 \log \{ \gamma /(1+\gamma) \}  + 4\log p - \log (\log p) $.
\end{proposition}

\subsection{Support Recovery of Covariance Matrices}\label{subsec:support}

The  primary interest of this section is on the recovery of $S(\sg_0)$, where $S(\sg_0) \subseteq \big\{ (i,j): 1\le i <j\le p \big\}$ is the nonzero index set of the true covariance matrix $\sg_0$. We call $S(\sg_0)$ the {\it support} of $\sg_0$. 
Estimating $S(\sg_0)$ corresponds to graph selection in covariance graph models \citep{cox1993linear}.
Despite its importance, few Bayesian articles have investigated this problem. 
\cite{kundu2019efficient} proposed the regularized inverse Wishart prior, which can be viewed as a group Lasso penalty \citep{yuan2006model} on the Cholesky factor.
They showed the consistency of their selection procedure for the support of precision matrices when the dimension $p$ is fixed.
Recently, \cite{gan2018bayesian} adopted the spike-and-slab Lasso prior \citep{rovckova2016fast,rovckova2018bayesian} for off-diagonal entries of the precision matrix.
Their proposed graph selection procedure for the precision matrix also yields selection consistency.
To the best of our knowledge, in the Bayesian literature, a consistent support recovery result for covariance matrices has not been established.
Although \cite{leday2018fast} proposed a graph selection procedure based on Bayesian modeling, their procedure relies on $p$-values and they do not show consistency.

To tackle this gap, we propose a scalable graph selection scheme for high-dimensional covariance matrices based on  pairwise Bayes factors.
Looking closely at the proof of Theorem \ref{thm:diag_BF}, each pairwise Bayes factor $\tilde{B}_{10}(\tilde{X}_i, \tilde{X}_j)$ can consistently determine whether the corresponding covariance element $\sigma_{0,ij}$ is zero or not.
Thus, we suggest using the estimated index set 
\bean\label{mxPBF_selection}
\what{S}_{pair, C_{sel}} &=& \Big\{ \, (i,j): \,\, 2  \log \tilde{B}_{10}(\tilde{X}_i, \tilde{X}_j) > C_{sel}   , \quad 1\le i<j \le p \,\,   \Big\}
\eean
for some constant $C_{sel} >0$.
Although any threshold $C_{sel}$ can be used for consistent selection asymptotically, the choice  is crucial in practice. 
As a default method, we suggest using cross-validation to select $C_{sel}$, as described in detail in Section \ref{subsec:real_covsel}.
The Supplemental Materials presents a simulation study  investigating the quality of support recovery for various threshold values.

In the frequentist literature, \cite{drton2004model,drton2007multiple} proposed selection procedures using a related idea to \eqref{mxPBF_selection}, which select a graph by multiple hypothesis testing on each edge.
However, they considered only the low-dimensional setting, $n \ge p +1$.

For the consistency of $\what{S}_{pair, C_{sel}}$, we introduce the following condition for some constants $0<C_3<1, C_4>1$ and $C_5>2$: \vspace{.2cm} \\
(A5) For a given pair $(i,j)$ such that $i\neq j$,
\bea
\sigma_{0,ij}^2 &\ge& \frac{C_4 \sigma_{0,jj} }{1- 2\epsilon_0 {C_5}^{1/2}}  \left[ \frac{9 C_5 \tau_{0,ij}^2}{(1-C_3)^2} \vee \frac{ \alpha(1+\gamma)   (1+4\epsilon_0{C_5}^{1/2}) \sigma_{0,ii}  }{C_3 } \right]   \frac{\log (n\vee p)}{n} . \quad\quad
\eea
The beta-min condition (A5) is almost the same as  (A4) except using $C_5>2$ instead of $C_1>0$ to control the probabilities of small events on which the pairwise Bayes factor might not be consistent.
Theorem \ref{thm:select} states that \eqref{mxPBF_selection} achieves model selection consistency if condition (A5) holds with $(i,j)$ or $(j,i)$ for any $(i,j) \in S(\sg_0)$.

\begin{theorem}\label{thm:select}
	Consider model \eqref{model} and prior \eqref{prior} with $\alpha > 4/(1- {2}^{1/2} 3 \epsilon_0)$ for some small constant $0<\epsilon_0< ({2}^{1/2} 3)^{-1}$ and each pair $(i,j)$ such that $i\neq j$. 
	Assume that $\log p \le \epsilon_0^2 \, n$ for all large $n$ and  condition (A5) holds with $(i,j)$ or $(j,i)$ for any $(i,j) \in S(\sg_0)$. 
	Then, we have  
	\bea
	\lim_{n\to\infty} \normalfont{\bbP_0}  \big(  \, \what{S}_{pair, C_{sel}} =S(\sg_0) \, \big)  &=& 1 .
	\eea
\end{theorem}

We note that $\what{S}_{pair, C_{sel}}$ consistently recovers the support of the true covariance matrix $\sg_0$ regardless of the true sparsity as long as $\log p \le \epsilon_0^2 n$ and nonzero entries satisfy the beta-min condition (A5).
\cite{rothman2009generalized} proved a similar support recovery result for generalized thresholding of the sample covariance matrix while assuming  $\log p = o(n)$, $\max_i \sigma_{0,ii} \le M$ for some $M>0$ and $\min_{(i,j)\in S(\sg_0) }\sigma_{0,ij}^2 \ge M' \log p/n$ for some sufficiently large $M'>0$.
\cite{cai2011adaptive} assumed $\log p = o(n^{1/3})$ and $\min_{(i,j)\in S(\sg_0) }\sigma_{0,ij}^2 \ge C \sigma_{0,ii} \sigma_{0,jj} \log p/n$  for some $C>0$ and  obtained  consistent support recovery using adaptive thresholding.
Our condition, $\log p \le \epsilon_0^2 n$, is much weaker than the conditions used in the literature.
The beta-min condition (A5) is similar to that in \cite{cai2011adaptive} and also has the same rate to that in \cite{rothman2009generalized} if we assume $\max_i \sigma_{0,ii} \le M$ for some $M>0$.
Thus, the required condition in Theorem \ref{thm:select} is weaker or comparable to the conditions used in the literature.



\section{Numerical Results}\label{sec:sim}

\subsection{Simulation Study: One-sample Covariance Test}\label{subsec:sim_one}

In this section, we demonstrate the performance of our one-sample covariance test in various simulation cases.
For the hyperparameters, we suggest using $a_0 = 2 + K^{-2}$ and $b_{0,ij} = \what{\tau}_{ij, \gamma=0}^2(a_0-1)$ for some large constant $K>0$, which leads to $\E^{\pi}(\tau_{ij}^2) = \what{\tau}_{ij, \gamma=0}^2$ and  a prior coefficient of variation $\{\V^\pi(\tau_{ij}^2)\}^{1/2} / \E^{\pi}(\tau_{ij}^2) = K$.
In the simulation studies, $K=100$ was used and the results are not sensitive to the choice of $K$.
The hyperparameter $\alpha$ was chosen as  $\alpha= 8.01 (1- 1/\log n)$.
If we assume a small $\epsilon_0>0$, the above choice of $\alpha$ asymptotically satisfies $\alpha>8(1+{2}^{1/2}\epsilon_0)^2 /\{1-{2}^{3/2}\epsilon_0(1+{2}^{1/2}\epsilon_0)  \}$.
We compare our one-sample covariance test with  frequentist tests, proposed by \cite{cai2013optimal}, \cite{srivastava2014tests} and \cite{gupta2014exact}.
The test suggested by \cite{srivastava2014tests} is based on estimating the squared Frobenius norm, and has a similar perspective to the test proposed by \cite{cai2013optimal}.
\cite{gupta2014exact} proposed an exact one-sample covariance test based on fixed columns of the sample covariance matrix.

We first generated 100 data sets from the null hypothesis $H_0:\sg_n=I_p$ for various choices of $n$ and $p$.
We considered two structures for the alternative hypothesis $H_1: \sg_n \neq I_p$.
First, we chose $\sg_0 = (\sigma_{0,ij})$  to have  a compound symmetry structure
\bean\label{al1}
\sigma_{0,ij} &=& I(i=j) \,+\, \rho  I (i \neq j) 
\eean
for some signal strength constant $\rho$ ranging from $0.05$ to $0.15$ by $0.025$. In this case, the difference between $\sg_0$ and $I_p$ is {\it dense}.  
As a second case for $\sg_0$, we let
\bean\label{al2}
\sigma_{0,ij} &=& I(i=j) \,+\,  \rho  I (i=1,j=2) \,+\, \rho  I (i=2,j=1) ,  
\eean
for some constant $\rho$ ranging from $0.3$ to $0.8$ by $0.025$.
Because \eqref{al2} has signals at only two locations,  the difference between $\sg_0$ and $I_p$ is {\it sparse}.
We generated 100 simulated data from $N_p(0, \sg_0)$ for each setting.

\begin{figure*}[!tb]
	\centering
	\includegraphics[width=5.cm,height=4.7cm]{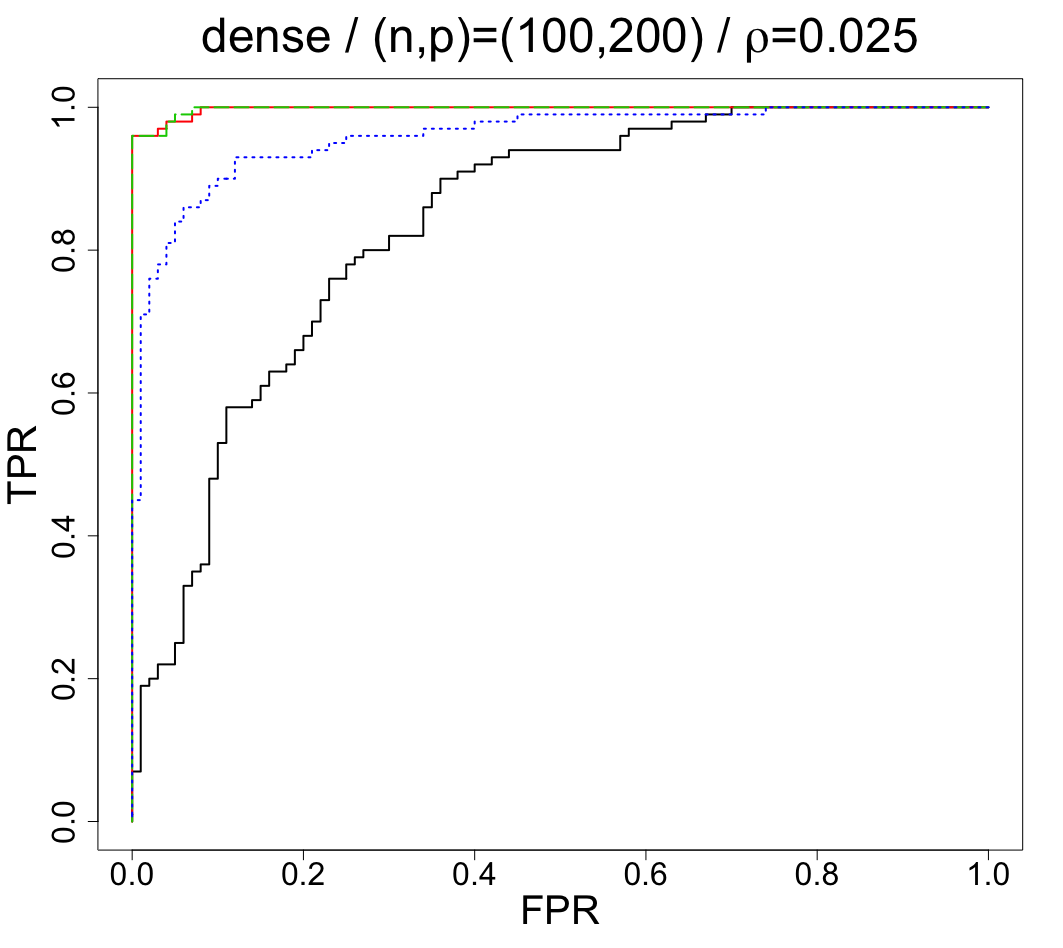}
	\includegraphics[width=5.cm,height=4.7cm]{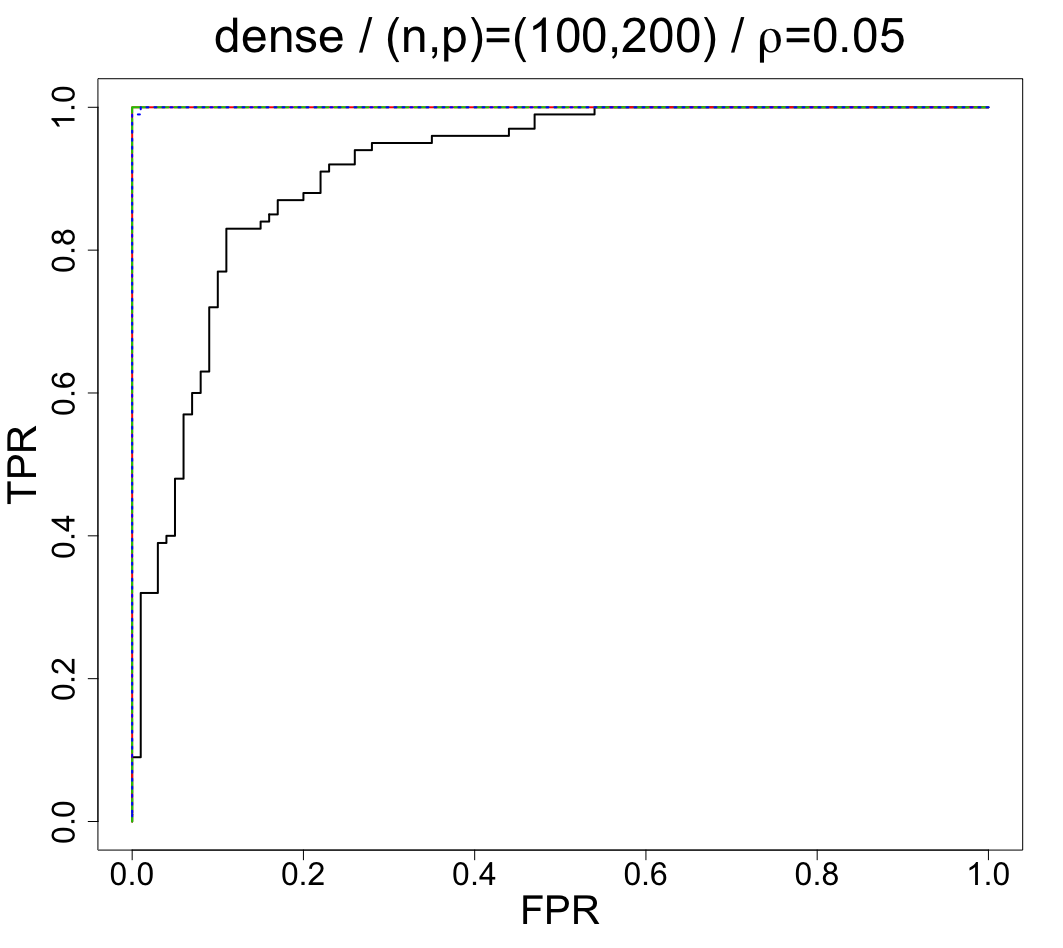}
	\includegraphics[width=5.cm,height=4.7cm]{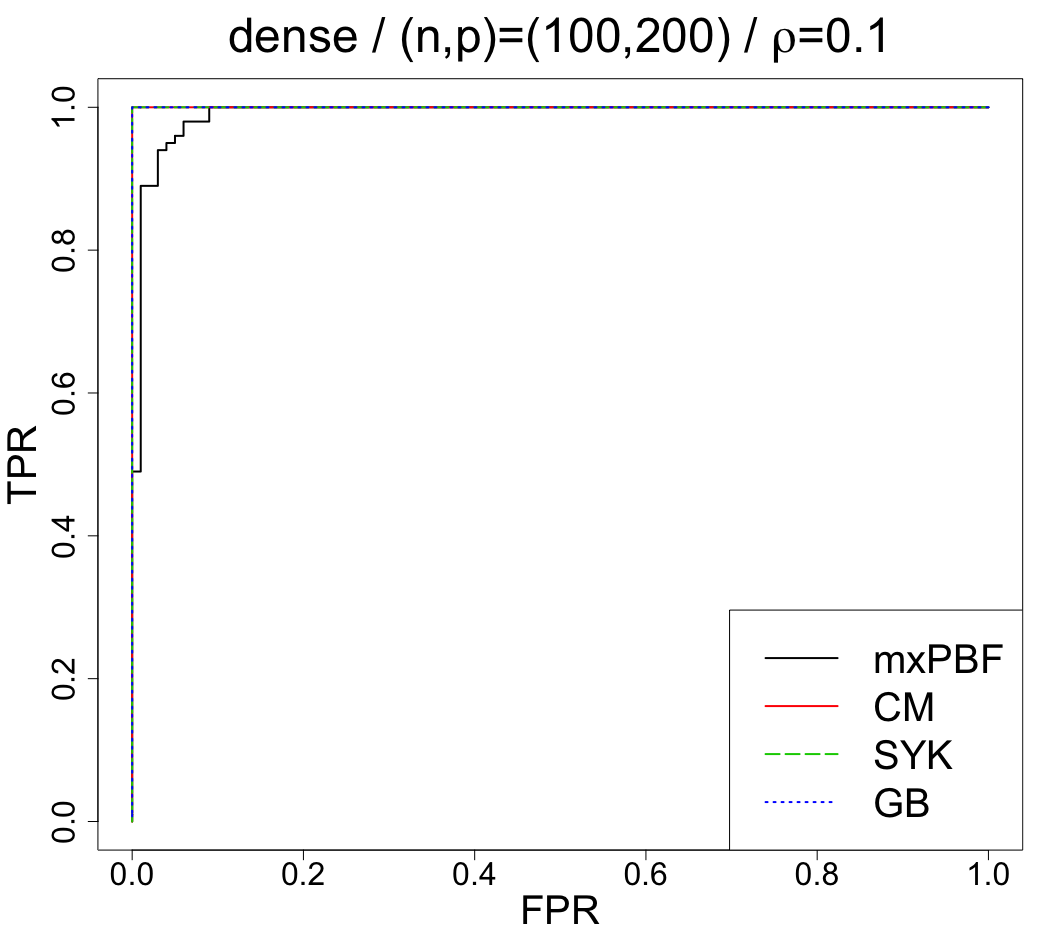}
	\includegraphics[width=5.cm,height=4.7cm]{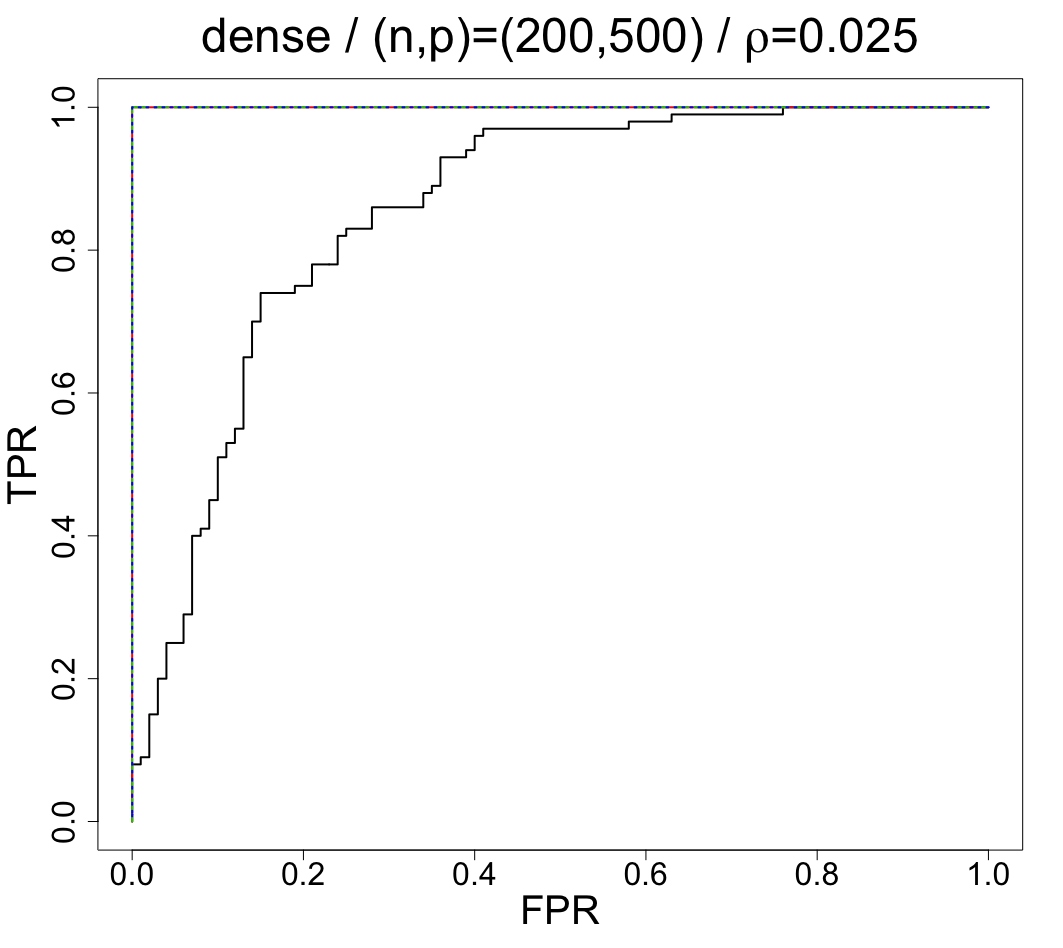}
	\includegraphics[width=5.cm,height=4.7cm]{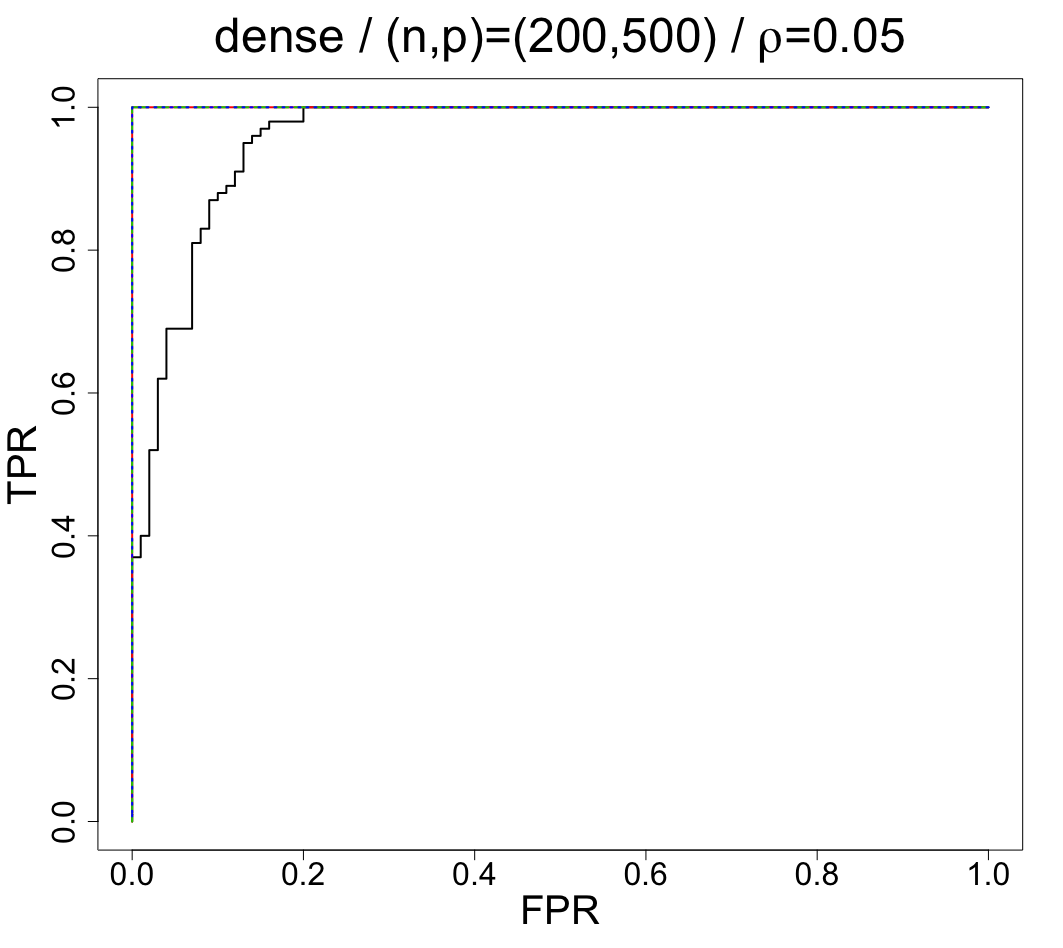}
	\includegraphics[width=5.cm,height=4.7cm]{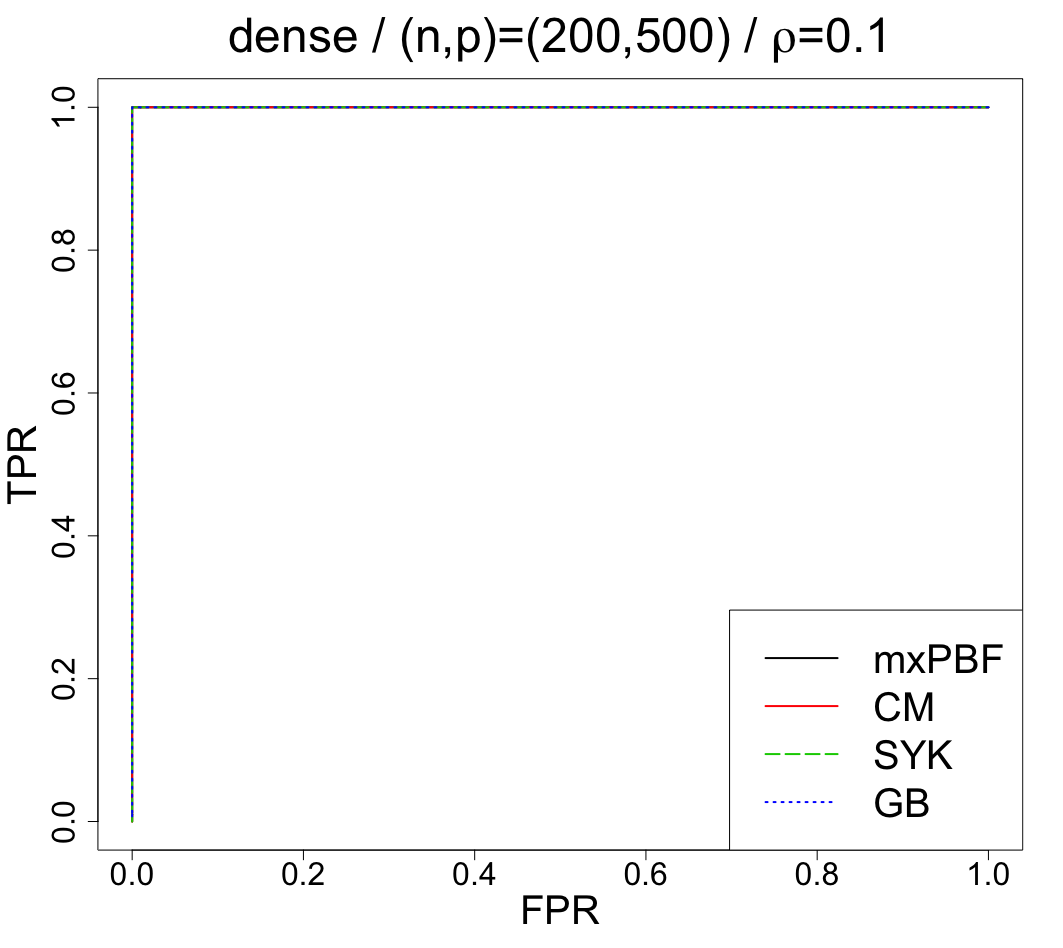}
	\vspace{-.2cm}
	\caption{
		Receiver operating characteristic curves are represented for the three tests based on 100 simulated data sets for each hypothesis  $H_0: \sg_n=I_p$ and $H_1: \sg_n \neq I_p$, where \eqref{al1} was used for $H_1$.
		mxPBF, CM, SYK and GB represent the test proposed in this paper, \cite{cai2013optimal}, \cite{srivastava2014tests} and \cite{gupta2014exact}, respectively.
	}
	\label{fig:roc1}
\end{figure*}
\begin{figure*}[!tb]
	\centering
	\includegraphics[width=5.cm,height=4.7cm]{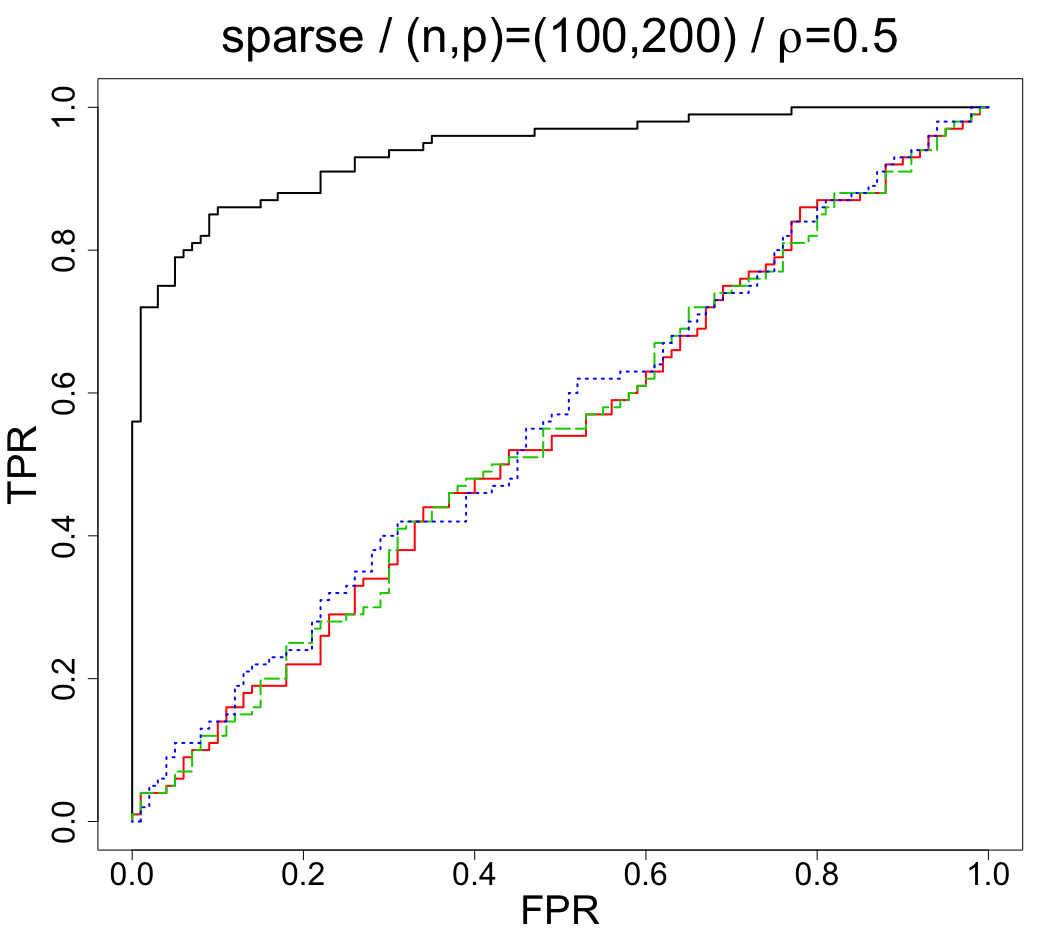}
	\includegraphics[width=5.cm,height=4.7cm]{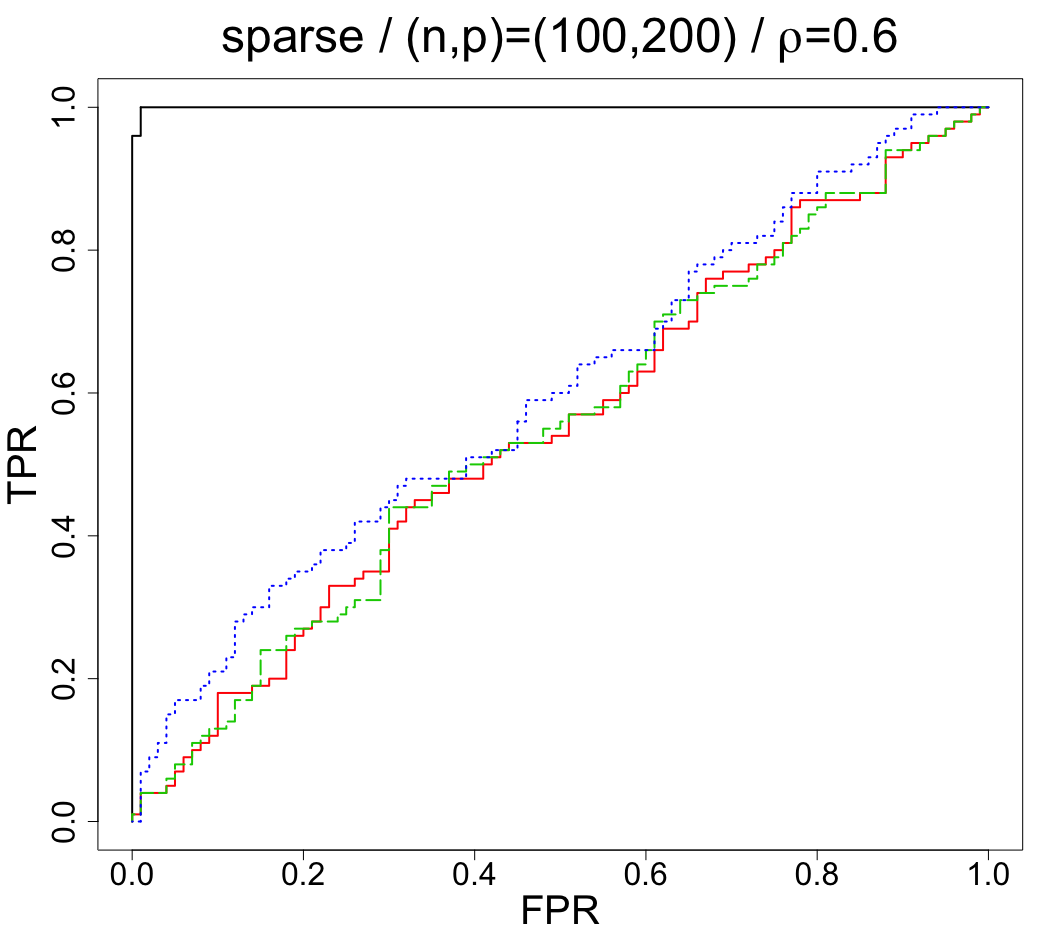}
	\includegraphics[width=5.cm,height=4.7cm]{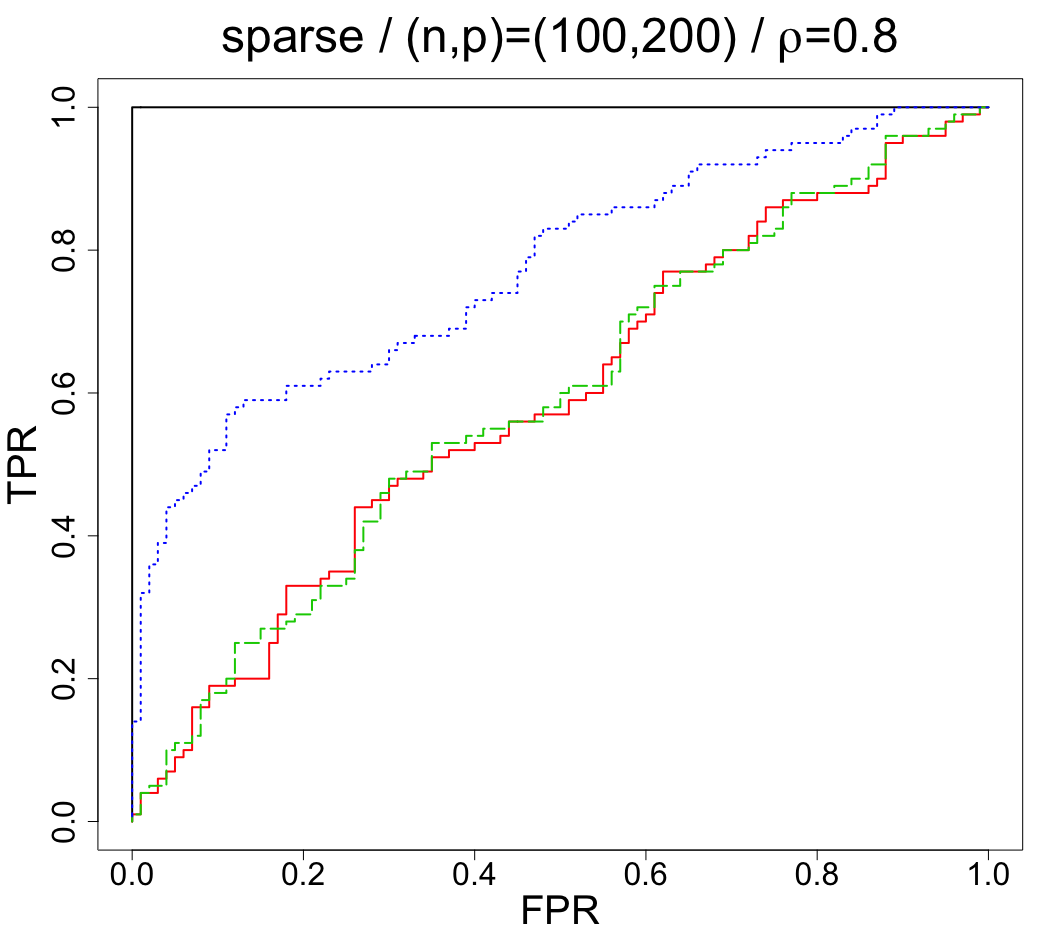}
	\includegraphics[width=5.cm,height=4.7cm]{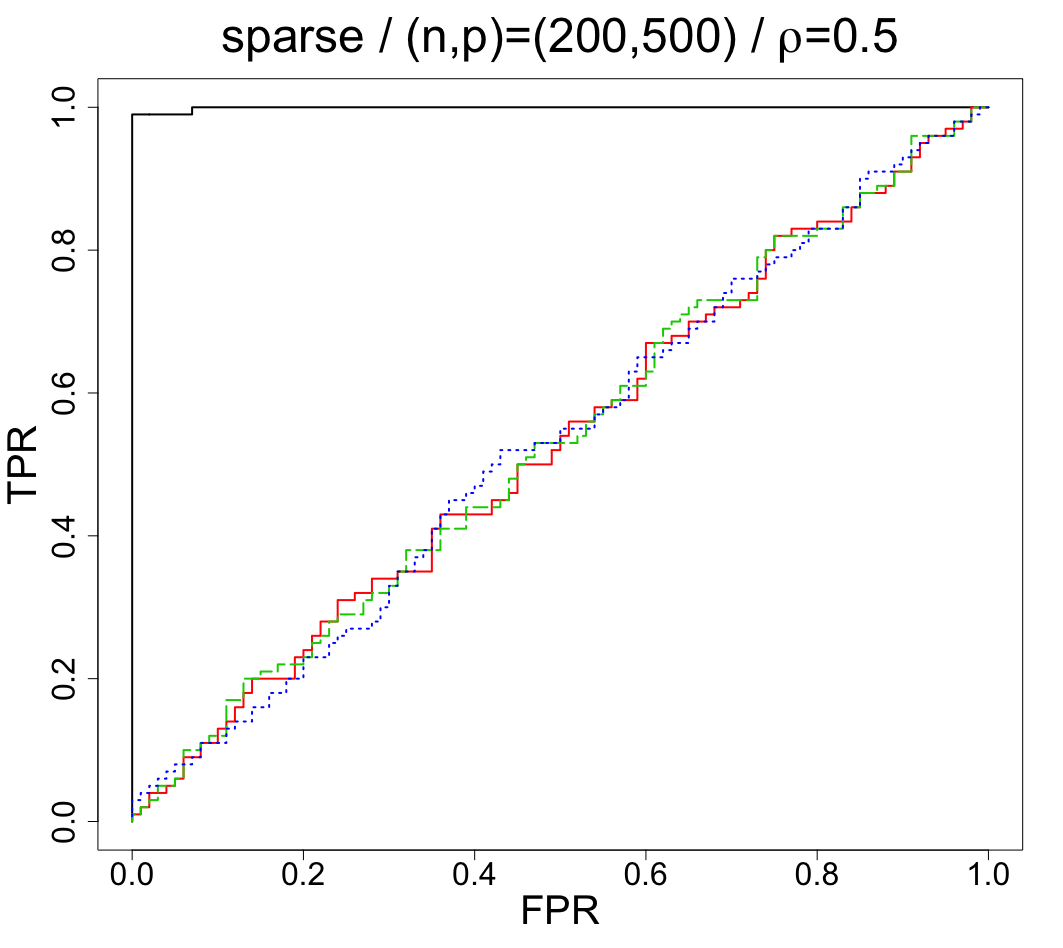}
	\includegraphics[width=5.cm,height=4.7cm]{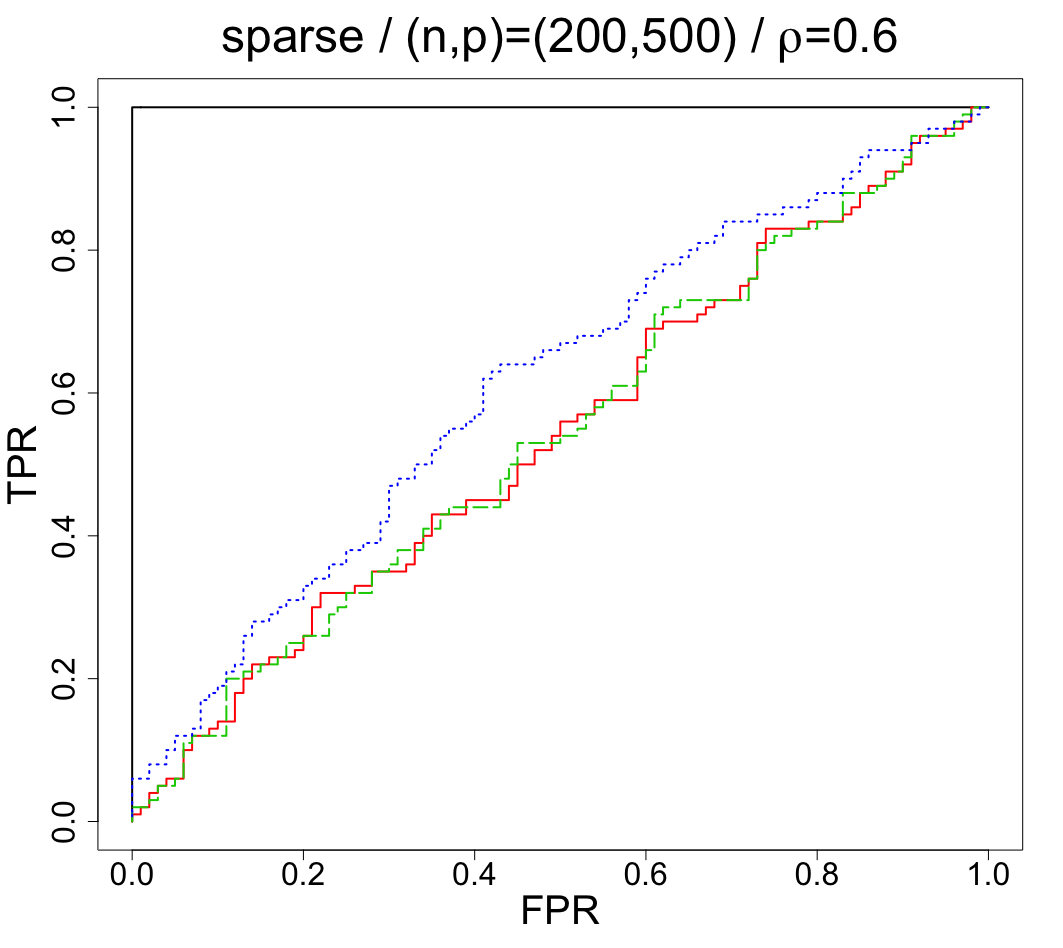}
	\includegraphics[width=5.cm,height=4.7cm]{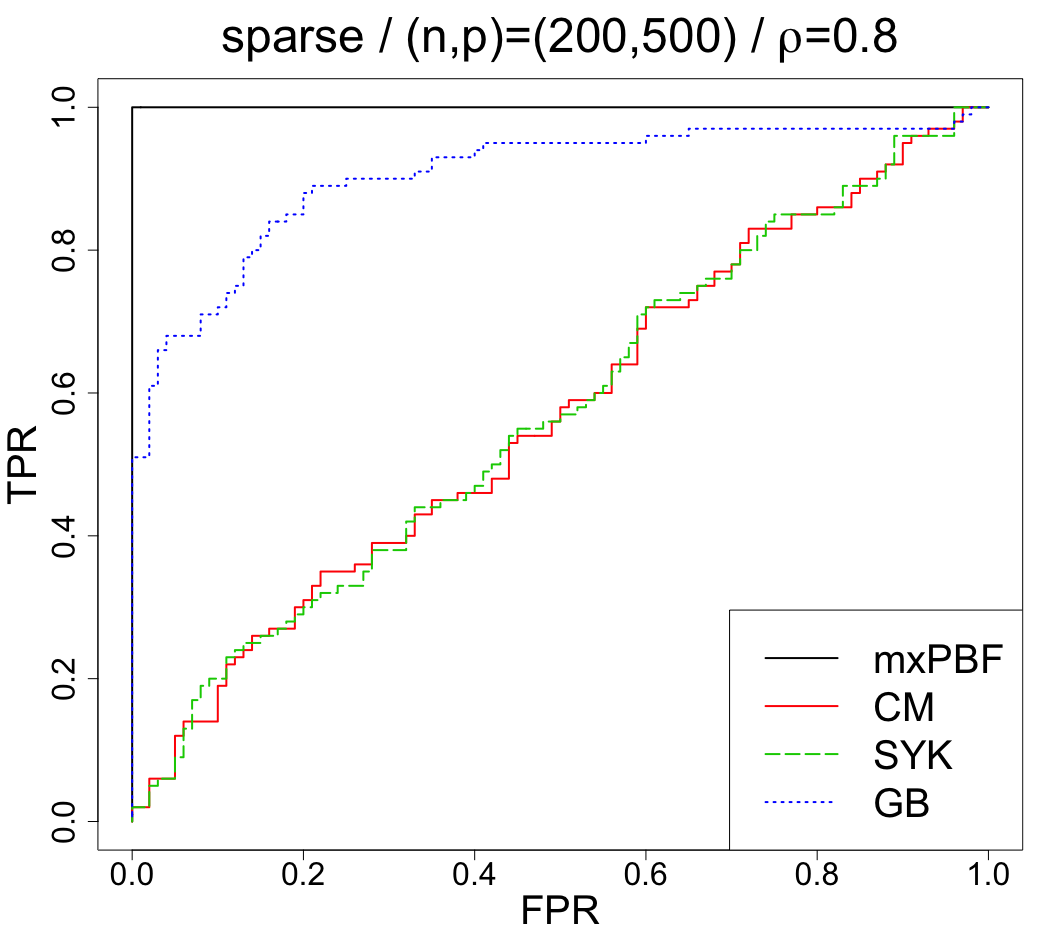}
	\vspace{-.2cm}
	\caption{
		Receiver operating characteristic curves are represented for the three tests based on 100 simulated data sets for each hypothesis  $H_0: \sg_n=I_p$ and $H_1: \sg_n \neq I_p$, where \eqref{al2} was used for $H_1$.
		mxPBF, CM, SYK and GB represent the test proposed in this paper, \cite{cai2013optimal}, \cite{srivastava2014tests} and \cite{gupta2014exact}, respectively.
	}
	\label{fig:roc2}
\end{figure*}
We calculated receiver operating characteristic curves to illustrate and compare the performance of the tests.
For each setting, points of the curves were obtained based on various thresholds and significance levels for  $B_{\max, 10}(X)$  and the frequentist tests, respectively.
We tried $n=100,200,300$ and $p=200, 500$ for each setting.
Figure \ref{fig:roc1} shows results based on 100 simulated data from $N_p(0, I_p)$ ($H_0$) and 100 simulated data from $N_p(0,\sg_0)$ with a compound symmetry structured $\sg_0$ ($H_1$) given in \eqref{al1}, for $(n,p) = (100, 200)$ and $(n,p)=(200, 500)$.
The false positive rate corresponds to the rate of $H_0$'s falsely detected as $H_1$'s.
Similarly, the true positive rate is the rate of $H_1$'s correctly detected as $H_1$'s.  
In this setting, as expected, the tests in \cite{cai2013optimal}, \cite{srivastava2014tests} and \cite{gupta2014exact} work better than the test proposed in this paper.
However, as we can see from the second and third columns in Figure \ref{fig:roc1}, $B_{\max, 10}(X)$ also performs well so long as there is a {\it meaningful signal} somewhere.  
The only case when our method is not as powerful is when weak signals are spread through the alternative covariance matrix, in which case one may question the meaningfulness of the signals.

Figure \ref{fig:roc2} shows results based on 100 simulated data from $N_p(0, I_p)$ and 100 simulated data from $N_p(0,\sg_0)$ with $\sg_0$ given in \eqref{al2}, when $(n,p) = (100, 200)$ and $(n,p)=(200, 500)$.
As expected, $B_{\max, 10}(X)$ is much more powerful than the frequentist tests when $\sg_0 - I_p$ is sparse. 
Furthermore, the performances of the frequentist tests based on the Frobenius norm are almost the same for every setting, while $B_{\max, 10}(X)$ has better performance when $(n,p)=(200,500)$ than $(n,p)=(100,200)$.
Interestingly, the performance of the test in \cite{gupta2014exact} improves as the signal strength $\rho$ increases.
Thus, the test in \cite{gupta2014exact} is more sensitive to sparse changes than other tests based on the Frobenius norm difference.
This makes sense because it focuses on the changes in a column of the covariance matrix rather than in the whole covariance matrix.

\subsection{Simulation Study: Testing Diagonality}\label{subsec:sim_diag}

We conducted a simulation study to illustrate the performance of our proposed diagonality test. 
The hyperparameter $\alpha$ was chosen as $\alpha = 4.01(1-1/\log n)$.
We generated 100 data sets from the null $H_0: \sigma_{ij}=0$ for any $i\neq j$ using $\sg_0 = I_p$.
The two structures of $\sg_0$ under $H_1$ used in the previous section, \eqref{al1} and \eqref{al2}, were considered.
For each setting, 100 data sets were generated.

We compare our test with some existing frequentist tests.
\cite{cai2011limiting} proposed a diagonality test based on the maximum of sample correlations.  
Here  $\what{\tau}_{ij, \gamma}^2$ in the pairwise Bayes factor $\tilde{B}_{10}(\tilde{X}_i, \tilde{X}_j)$ is a decreasing function of the sample correlation between $\tilde{X}_i$ and $\tilde{X}_j$. 
\cite{lan2015testing} developed a test in the regression setting based on the squared Frobenius norm of a sample covariance matrix.
Their test should be powerful against dense alternatives.
We also conducted  maximum-type tests based on Kendall's tau and Spearman's rho \citep{han2017distribution}.
\cite{chen2018testing} assumed $p$-dimensional observations from a common multivariate normal distribution and investigated the dependence {\it among samples}. 
Since their method can be applied to the diagonality test by considering $X^T$ instead of $X$, we included it as a contender. 
Their test requires $p = O(n)$ and the uniformly bounded condition for the eigenvalues of $\sg_0$ for theoretical properties, excluding the high-dimensional setting $p \gg n$ and some interesting covariance classes like compound symmetry.
Finally, we also considered frequentist union-intersection tests based on the p-values associated with the marginal independence tests.
A $t$-test for Pearson's correlation was conducted for testing $H_{0,ij}:\sigma_{ij}=0$ for each pair $i>j$, and the null hypothesis $H_0: \sigma_{ij} =0$ for any $i\neq j$ was rejected if at least one $H_{0,ij}$ was rejected.
To calculate the p-values, we used the \verb|cor0.test| function in the \verb|GeneNet| package.

\begin{figure*}[!tb]
	\centering
	\includegraphics[width=5.cm,height=4.7cm]{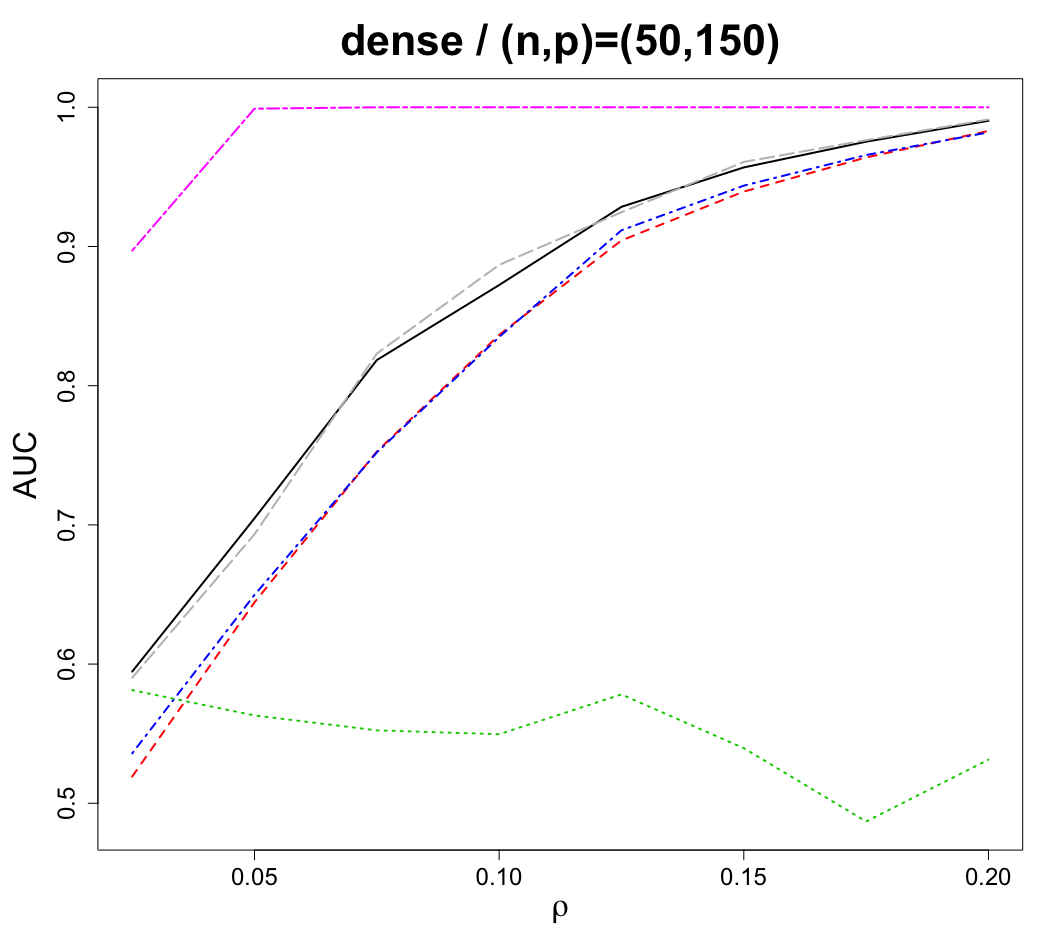}
	\includegraphics[width=5.cm,height=4.7cm]{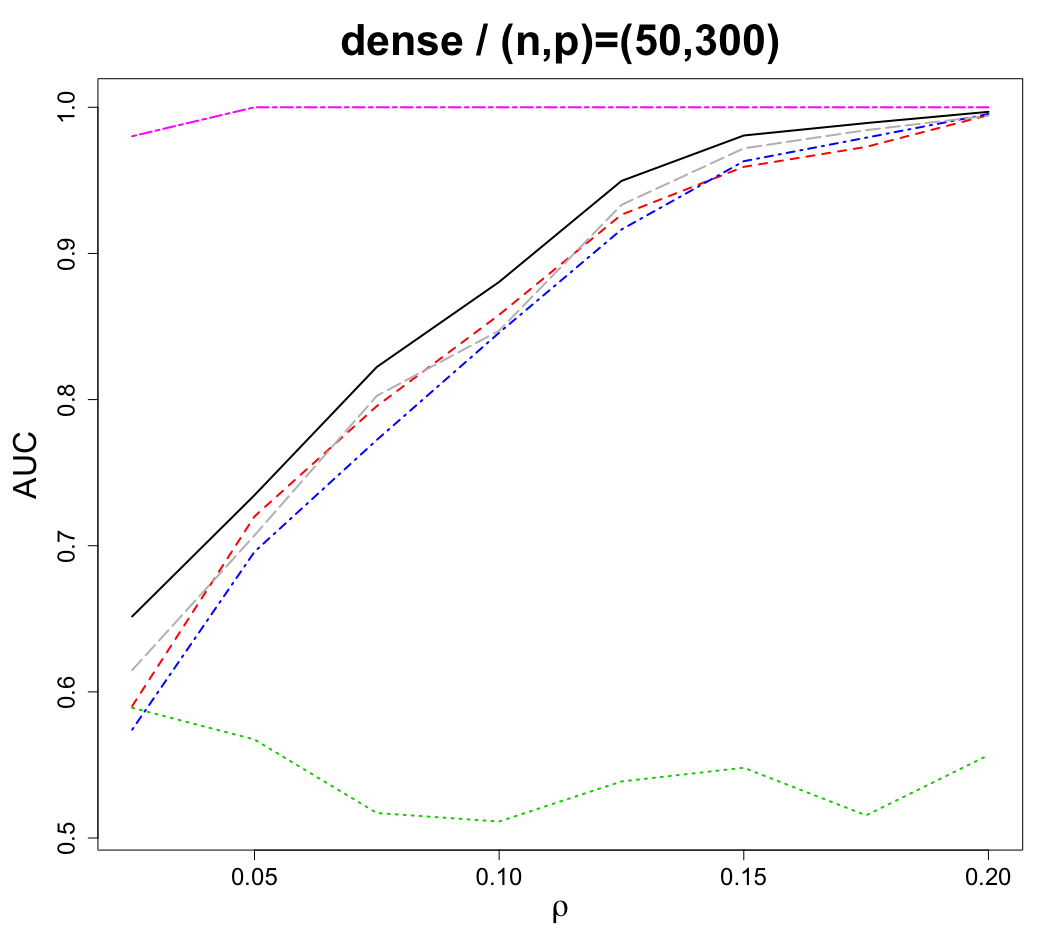}
	\includegraphics[width=5.cm,height=4.7cm]{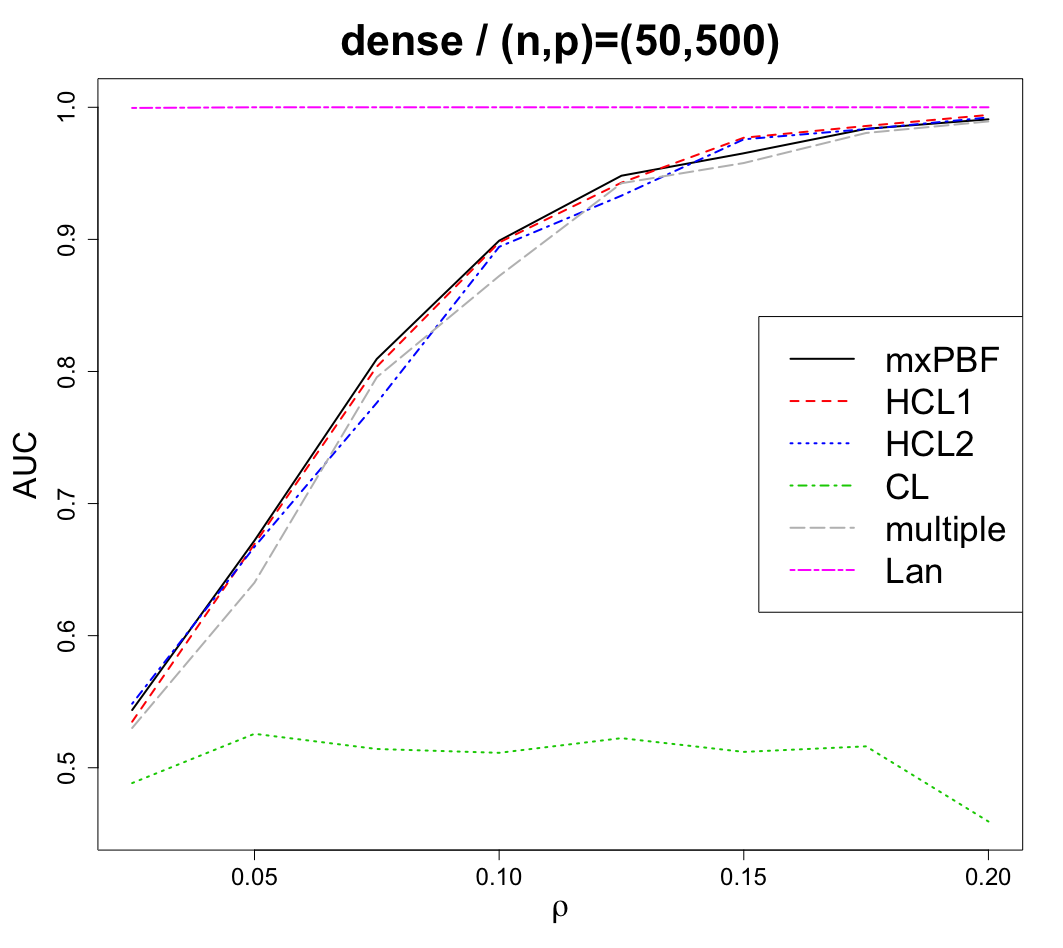}
	\includegraphics[width=5.cm,height=4.7cm]{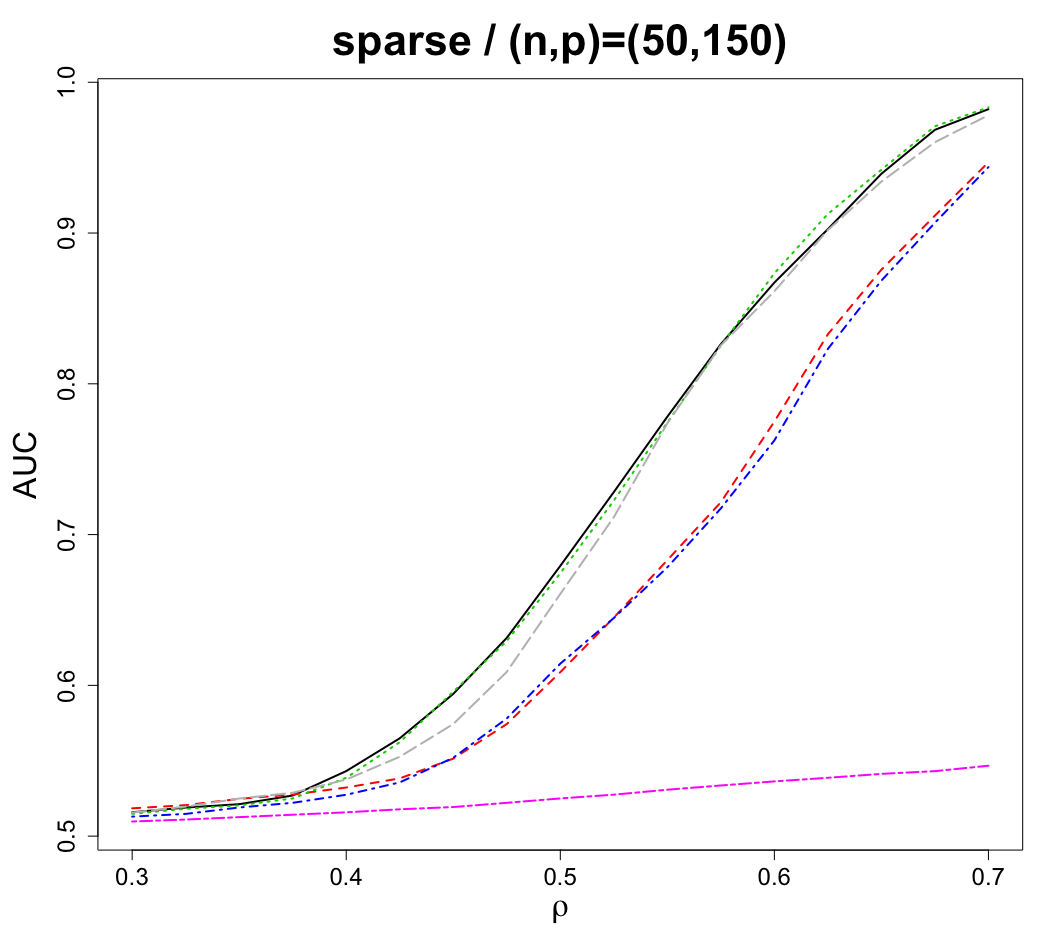}
	\includegraphics[width=5.cm,height=4.7cm]{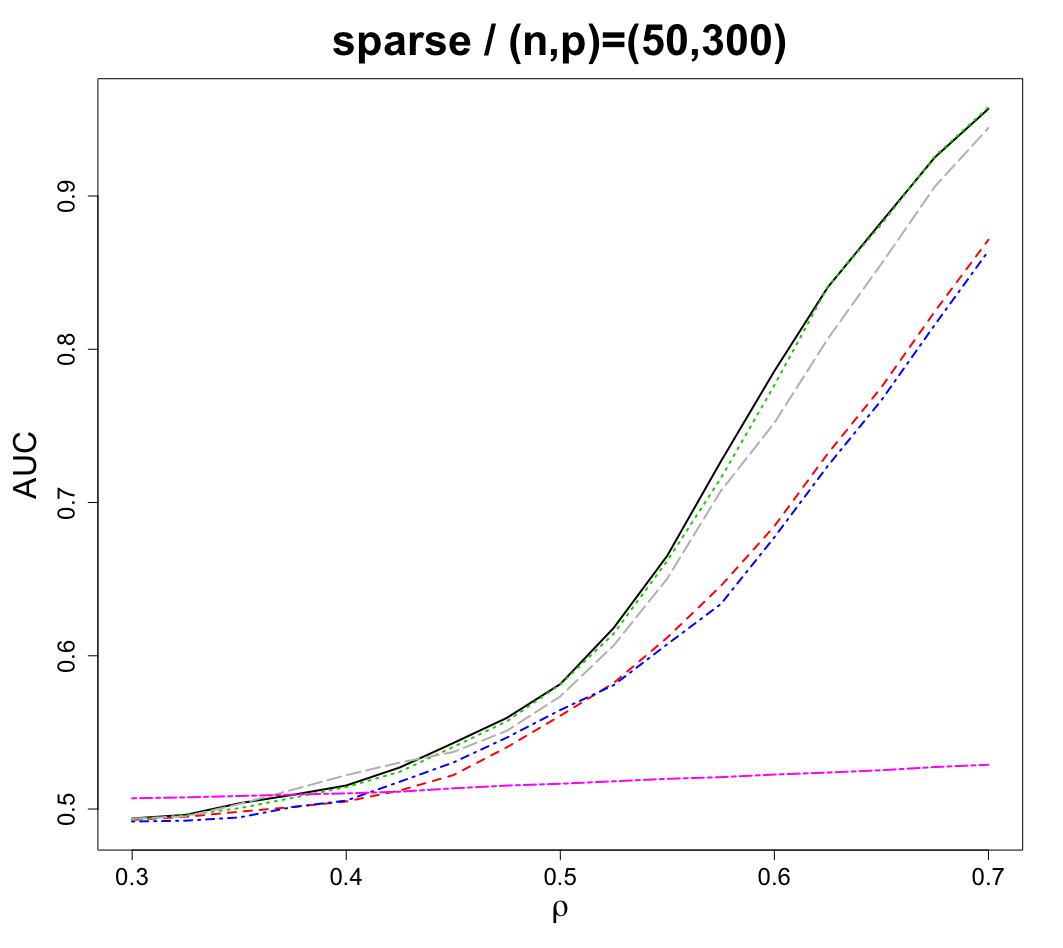}
	\includegraphics[width=5.cm,height=4.7cm]{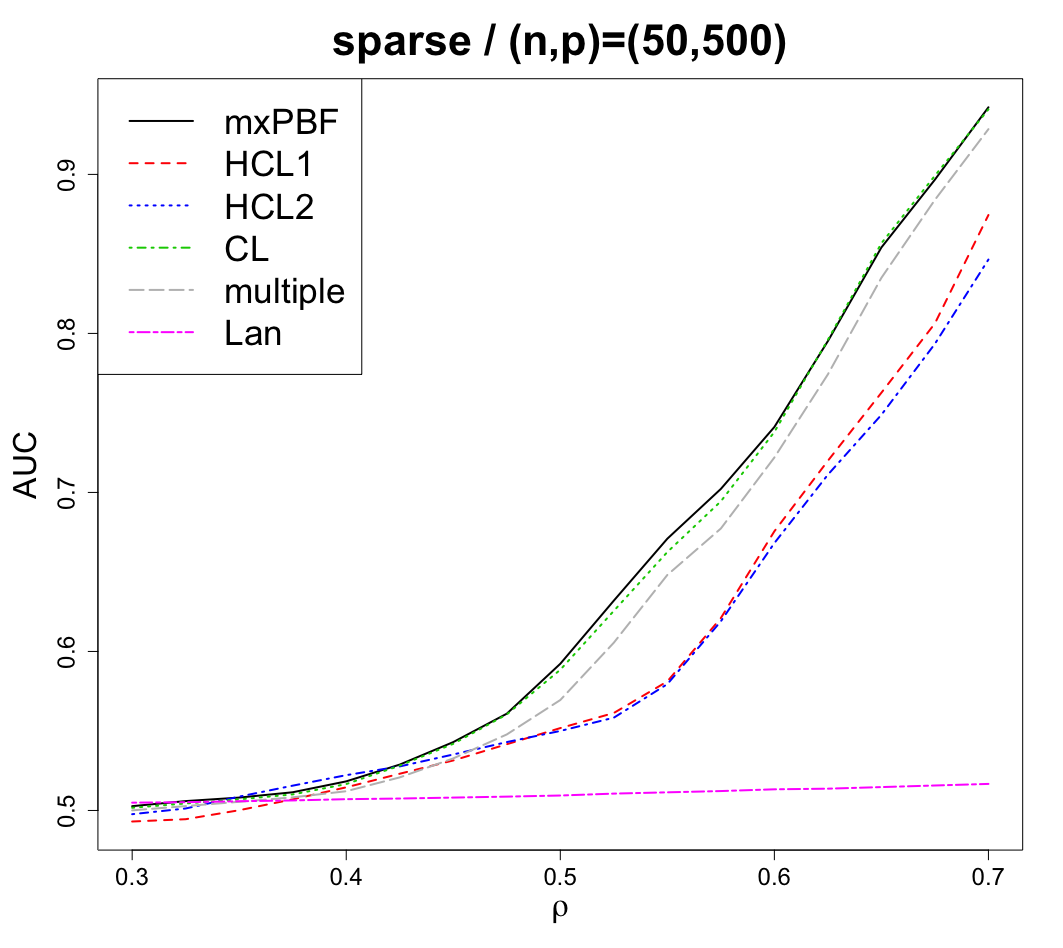}
	\vspace{-.2cm}
	\caption{
		Area under the curves are represented for the tests based on 100 simulated data sets for each hypothesis $H_0: \sigma_{ij}=0$ for any $i\neq j$ and $H_1:$ not $H_0$.
		``dense" and ``sparse" mean that the true covariance matrix $\sg_0$ under $H_1$ were generated from \eqref{al1} and \eqref{al2}, respectively.
		mxPBF, CL and Lan represent the tests proposed in this paper, \cite{chen2018testing} and \cite{lan2015testing}, respectively.
		HCL1 and HCL2 represent the test based on Kendall's tau and Spearman's rho, respectively.
		``multiple" means the frequentist union-intersection test.
	}
	\label{fig:roc3}
\end{figure*}
Figure \ref{fig:roc3} shows the area under the receiver operating characteristic curve for varying  signal strength $\rho$ for each fixed $(n,p)$.
We omit the results of \cite{cai2011limiting}, which were almost identical to our test in every setting. 
As expected, the test of  \cite{lan2015testing} is more powerful against dense alternatives. 
The other tests, except the test of \cite{chen2018testing}, have less power, but work reasonably well as the signal $\rho$ grows.
The test of \cite{chen2018testing} does not work well, likely because \eqref{al1} violates their assumptions.
When $\sg_0 - I_p$ is sparse, the test of \cite{lan2015testing} does not work well even when $\rho$ is large.
The other tests show good results against sparse alternatives, but our test has better performance.

\subsection{Support Recovery using Gene Expression Data}\label{subsec:real_covsel}

To describe the practical performance of the support recovery procedure \eqref{mxPBF_selection}, $\what{S}_{pair}$, we analyzed a data set  from a small  round blue-cell tumor microarray experiment \citep{khan2001classification}.
The data set originally had 6,567 gene expression values, and 2,308 gene expressions were selected by an initial filtering \citep{khan2001classification}.
For comparison purposes, we focus on the preprocessed data used in \cite{rothman2009generalized} and \cite{cai2011adaptive}, consisting of $p=200$ gene expression values for each of $n=64$ training tissue samples. 
There are four types of tumors represented in these tissue samples.
Data were centered prior to analysis.

For pairwise Bayes factors, the hyperparameter was set at $\alpha = 4.01 (1 - 1/\log n)$. 
We used cross-validation to select $C_{sel}$. 
Let $n$ be the number of observations for a given data set.
We randomly divided the data 50 times into two subsamples with size $n_1=\lceil n/3 \rceil$ and $n_2 = n-n_1$ as a test set and training set, respectively.
Denote $I_1$ and $I_2$ as indices for the test set and training set, respectively, thus, $|I_1|= n_1$, $|I_2|=n_2$ and $I_1 \cup I_2 = \{1,\ldots, n\}$.
Let $\hat{S}_j(C_{sel})$ be the estimated support for the $j$th column of the covariance matrix via pairwise Bayes factors, based on $\{ X_i \}_{i \in I_2}$ and a given threshold $C_{sel}$.
We calculated the averaged mean squared error
\bea
MSE(C_{sel}) &=& \sum_{j=1}^p \sum_{l \in \hat{S}_j(C_{sel}) } \Big\{ \sum_{i \in I_1} (X_{ij} -  X_{il} \what{\beta}_{jl} )^2 /(n_1-1) \Big\} /|\hat{S}_j(C_{sel})|,
\eea
where $\what{\beta}_{jl}$ is a least square estimate with respect to the dependent variable $\{X_{ij}\}_{i\in I_1}$ and covariate $\{X_{il} \}_{i\in I_1}$.
The threshold $C_{sel}$ was varied from $-7$ to $10$ with increment $0.2$, and we selected $\what{C}_{sel}$ which minimizes $50^{-1} \sum_{\nu=1}^{50} MSE_\nu(C_{sel})$, where $MSE_\nu(C_{sel})$ is the averaged mean squared error based on the $\nu$th split.

We compared our method with generalized thresholding estimators of \cite{rothman2009generalized} and \cite{cai2011adaptive}. 
\cite{rothman2009generalized} used a universal threshold $\lambda = \delta (\log p/n)^{1/2}$, while \cite{cai2011adaptive} used an individual threshold $\hat{\lambda}_{ij} = \delta ( \hat{\theta}_{ij}\log p/n)^{1/2}$ with a data-dependent $\hat{\theta}_{ij}$.
We denote thresholding estimators proposed by \cite{rothman2009generalized} and \cite{cai2011adaptive} by $\what{\sg}_\delta$ and $\what{\sg}_\delta^\star$, respectively.
We used the adaptive lasso thresholding rule, $s_\lambda(\sigma) = \sigma \max( 1- |\lambda/\sigma|^\eta , 0)$ with $\eta=4$, because it gave good support recovery results in simulation studies in \cite{rothman2009generalized} and \cite{cai2011adaptive}.
We adopted the cross-validation method described in Section 4 of \cite{cai2011adaptive} to select $\delta$ and denote the selected tuning parameter by $\hat{\delta}$.

\begin{figure*}[!tb]
	\centering
	\includegraphics[width=10cm,height=7cm]{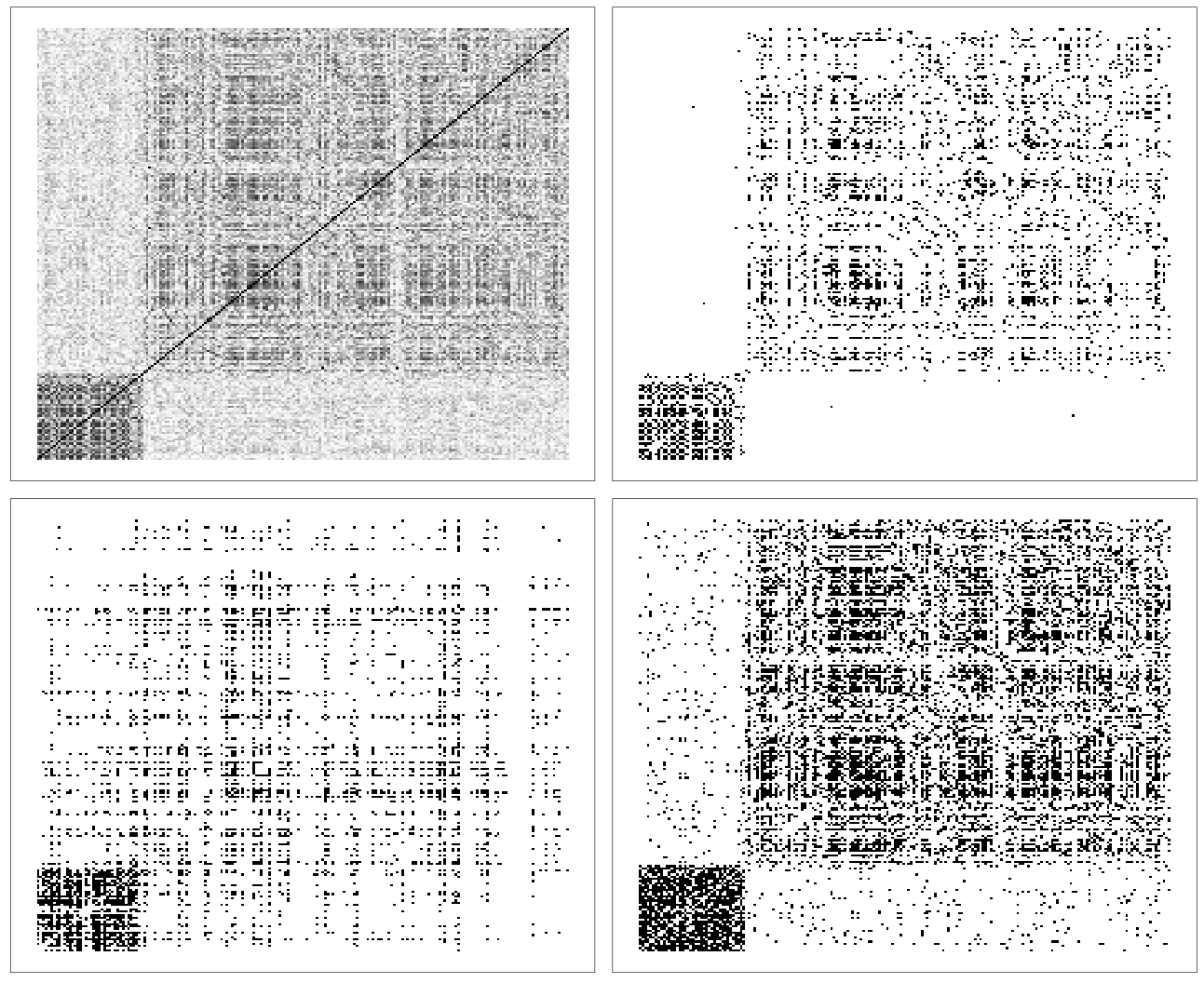}
	\caption{
		The absolute sample correlation matrix (top left) and estimated supports from various methods.
		Clockwise from the top right are plots for the estimated supports based on $\what{S}_{pair,\what{C}_{sel}}$, $\what{\sg}_{\hat{\delta}}^\star$ and $\what{\sg}_{\hat{\delta}}$, respectively.
	}
	\label{fig:estimated_supports}
\end{figure*}
\begin{figure*}[!tb]
	\centering
	\includegraphics[width=14cm,height=5.3cm]{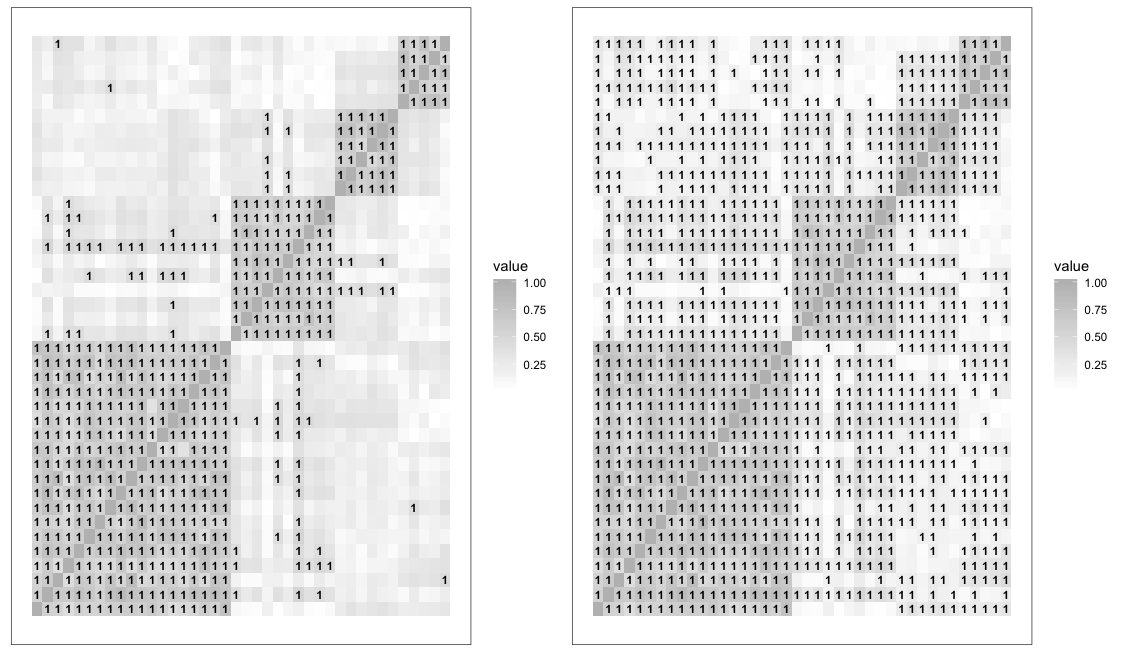}
	\caption{
		The ordered absolute sample correlation matrix and estimated supports for top 40 genes, with 1's representing the estimated supports from  $\what{S}_{pair,\what{C}_{sel}}$ (left) and  $\what{\sg}_{\hat{\delta}}^\star$ (right).
	}
	\label{fig:two_supports}
\end{figure*}
Figure \ref{fig:estimated_supports} shows the support recovery results and the absolute sample correlation matrix.
The estimated supports based on $\what{S}_{pair,\what{C}_{sel}}$, $\what{\sg}_{\hat{\delta}}^\star$ and $\what{\sg}_{\hat{\delta}}$ are represented.
One can see that $\what{S}_{pair,\what{C}_{sel}}$ and  $\what{\sg}_{\hat{\delta}}^\star$ show the clustering structure between informative (top 40) and non-informative (bottom 160) genes, while the structure is somewhat blurred in $\what{\sg}_{\hat{\delta}}$.
To compare $\what{S}_{pair,\what{C}_{sel}}$ and  $\what{\sg}_{\hat{\delta}}^\star$ in more detail, we further focused on the top 40 genes.
We applied hierarchical clustering to the genes based on the complete linkage method using \verb|R| function \verb|hclust|, and the genes were ordered according to the clustering result.
Figure \ref{fig:two_supports} shows the ordered absolute sample correlation matrix and estimated supports for the top 40 genes.
The clustering result suggests that there are four clusters, consistent with the four tumor types.
Both support recovery procedures detect significant blocks in the sample correlation matrix. 
However, our support recovery procedure shows the clustering structure much clearer, while $\what{\sg}_{\hat{\delta}}^\star$ gives a blurred structure due to a dense support estimate.
The estimated support based on pairwise Bayes factors has the advantage of producing a sparser, and hence potentially more interpretable, estimate of support.


\section{Discussion}\label{sec:disc}


We have focused on covariance matrix structure testing in this paper, but the  maximum pairwise Bayes factor idea can be easily applied to other related settings.
For example, testing differences across groups in  high-dimensional mean vectors is an interesting possibility.
When the two mean vectors are almost the same but  differ only at a few locations, a maximum pairwise Bayes factor approach should have relatively high power.
Similarly, it can be applied to the high-dimensional two-sample covariance test.
Two covariances from two populations may differ only in a small number of entries.

There are some possible generalizations of the pairwise Bayes factor idea.
To accelerate the speed of computation, a random subsampling method can be used instead of calculating the pairwise Bayes factor for every single pair $(i,j)$.
It should be interesting to develop a suitable random subsampling or random projection scheme achieving desirable theoretical properties.
Especially when $p$ is huge, it will effectively reduce the computational complexity.
The maximum pairwise Bayes factor approach is also trivially parallelizable.
Another possibility is considering alternative combining approaches to the max in merging the information from every pairwise Bayes factor.
If there are many weak non-zero covariances, then the average or summation may be preferable to the maximum.
A suitable modification to learn parameters in the combining operator can potentially make the test powerful to a broad class of alternative hypotheses.


\appendix

\section{Testing  pairwise independence}

As a by-product of Theorem 3, 
when pairwise independence testing (3.12) 
itself is of interest, we suggest a pairwise Bayes factor $\tilde{B}_{10} (\tilde{X}_i, \tilde{X}_j)$, which can be shown to be consistent.
For consistency under the alternative hypothesis, we assume 
\bean\label{betamin_pair}
\sigma_{0,ij}^2 &\ge&  \frac{C_4 \sigma_{0,jj} }{1- 2\epsilon_0 {C_1}^{1/2}} \left\{ \frac{9C_1 \tau_{0,ij}^2}{(1-C_3)^2} \vee \frac{ \alpha(1+\gamma)   (1+4\epsilon_0 {C_1}^{1/2}) \sigma_{0,ii}  }{C_3 } \right\} \frac{\log n}{n}  \quad\quad\quad
\eean
for constants $C_1>0, 0<C_3<1$ and $C_4>1$ defined in Section 3.1. 
If we substitute $\log n$ in the above condition with $\log (n\vee p)$, it coincides with condition (A4).
The proof of Corollary \ref{cor:pair_indep} follows from that of Theorem 3, 
and  thus it is omitted.

\begin{corollary}\label{cor:pair_indep}
	Consider model (1.1) 
	and a hypothesis testing problem (3.12) 
	for a given pair $(i,j)$  such that $i\neq j$.
	Suppose we use the prior $\pi(\tau_{ij}^2) \propto\tau_{ij}^{-2}$ under both $H_{0,ij}$ and $H_{1,ij}$, and the prior $\pi(a_{ij} \mid \tau_{ij}^2)$ defined in (2.5) 
	under $H_{1,ij}$ with $\gamma = n^{-\alpha}$ for some positive constant $\alpha$. 
	Then under $H_{0,ij}: \sigma_{ij} =0$, for some constant $c>0$,
	\bea
	\tilde{B}_{10} (\tilde{X}_i, \tilde{X}_j) &=&   O_p\big( n^{-c} \big) .
	\eea
	If, under $H_{1,ij}: \sigma_{ij} \neq 0$, at least one of $\sigma_{0,ij}$ and $\sigma_{0,ji}$ satisfies \eqref{betamin_pair}, for some constant $c'>0$,
	\bea
	\tilde{B}_{10} (\tilde{X}_i, \tilde{X}_j)^{-1} &=&   O_p\big( n^{-c'}\big) .  
	\eea
\end{corollary}

\section{Simulation Study: Support Recovery}

The proposed support recovery procedure, $\what{S}_{pair, C_{sel}}$, consistently recovers the true support $S(\sg_0)$ for any constant threshold $C_{sel}$ (Theorem 4). 
However, in practice, the choice of the threshold is crucial.
In this section, we investigate the quality of $\what{S}_{pair, C_{sel}}$ as a function of the threshold $C_{sel}$ to assess the importance of the choice of $C_{sel}$.
Furthermore, the performance of the cross-validation-based threshold $\what{C}_{sel}$ proposed in Section 4.3 
also will be demonstrated.

Two structures of covariance matrices were investigated.
In the first setting, we consider
\bea
\sigma_{0,ij}^* &=& 2 \max\Big(  1- \frac{|i-j|}{10}, \, 0 \Big) I \left\{ |i-j| \le 5, \, (i\vee j) \le \frac{p}{2}  \right\}
\eea
for $i \neq j$.
In the second setting, we consider
\bea
\sigma_{0,ij}^* &=& 2 \max\Big(  1- \frac{|i-j|}{10}, \, 0 \Big) I \left\{ |i-j| \le 5, \, i \le \frac{p}{2}  \right\} \\
&&+\,\, 2 \max\Big(  1- \frac{|i-j|}{20}, \, 0 \Big) I \left\{ |i-j| \le 10, \, i > \frac{p}{2}  \right\}
\eea
for $i<j$.
In both settings, we set $\sigma_{0,ii}^* = 1$ and $\sigma_{0,ij}^* = \sigma_{0,ji}^*$, where $\sg_0^* = (\sigma_{0,ij}^*)$. 
Let $\sg_0 = (\sigma_{0,ij})$ be the true covariance matrix.
If $\sg_0^*$ is positive definite, $\sg_0 = \sg_0^*$ is used, and if $\sg_0^*$ is not positive definite, $\sg_0 = (\sigma_{0,ij})$ is used where $\sigma_{0,ij} = \sigma_{0,ij}^*$ and $\sg_{0,ii} = \sg_{0,ii}^* - \lambda_{\min}(\sg_0)  + 0.01$ for all $1 \le i \neq j \le p$.
The data were generated from $N_p(0, \sg_0)$ with $n=50, 100$ and $p=100, 300$.
To illustrate the performance of the estimated support $\what{S}_{pair, C_{sel}}$, we consider (i) the Matthews correlation coefficient,
\bea
MCC &=& \frac{TP \times TN - FP \times FN}{\{(TP+FP)(TP+FN)(TN+FP)(TN+FN)\}^{1/2}},
\eea
and (ii) the  number of errors, $FP+FN$, where $TP, TN, FP$ and $FN$ are true positive, true negative, false positive and false negative, respectively.
\begin{figure*}[!tb]
	\centering
	\includegraphics[width=4.cm,height=4.5cm]{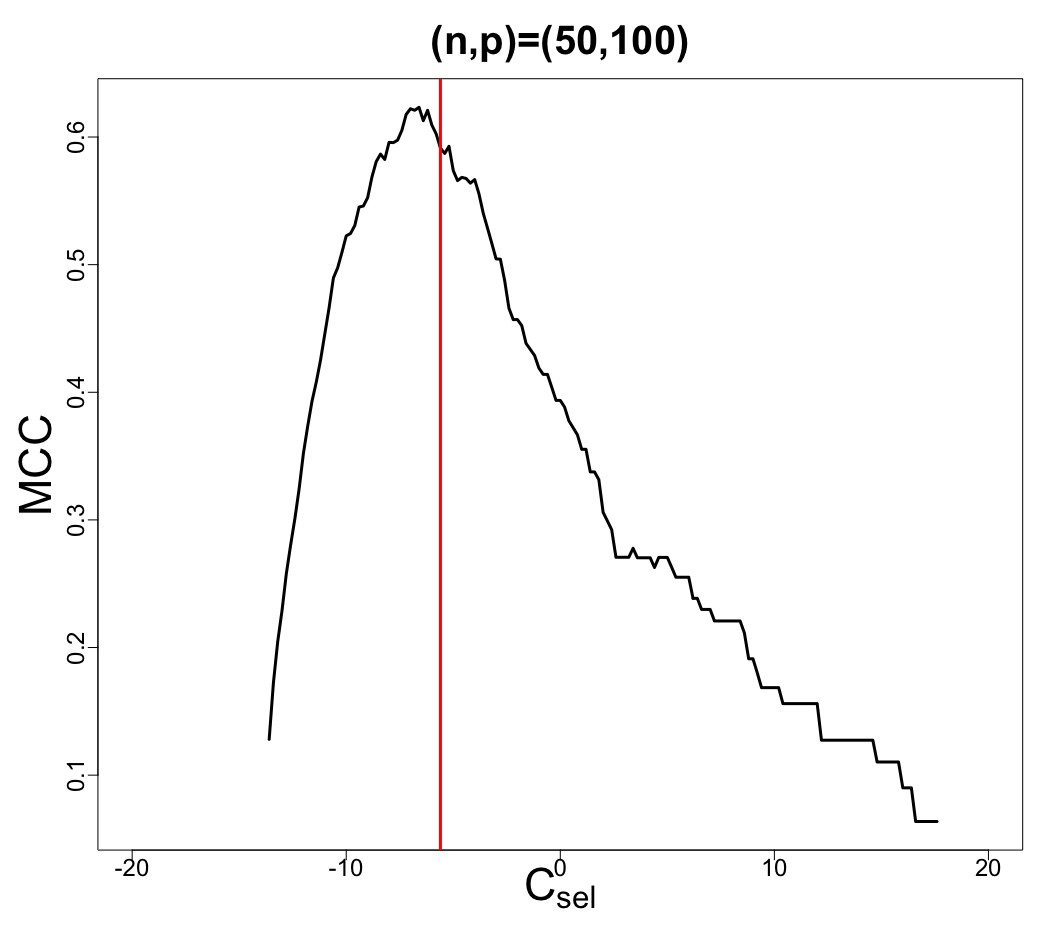}
	\includegraphics[width=4.cm,height=4.5cm]{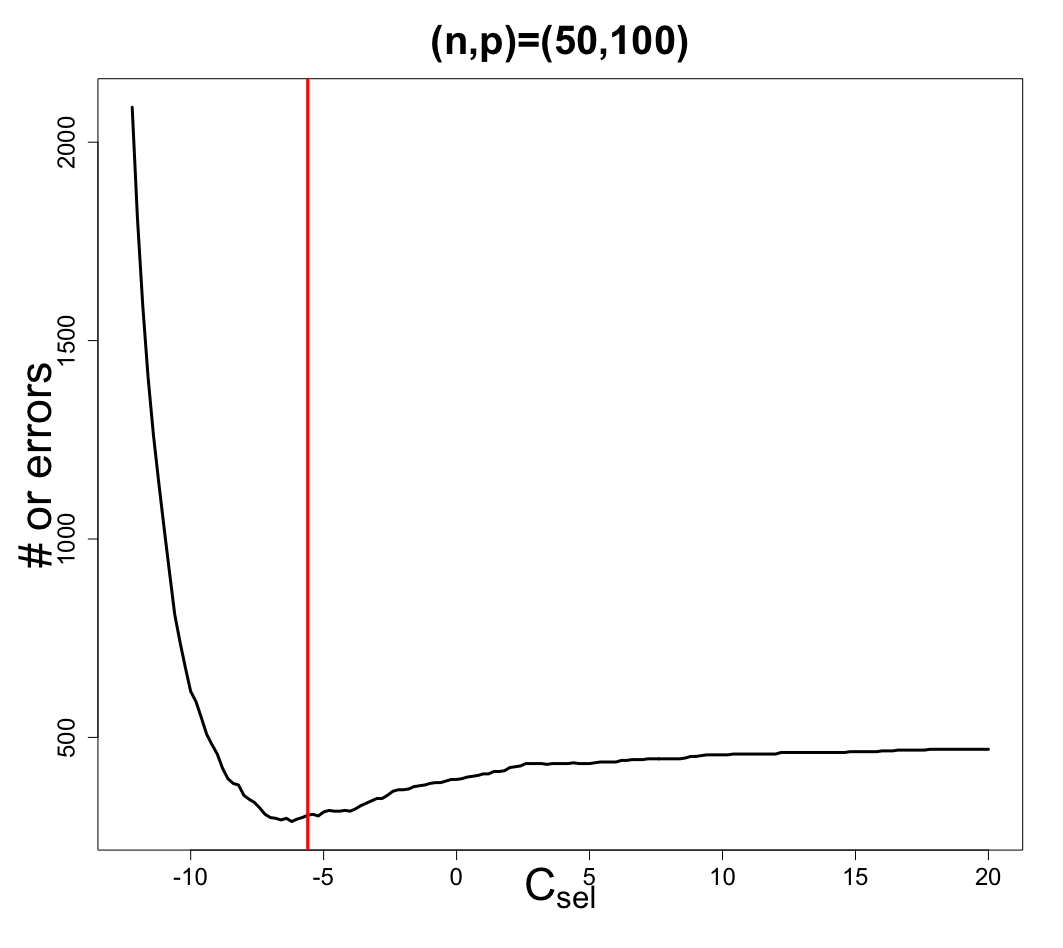}
	\includegraphics[width=4.cm,height=4.5cm]{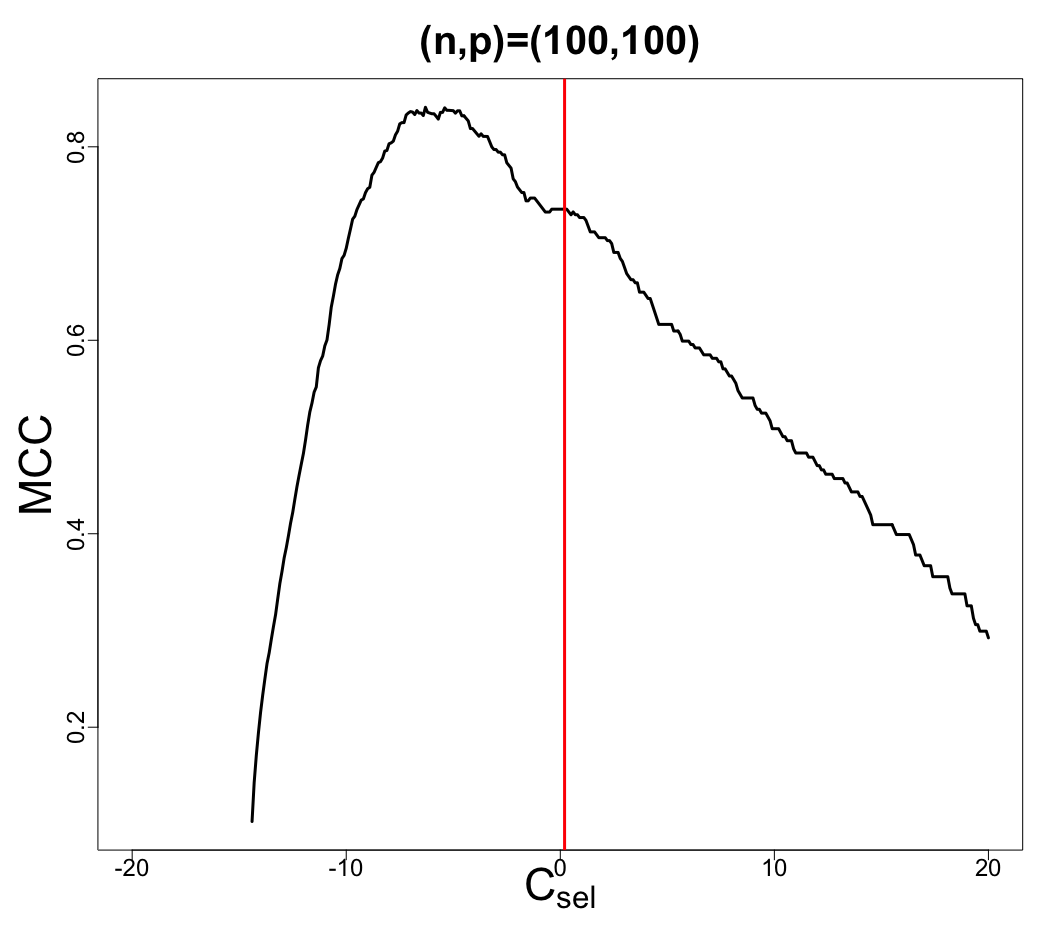}
	\includegraphics[width=4.cm,height=4.5cm]{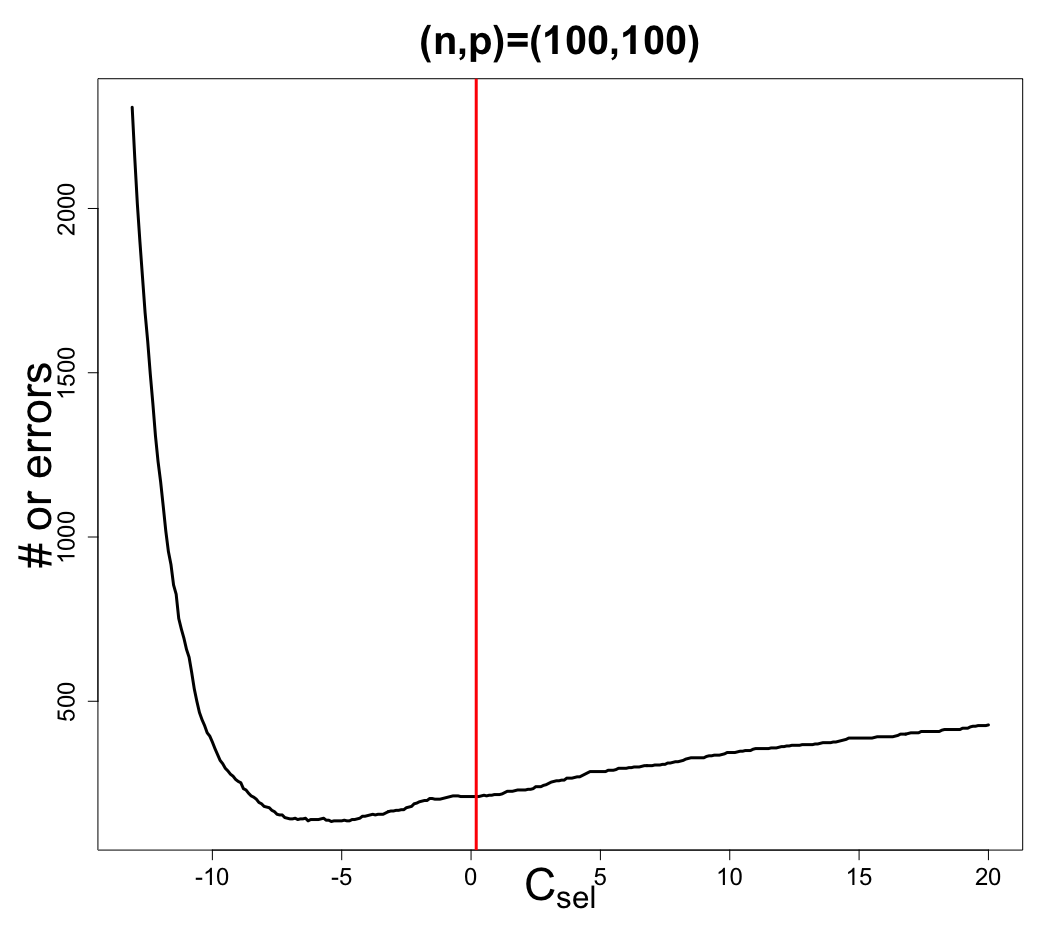}
	\includegraphics[width=4.cm,height=4.5cm]{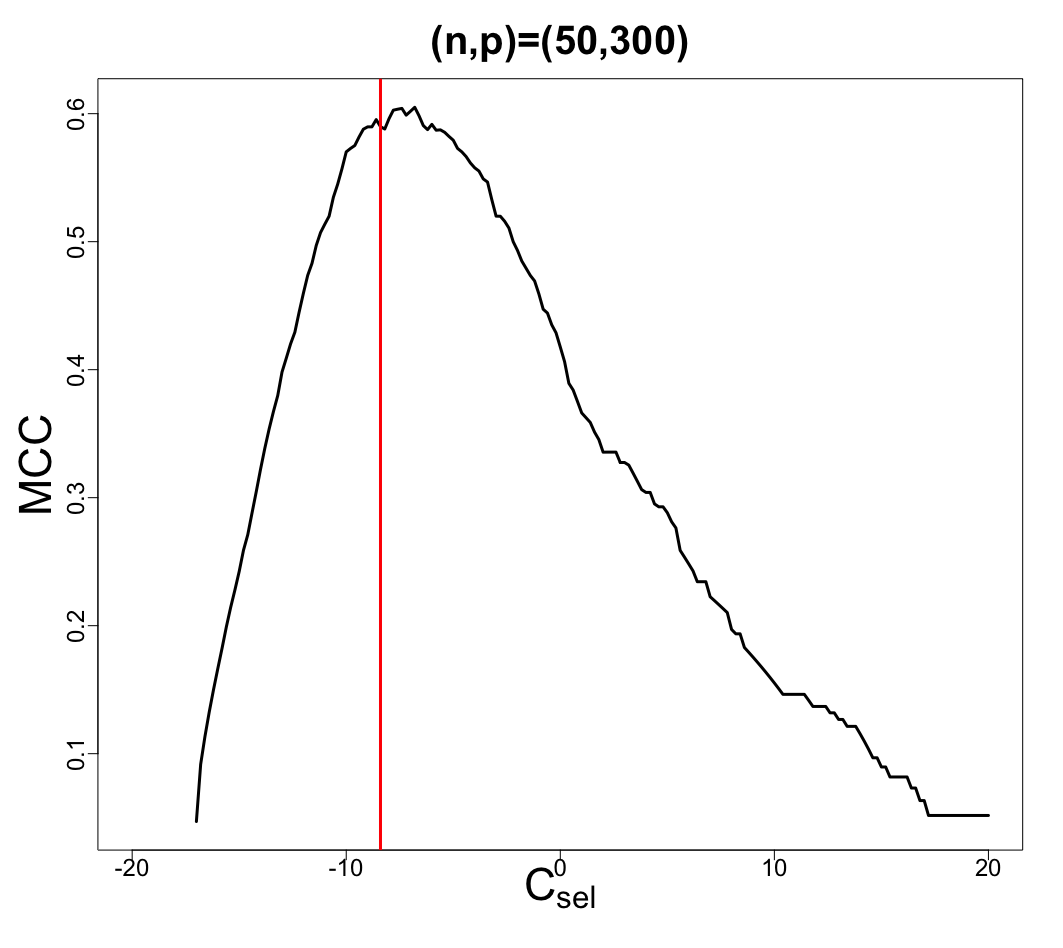}
	\includegraphics[width=4.cm,height=4.5cm]{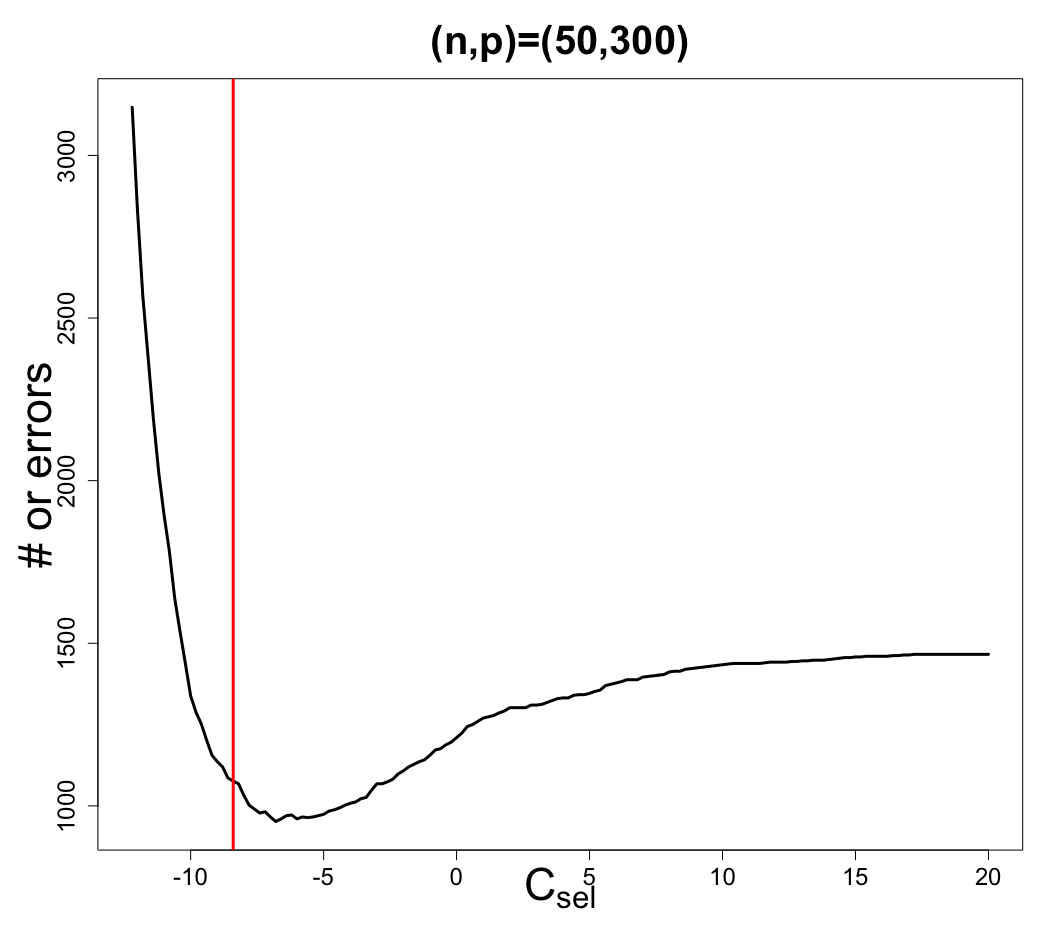}
	\includegraphics[width=4.cm,height=4.5cm]{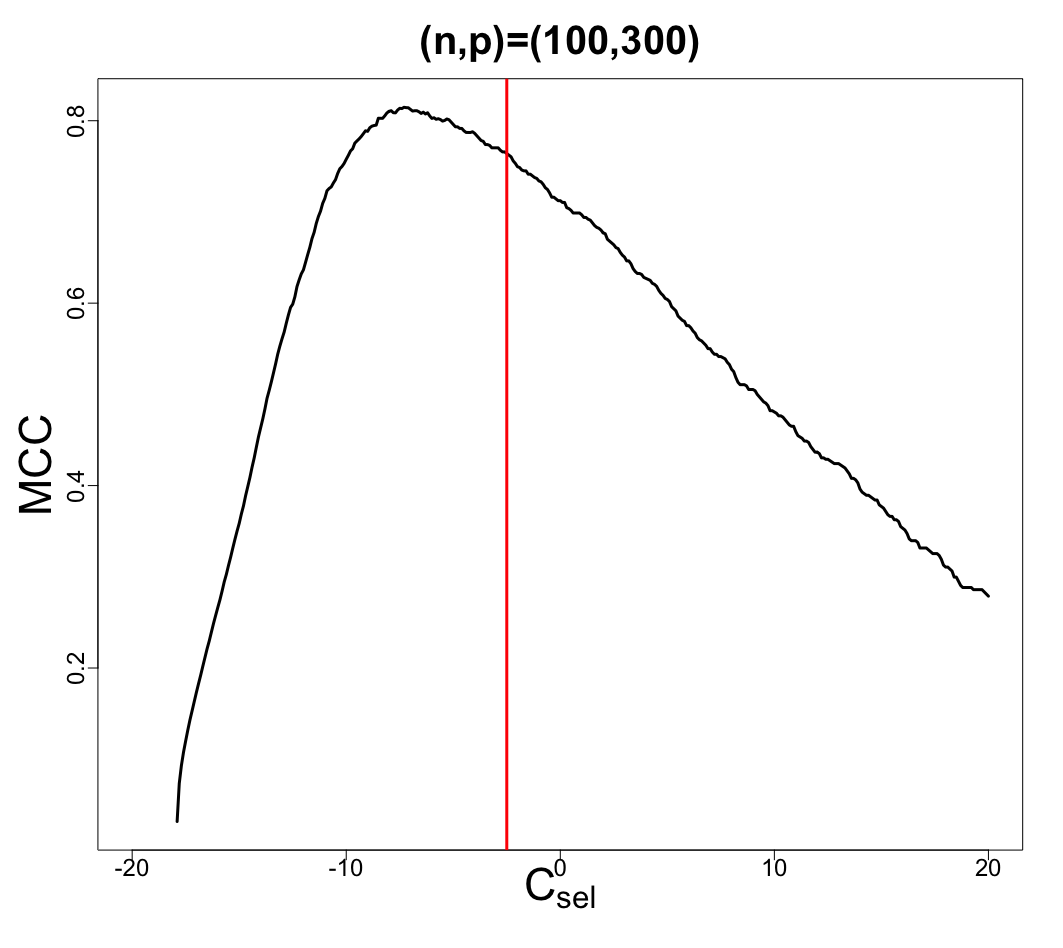}
	\includegraphics[width=4.cm,height=4.5cm]{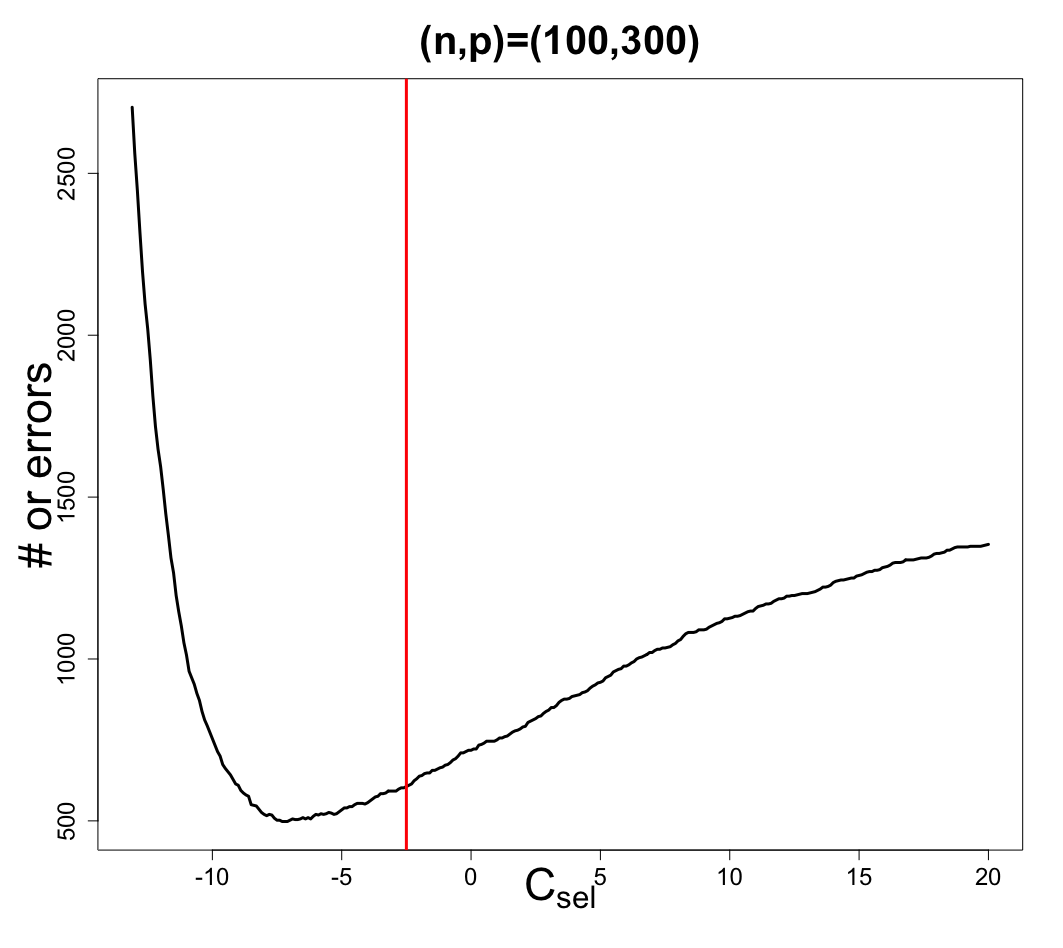}
	\vspace{-.2cm}
	\caption{
		MCC (Matthews correlation coefficient) and the number of errors for the first setting.
		The red vertical line is the cross-validation-based threshold $\what{C}_{sel}$.
	}
	\label{fig:supp1}
\end{figure*}
\begin{figure*}[!tb]
	\centering
	\includegraphics[width=4.cm,height=4.5cm]{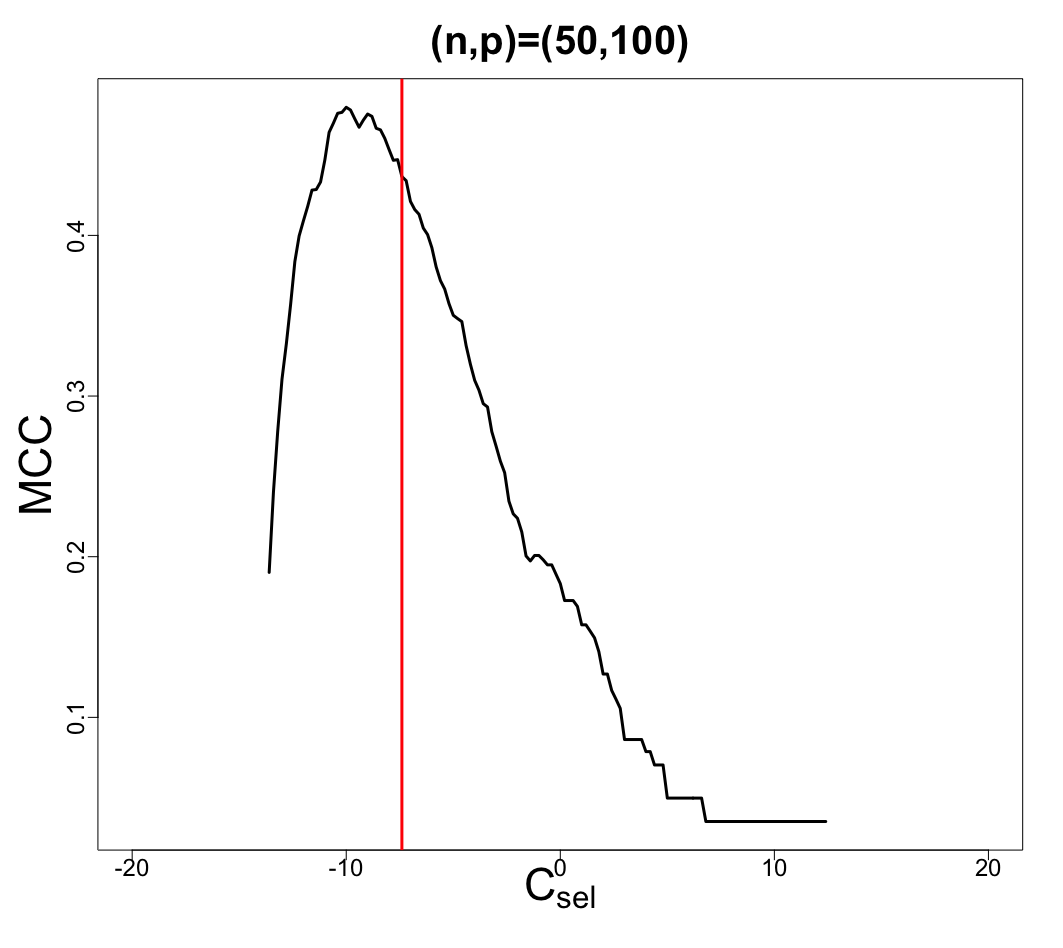}
	\includegraphics[width=4.cm,height=4.5cm]{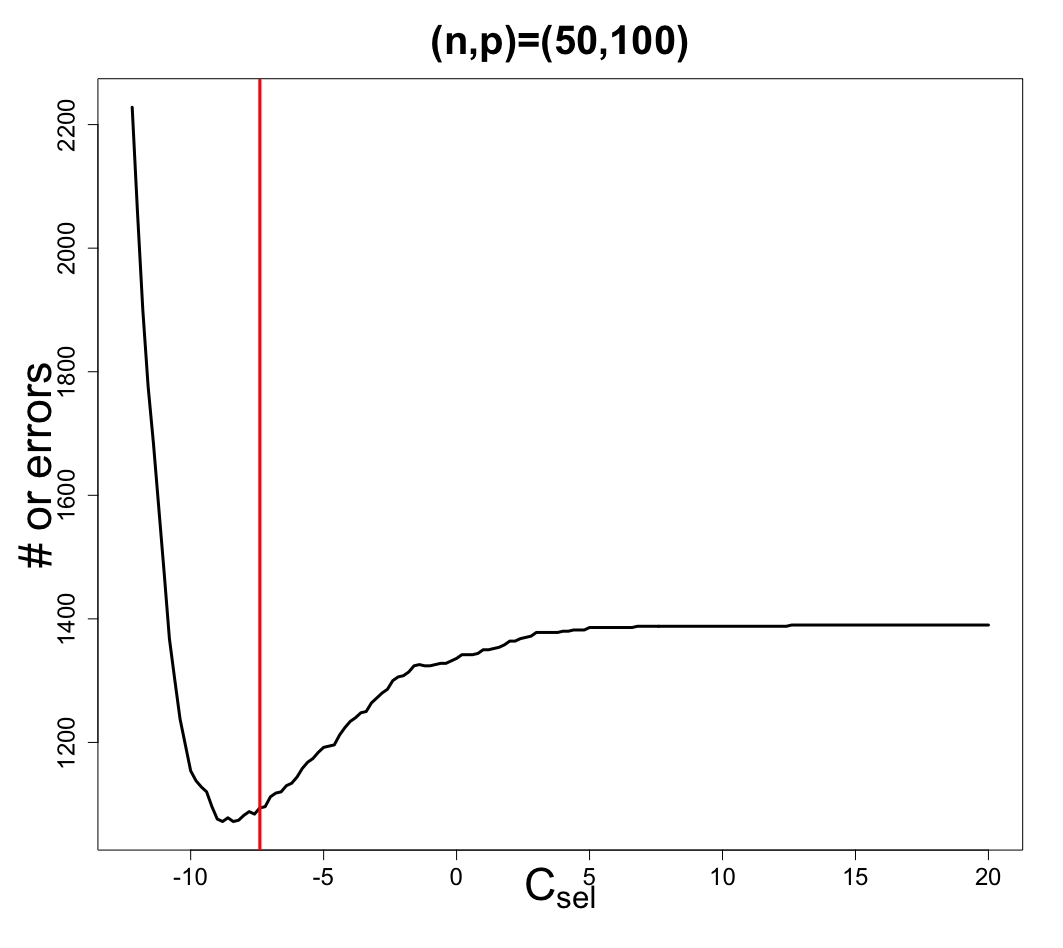}
	\includegraphics[width=4.cm,height=4.5cm]{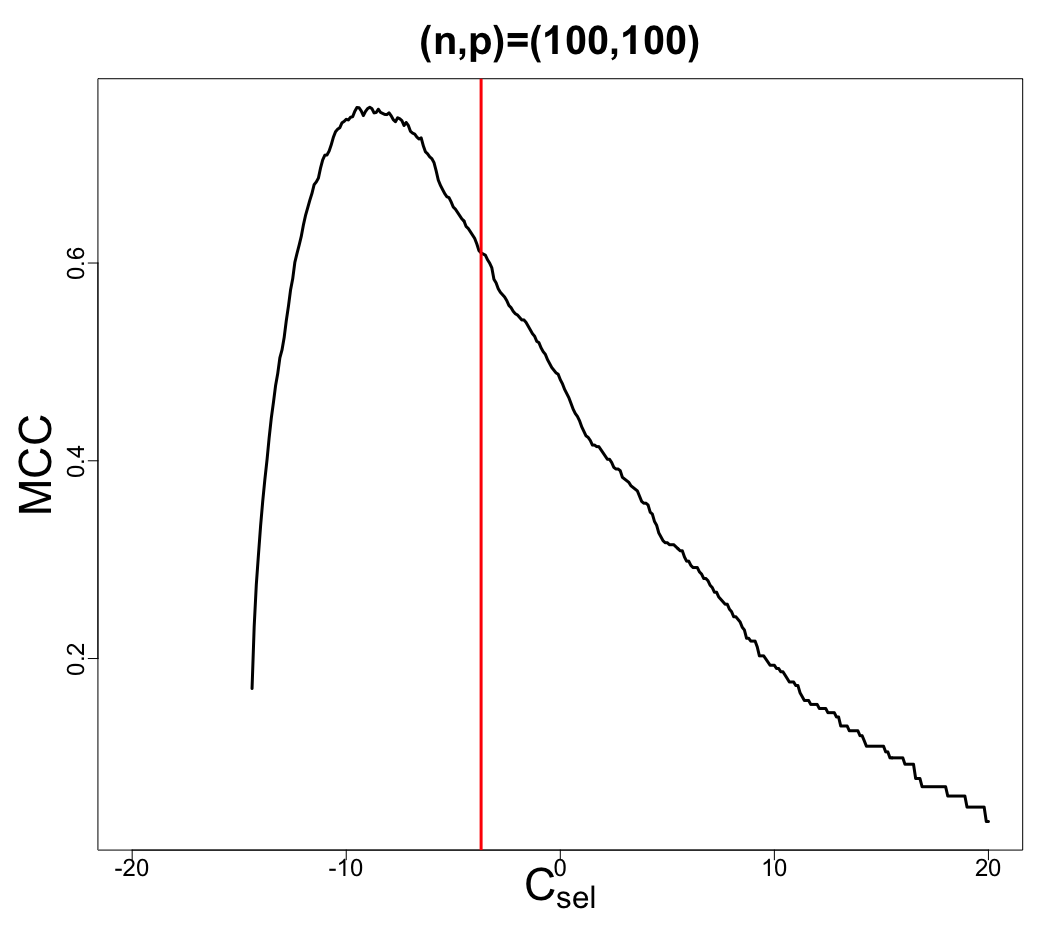}
	\includegraphics[width=4.cm,height=4.5cm]{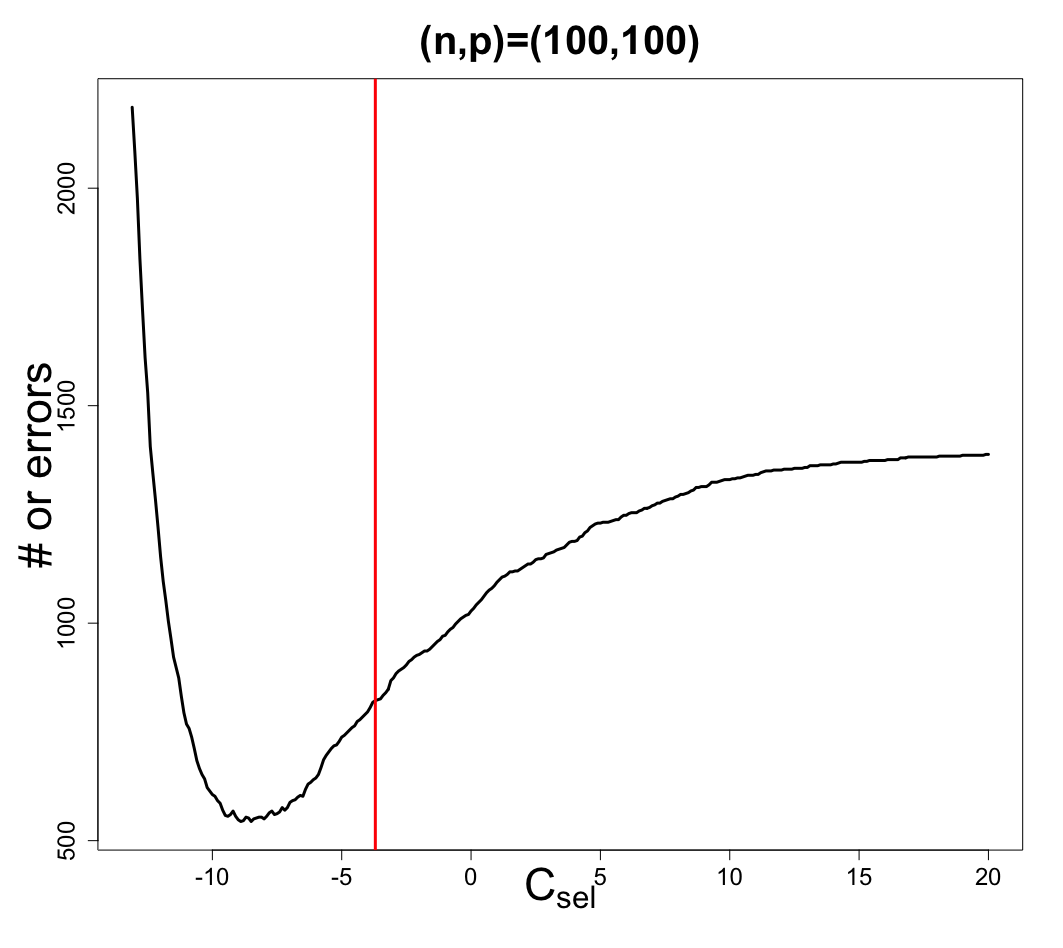}
	\includegraphics[width=4.cm,height=4.5cm]{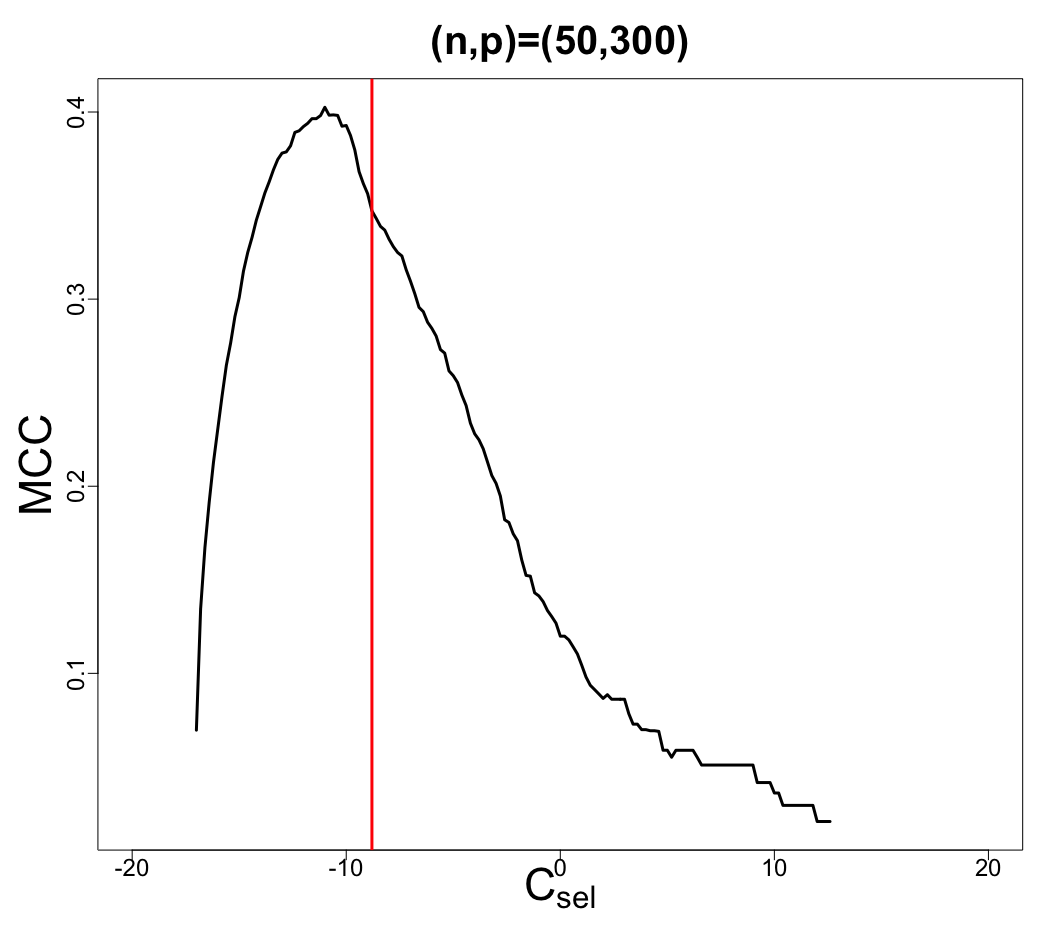}
	\includegraphics[width=4.cm,height=4.5cm]{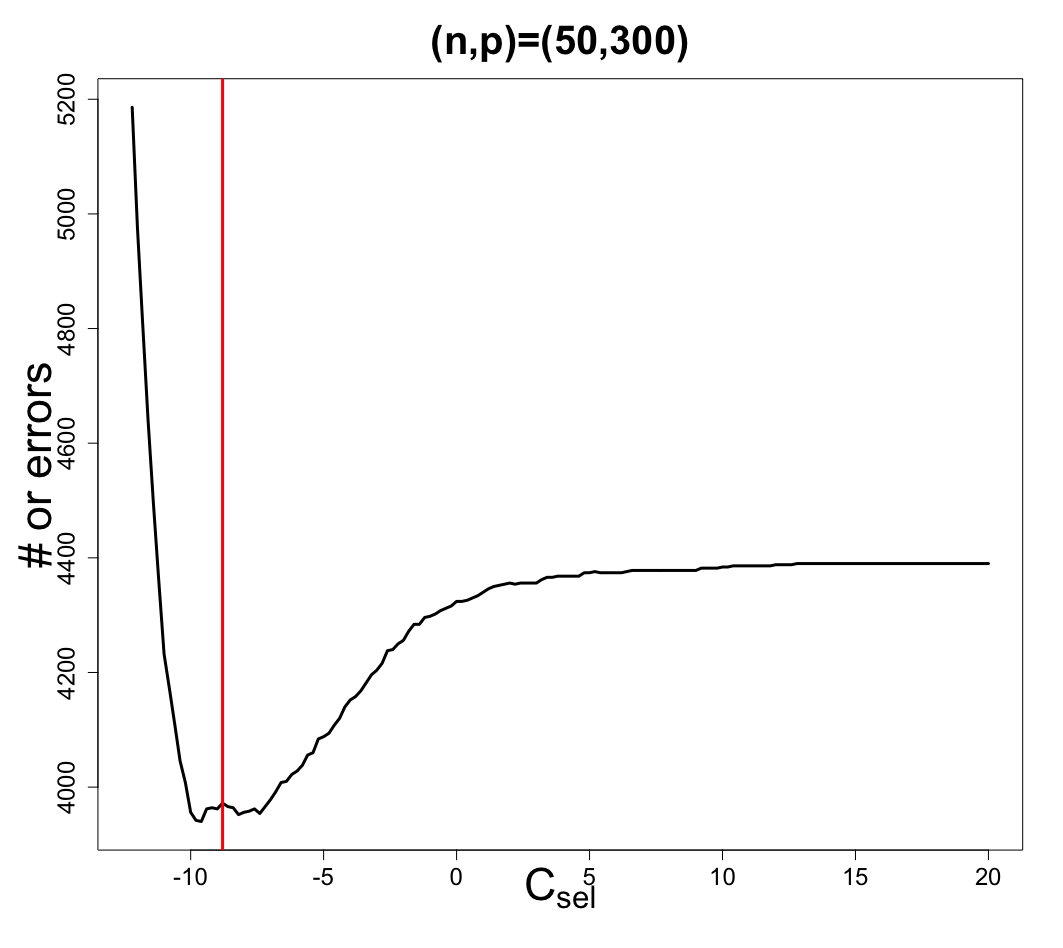}
	\includegraphics[width=4.cm,height=4.5cm]{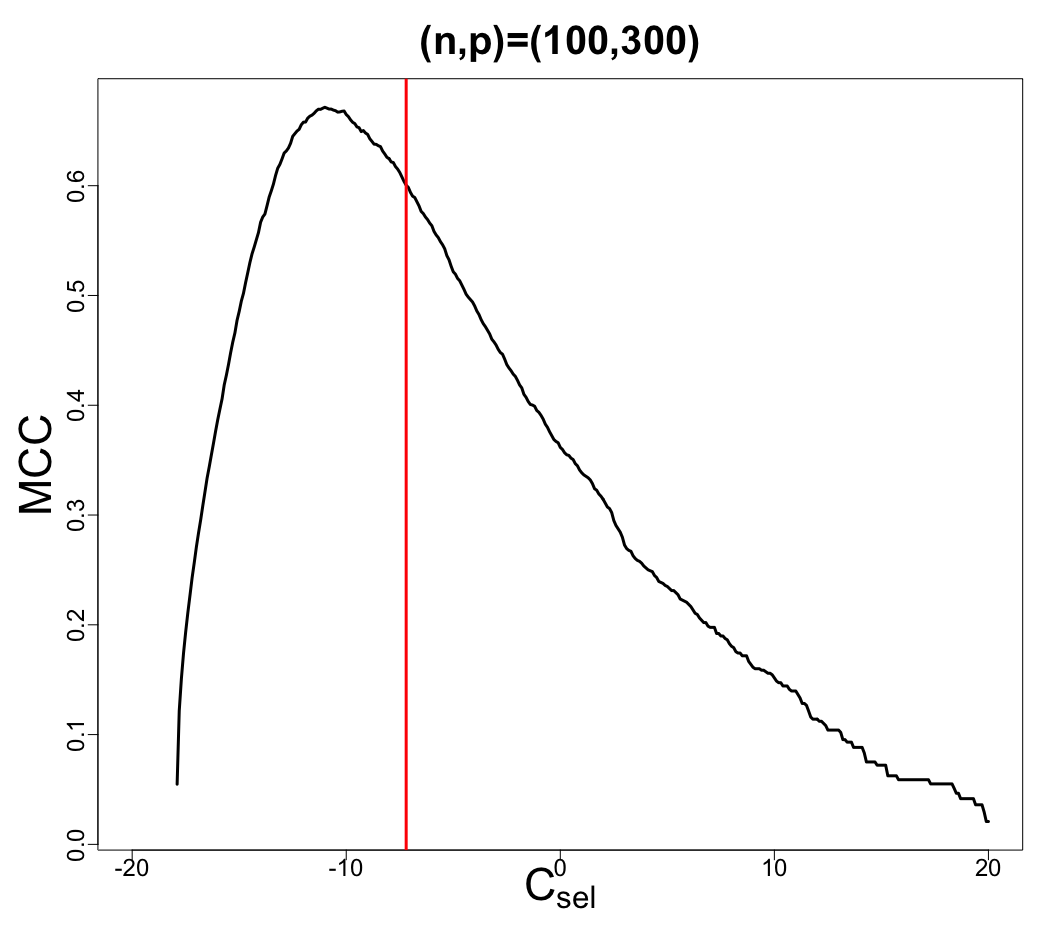}
	\includegraphics[width=4.cm,height=4.5cm]{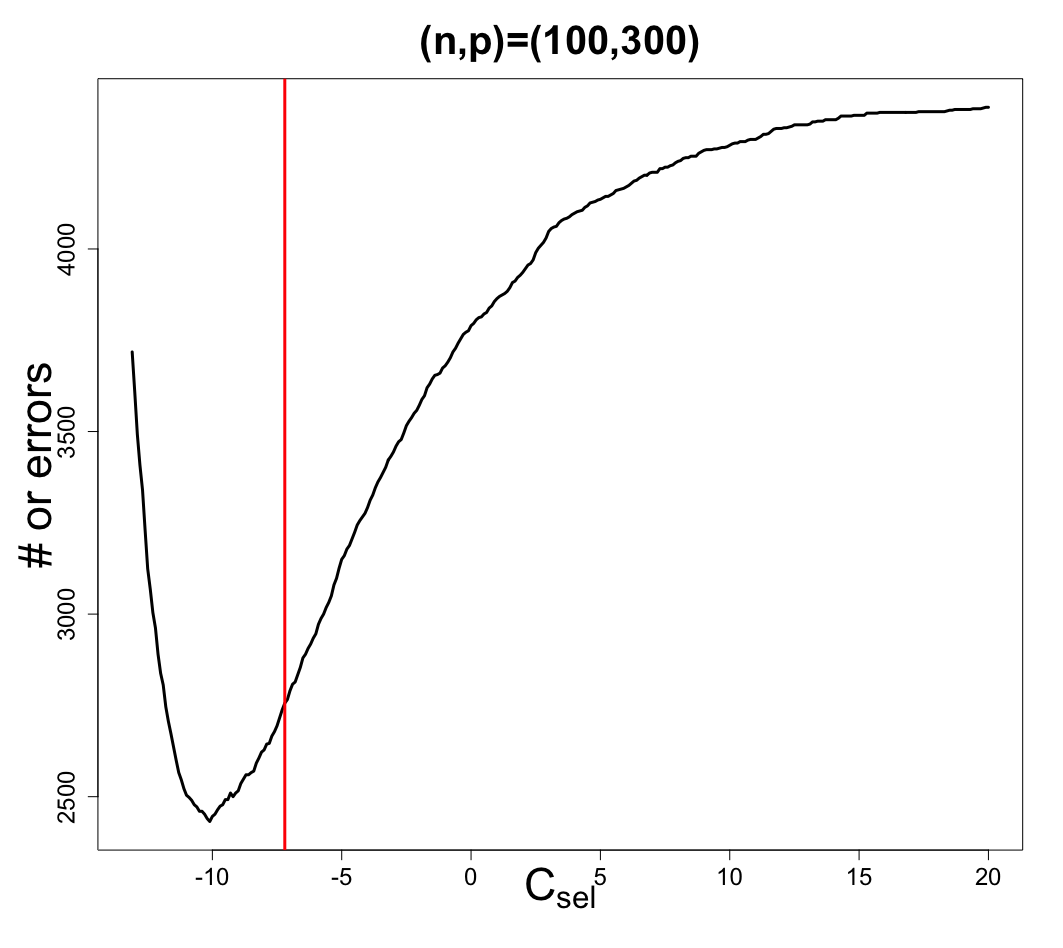}
	\vspace{-.2cm}
	\caption{
		MCC (Matthews correlation coefficient) and the number of errors for the second setting.
		The red vertical line is the cross-validation-based threshold $\what{C}_{sel}$.
	}
	\label{fig:supp2}
\end{figure*}

Figures \ref{fig:supp1} and \ref{fig:supp2} show the performance of $\what{S}_{pair, C_{sel}}$ for the first setting and the second setting, respectively.
Each plot represents the performance of $\what{S}_{pair, C_{sel}}$ as a function of the threshold $C_{sel}$, where the red vertical line is the cross-validation-based threshold $\what{C}_{sel}$.
We found that the quality of a fixed threshold changes with the sample size $n$, the dimension $p$ and the structure of $\sg_0$.
The estimated $\what{C}_{sel}$ has reasonable performance in terms of the two criteria.
Therefore, based on the simulation results, the cross-validation-based threshold $\what{C}_{sel}$ seems adequate for adaptive selection of the threshold.
\begin{figure*}[!tb]
	\centering
	\includegraphics[width=6.2cm,height=6.5cm]{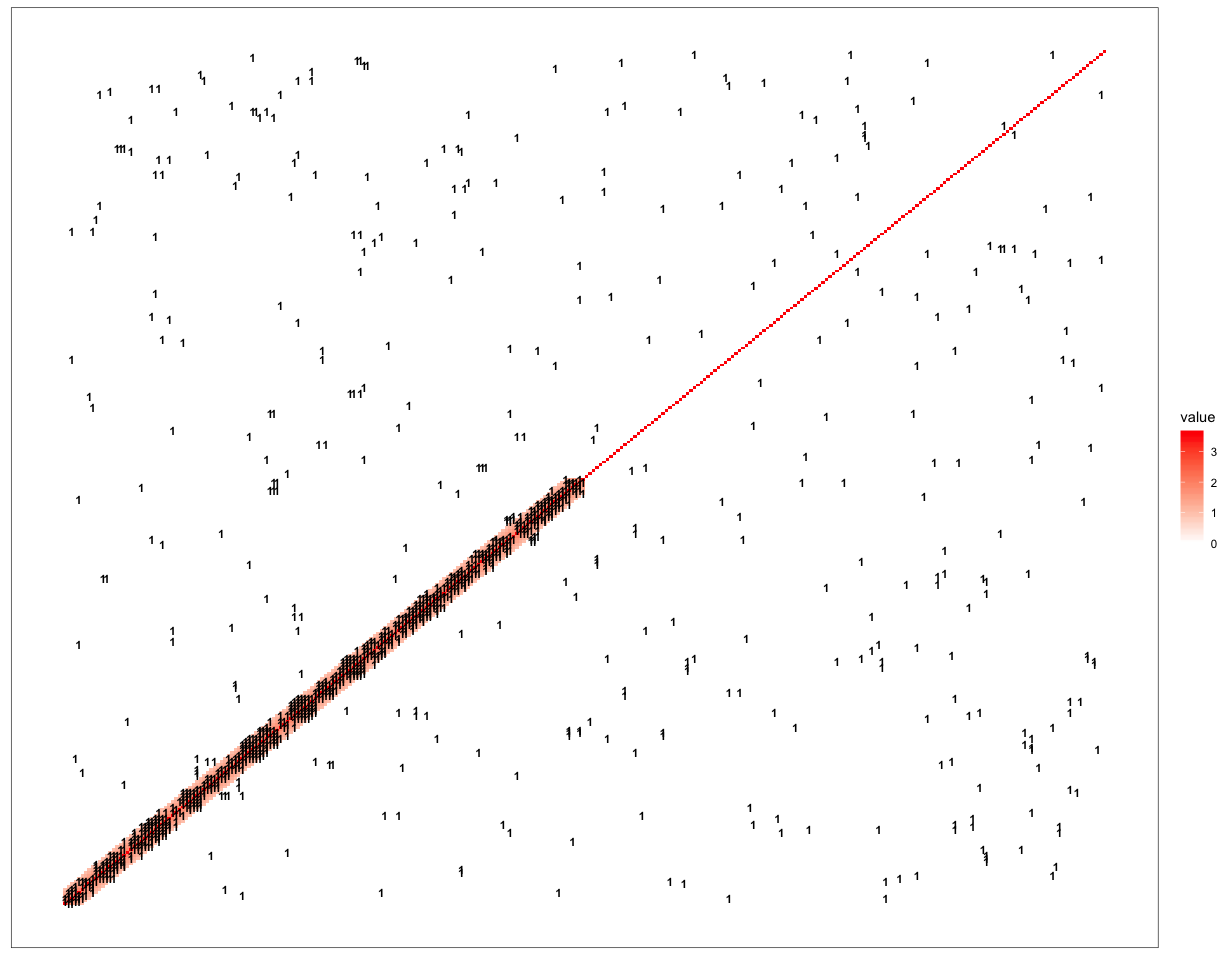}
	\includegraphics[width=6.2cm,height=6.5cm]{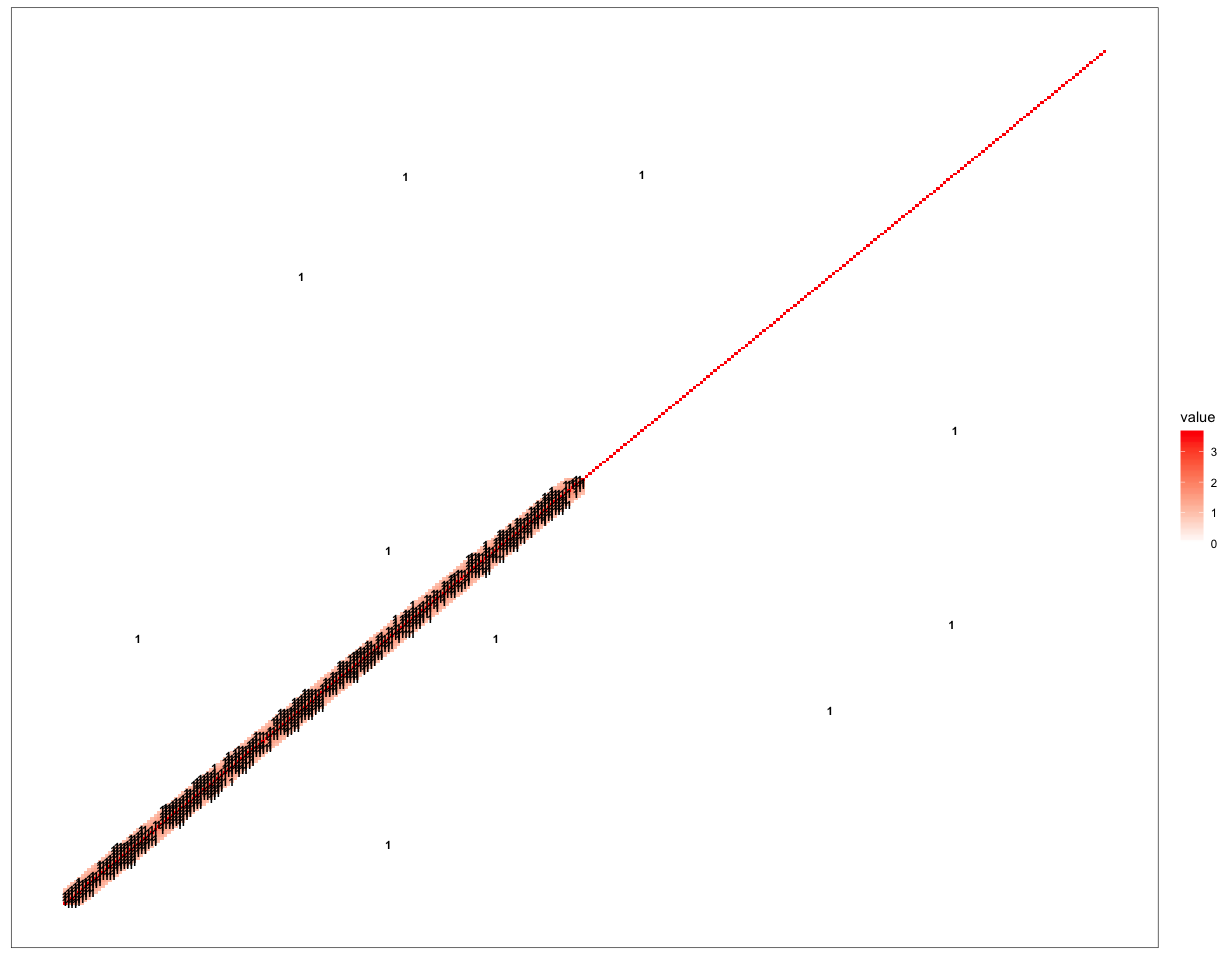}
	\vspace{-.2cm}
	\caption{
		The character $1$ indicates the estimated supports based on $\what{C}_{sel}$ for the first setting with $p=300$ when $n=50$ (left) and $n=100$ (right).
		The values in red represent the entries of the true covariance matrix.
	}
	\label{fig:supp3}
\end{figure*}
Figure \ref{fig:supp3} represents the estimated support $\what{S}_{pair, \what{C}_{sel}}$ for the first setting with $p=300$.
It shows that the cross-validation-based threshold $\what{C}_{sel}$ has reasonable performance and the quality of the support recovery increases as the sample size gets larger.

\section{Proof of Theorem 1} 
\begin{proof}
	For a given pair $(i,j)$,
	\bean
	&&\log B_{10}(\tilde{X}_i, \tilde{X}_j) \nonumber\\ 
	&=& a_0 \log b_{0,ij} + \frac{1}{2} \log \Big(\frac{\gamma}{1+\gamma} \Big) \label{BF_part1} \\
	&+& \Big\{ \log \Gamma\big(\frac{n}{2}+a_0 \big) - \log \Gamma(a_0) - \big(\frac{n}{2}+a_0 \big)\log \frac{n}{2} + \frac{n}{2}  \Big\}  \label{BF_part2} \\
	&+& \Big\{ \frac{n}{2}\what{\tau}_i^2 - \big(\frac{n}{2}+a_0 \big) \log \big( \what{\tau}_{ij,\gamma}^2 + \frac{2 b_{0,ij}}{n} \big) - \frac{n}{2}  \Big\}.   \label{BF_part3}
	\eean
	Equation \eqref{BF_part2} can be written as 
	\bea
	\log \Gamma\big(\frac{n}{2}+a_0 \big)  -\log  \Gamma\big( 1 +a_0 \big)  + a_0 - \big(\frac{n}{2}+a_0 \big) \log \frac{n}{2} + \frac{n}{2},
	\eea
	where by Theorem 1 of \cite{kevckic1971some},
	\bea
	&& \big( \frac{n}{2} +a_0 -1 \big) \log \big( \frac{n}{2}+a_0 \big) -a_0 \log(1+ a_0) - \frac{n}{2} +1 \,\,\le\,\, \log \Gamma\big(\frac{n}{2}+a_0 \big)  -\log  \Gamma\big( 1 +a_0 \big) \\
	&\le& \big( \frac{n}{2} +a_0 -\frac{1}{2} \big) \log \big( \frac{n}{2}+a_0 \big) -  (\frac{1}{2}+ a_0 )\log(1+ a_0) - \frac{n}{2} +1  .
	\eea
	Thus, it is easy to see that \eqref{BF_part2} is equal to
	\bean\label{part2-1}
	- C \log \Big(\frac{n}{2}+ a_0 \Big) + \Big(a_0 + \frac{2a_0^2}{n} \Big) \log \Big(1+ \frac{2a_0}{n} \Big)^{n/2a_0} + C'
	\eean
	for some constants $1/2 < C<1$ and $a_0 \{ 1 - \log(1+a_0) \} + 1 - 0.5 \log (1+a_0) < C' < a_0\{ 1 - \log(1+a_0) \} + 1$, which is of order $O(- \log n)$.  
	Since $\gamma =  (n\vee p)^{-\alpha}$, \eqref{BF_part1} is equal to
	\bean
	&&a_0 \log b_{0,ij}   - \frac{\alpha}{2} \log (n\vee p) - \frac{1}{2} \log \big\{ 1+  (n\vee p)^{-\alpha} \big\} \nonumber \\
	&=& - \frac{\alpha}{2} \log (n\vee p) + C  \label{part1-1}
	\eean
	for some constant $C$.
	Thus, we only need to focus on the behavior of \eqref{BF_part3}.

	For the true covariance matrix $\sg_0$, one of the following cases holds:
	\vspace{-0.1cm}
	\begin{description}
		\item [Case (i).] $\sg_0 = I_p$, \vspace{-0.05cm}
		\item [Case (ii).] $\sigma_{0,ii} $ satisfies condition (A1) for some pair $(i,j)$,  \vspace{-0.05cm}
		\item [Case (iii).] $\tau_{0,ij}^2$ satisfies condition (A2) for some pair $(i,j)$, \vspace{-0.05cm}
		\item [Case (iv).] $\sigma_{0,ij}$ satisfies condition (A3) for some pair $(i,j)$. \vspace{-0.1cm}
	\end{description}
	Note that \eqref{BF_part3} can be expressed as
	\bean
	&& \frac{n}{2}\Big\{ \what{\tau}_{ij,\gamma}^2 + \frac{2 b_{0,ij}}{n} - \log \big( \what{\tau}_{ij,\gamma}^2 + \frac{2 b_{0,ij}}{n} \big) - 1 \Big\}  \label{part3_1}  \\
	&+&  \frac{n}{2}\big( \what{\tau}_{i}^2 - \what{\tau}_{ij,\gamma}^2 \big)  \label{part3_2} \\ 
	&-& \Big\{ a_0 \log \big( \what{\tau}_{ij,\gamma}^2 + \frac{2 b_{0,ij}}{n} \big) + b_{0,ij}  \Big\}  .  \label{part3_3}
	\eean
	We will calculate the rate of the above three terms \eqref{part3_1}-\eqref{part3_3} for every possible case.
	More precisely, we will show that for all sufficiently large $n$, $\log B_{10}(\tilde{X}_i, \tilde{X}_j) \le - C\log (n\vee p)$ with probability at least $1-(n\vee p)^{-c}$ for some constants $C>0$ and $c>2$ under Case (i), and $\log B_{10}(\tilde{X}_i, \tilde{X}_j) \ge C' \log (n\vee p)$ with probability at least $1-(n\vee p)^{-c'}$ for some constants $C'>0$ and $c'>0$ under Cases (ii)--(iv).
	Then, we have
	\bea
	\bbP_0 \Big\{ \log B_{\max, 10}(X ) \le - C\log (n\vee p) \Big\}
	&=& 1- \bbP_0 \left\{ \max_{(i,j): i \neq j} \log B_{10}(\tilde{X}_i, \tilde{X}_j) > - C\log (n\vee p) \right\} \\
	&\ge& 1- \sum_{(i,j): i \neq j} \bbP_0 \left\{  \log B_{10}(\tilde{X}_i, \tilde{X}_j) > - C\log (n\vee p) \right\} \\
	&\ge& 1 - (n\vee p)^{-c +2}
	\eea
	if Case (i) holds, and
	\bea
	\bbP_0 \Big\{ \log B_{\max, 10}(X ) \ge  C'\log (n\vee p) \Big\}
	&\ge& \max_{(i,j): i \neq j} \bbP_0 \Big\{ \log B_{10}(\tilde{X}_i, \tilde{X}_j) \ge  C'\log (n\vee p) \Big\} \\
	&\ge& 1- (n\vee p)^{-c'}
	\eea
	if there exists at least one pair $(i,j)$ satisfying one of Cases (ii)--(iv).

	\noindent{\bf{Case (i).}}
	Define $n\what{\tau}_{ij}^2 = n \what{\tau}_{ij, \gamma=0}^2 = \tilde{X}_i^T (I_n - H_j) \tilde{X}_i$, then we have $n \what{\tau}_{ij,\gamma}^2 = n \what{\tau}_{ij}^2 + \gamma (1+\gamma)^{-1} \big(\tilde{X}_j^T \tilde{X}_i \big)^2 \| \tilde{X}_j\|_2^{-2}$.
	Note that $\big(\tilde{X}_j^T \tilde{X}_i \big)^2 \le \|\tilde{X}_j\|_2^2  \|\tilde{X}_i\|_2^2$ and $\|\tilde{X}_i\|_2^2 \sim \chi_n^2$ under Case (i).
	By Lemma 1 in \cite{laurent2000adaptive}, we have $P\{ k^{-1} \chi_k^2 -1 \ge 2(k^{-1}x)^{1/2} + 2k^{-1}x \} \le 2 \exp(-x)$ and $P\{1 - k^{-1} \chi_k^2 \ge 2{(k^{-1}x)^{1/2}}  \} \le 2 \exp(-x)$ for all $x>0$, which implies 
	\bea
	\what{\tau}_{ij}^2 &\le& \what{\tau}_{ij,\gamma}^2 \,\,\le\,\, \what{\tau}_{ij}^2 + \frac{\gamma }{1+\gamma}  \Big[ 1 + 2 \Big\{  \frac{C\log (n\vee p)}{n} \Big\}^{1/2} + \frac{2C\log (n\vee p)}{n} \Big]
	\eea
	with probability at least $1- 2 (n\vee p)^{-C}$ for some constant $C>2$.
	Since $\gamma =  (n \vee p)^{-\alpha}$ under the conditions on $\alpha$, we have
	\bean\label{sig_ij_and_gamma}
	\big| \what{\tau}_{ij,\gamma}^2 - \what{\tau}_{ij}^2 \big| &\le& (n \vee p)^{-2}
	\eean	
	with probability at least $1- 2 (n\vee p)^{-C}$ for some constant $C>2$ and all large $n$.
	It is easy to check that $n\what{\tau}_{ij}^2 \sim \chi_{n-1}^2$ and $n(\what{\tau}_i^2 - \what{\tau}_{ij}^2) \sim \chi_1^2$ under Case (i).
	Thus, we have
	\bean
	\bbP_0 \Big\{\, \big|n(\what{\tau}_i^2 - \what{\tau}_{ij}^2 )-1 \big| \le 2 C' c \log (n\vee p)  \,\Big\} &\ge& 1- 2 (n\vee p)^{-c} , \quad\quad \label{i-1} \\
	\bbP_0 \left[\, \Big| \frac{n}{n-1}\what{\tau}_{ij}^2 -1  \Big| \le 2C ' \Big\{\frac{c \log (n\vee p)}{n-1} \Big\}^{1/2} \,\right] &\ge& 1- 2 (n\vee p)^{-c}  \quad\quad \label{i-2}
	\eean
	for any constants $c>2$, $C'>1 + {2}^{1/2}\epsilon_0$ and all large $n$.
	The inequalities \eqref{sig_ij_and_gamma} and \eqref{i-2} imply
	\bean
	\bbP_0 \left[\, \Big| \what{\tau}_{ij,\gamma}^2 + \frac{2 b_{0,ij}}{n} -1  \Big| < C \Big\{\frac{ \log (n\vee p)}{n} \Big\}^{1/2} \,\right] &\ge& 1- 4 (n\vee p)^{-c}  \label{i-3}
	\eean
	for any constant $C > 2 ({2}^{1/2} + 2 \epsilon_0)$ and all large $n$.
	By the Taylor expansion of $\log (1+x)$, 
	\bea
	x - \log (1+x) &=& x - \Big( x - \frac{1}{2}x^2 + \frac{1}{3}x^3 - \frac{1}{4}x^4 + \cdots \Big) \\
	&=& \frac{1}{2}x^2 - \frac{1}{3}x^3 + \frac{1}{4}x^4 - \cdots   \,\,\,\,\le\,\,\, \frac{1}{2} x^2  \big( 1 - |x| \big)^{-1}
	\eea
	for small $|x|$.
	Thus, on the event in \eqref{i-3}, \eqref{part3_1} is bounded above by $4^{-1} C^2 ( 1- C\epsilon_0)^{-1} \log (n\vee p)$ for any constant $ 2 ({2}^{1/2} + 2 \epsilon_0) < C < \epsilon_0^{-1}$.
	Since \eqref{part3_3} is of order $O(1)$ on the event in \eqref{i-3}, we have
	\bea
	\log B_{10}(\tilde{X}_i, \tilde{X}_j ) &\le& - 2^{-1}\alpha \log (n\vee p) - 2^{-1} \log n + 2^{-1} C^2 ( 1- C\epsilon_0)^{-1} \log (n\vee p) \\
	&\le& - 2^{-1}\big\{ \alpha -  C^2 ( 1- C\epsilon_0)^{-1} \big\} \log (n\vee p)
	\eea
	with probability at least $1 - 4 (n\vee p)^{-c}$ for any constants $2 (\sqrt{2} + 2 \epsilon_0) < C < \epsilon_0^{-1}$ and all large $n$.
	Thus, if $\alpha >8 (1 + \sqrt{2} \epsilon_0)^2/ \{1- 2\sqrt{2}\epsilon_0 (1 + \sqrt{2}\epsilon_0)\}$, 
	\bea
	\log B_{10}(\tilde{X}_i, \tilde{X}_j ) &\le& - 2^{-1} C\log (n\vee p)
	\eea
	with probability at least $1- 4(n\vee p)^{-c}$ for any constants $0<C < \alpha - 8 (1 + \sqrt{2} \epsilon_0)^2/ \{1- 2^{3/2}\epsilon_0 (1 + {2}^{1/2}\epsilon_0)\}$, $c>2$ and all large $n$.
	This completes the proof for Case (i).

	\noindent{\bf{Case (ii).}}
	Now assume that  $\sigma_{0,ii}$ satisfies condition (A1) for some pair $(i,j)$.
	Note that \eqref{BF_part3} can be expressed as
	\bean
	&&\frac{n}{2} \Big\{ \what{\tau}_i^2  + \frac{2 b_{0,ij}}{n} - \log \big(\what{\tau}_i^2  + \frac{2 b_{0,ij}}{n} \big) - 1 \Big\}   \label{iv-4} \\
	&+& \Big(\frac{n}{2} + a_0 \Big) \log \left( \frac{\what{\tau}_i^2  + \frac{2 b_{0,ij}}{n}}{\what{\tau}_{ij,\gamma}^2  + \frac{2 b_{0,ij}}{n}}  \right)  \label{iv-5}  \\
	&-& a_0 \log \Big(\what{\tau}_i^2  + \frac{2 b_{0,ij}}{n} \Big) - b_{0,ij} .  \label{iv-6}
	\eean
	We will show that for given constants $C_1>0$ and $C_2 > 2{(\alpha +2)^{1/2}}$,
	\bean\label{iv_key}
	\bbP_0 \left\{ \, \Big| \what{\tau}_{i}^2 + \frac{2 b_{0,ij}}{n} -1  \Big| \ge C_2 \Big({\frac{\log (n\vee p)}{n}}\Big)^{1/2}  \, \right\} &\ge& 1 - 2 (n\vee p)^{-C_1} . \quad\quad\,\,
	\eean
	On this event, we can show that \eqref{iv-4} is larger than $4^{-1} C_2^2 (1 - 2 C_2 \epsilon_0/3) \log (n\vee p)$ for all large $n$ by the Taylor expansion of $\log(1+x)$ and the fact that $x-\log x-1$ is increasing in $|x-1|$.
	Note that \eqref{iv-5} is positive and \eqref{iv-6} is negligible compared to \eqref{iv-4}.
	Then, by \eqref{part2-1} and \eqref{part1-1}, 
	\bea
	\log B_{10}(\tilde{X}_i, \tilde{X}_j )  
	&>& - \frac{\alpha}{2}\log (n\vee p) - \log n + 4^{-1} C_2^2 (1 - 2 C_2 \epsilon_0/3) \log (n\vee p) + C \\
	&\ge& 8^{-1} \big( C_2^2 - 4\alpha -8  \big) \log (n\vee p) + C
	\eea
	with probability at least $1 - 4 (n\vee p)^{-C_1}$ for some constant $C$ and all large $n$, by the condition on $\epsilon_0$.

	Now, we only need to show \eqref{iv_key}.
	By Lemma 1 in \cite{laurent2000adaptive}, one can show that
	\bea
	&&\bbP_0 \Bigg[ \sigma_{0,ii} - 1 + \frac{2 b_{0,ij}}{n} - 2 \sigma_{0,ii} \Big\{{\frac{C_1 \log (n\vee p)}{n}} \Big\}^{1/2}
	\,\,\le\,\, \what{\tau}_i^2 +\frac{2 b_{0,ij}}{n} -1    \\
	&& \quad\quad\quad\quad\quad\quad\quad\quad\quad \le   \sigma_{0,ii} - 1 + \frac{2 b_{0,ij}}{n} + 4 \sigma_{0,ii} \Big\{{\frac{C_1 \log (n\vee p)}{n}} \Big\}^{1/2}  \,\,\Bigg]   \,\,\ge\,\, 1- (n\vee p)^{-2C_1},
	\eea
	because $n\what{\tau}_{i}^2 / \sigma_{0,ii} \sim \chi_{n}^2$.
	Thus, it suffices to prove 
	\bea
	\sigma_{0,ii} - 1 + \frac{2 b_{0,ij}}{n} - 2 \sigma_{0,ii} \Big\{{\frac{C_1 \log (n\vee p)}{n}} \Big\}^{1/2} 
	&\ge& C_2  \Big\{{\frac{ \log (n\vee p)}{n}} \Big\}^{1/2}
	\eea
	or
	\bea
	\sigma_{0,ii} - 1 + \frac{2 b_{0,ij}}{n} + 4 \sigma_{0,ii} \Big\{{\frac{C_1 \log (n\vee p)}{n}} \Big\}^{1/2} 
	&\le& - C_2  \Big\{{\frac{ \log (n\vee p)}{n}} \Big\}^{1/2},
	\eea
	which is satisfied by (A1).

	\noindent{\bf{Case (iii).}}
	If $\sigma_{0,ii}$ satisfies (A1) for some pair $(i,j)$, the previous case gives the desired result.
	Here we assume that $\sigma_{0,ii}$ does not satisfy (A1) for all $i$, and  $\tau_{0,ij}^2$ satisfies condition (A2) for some pair $(i,j)$.
	Similar to Case (ii), we will show that for given constants $C_1>0$ and $C_2> 2({\alpha +2})^{1/2}$, 
	\bean\label{ii-1}
	\bbP_0 \left[ \, \Big| \what{\tau}_{ij,\gamma}^2 + \frac{2 b_{0,ij}}{n} -1  \Big| \ge C_2 \Big\{{\frac{ \log (n\vee p)}{n}} \Big\}^{1/2}  \, \right] &\ge& 1 - 4 (n\vee p)^{-C_1} , \quad\quad\quad
	\eean
	which gives the desired result by \eqref{part3_1}--\eqref{part3_3}.
	Note that we have $n\what{\tau}_{ij}^2/ \tau_{0,ij}^2 \sim \chi_{n-1}^2$.
	Then, similar to \eqref{i-2}, 
	\bea
	\bbP_0 \left[\,  \Big|\what{\tau}_{ij,\gamma}^2 - \frac{n-1}{n} \tau_{0,ij}^2 \Big| \le 4 \tau_{0,ij}^2 \Big\{{\frac{ C_1\log (n\vee p)}{n}} \Big\}^{1/2} \,\right] &\ge& 1 - 4 (n\vee p)^{-C_1} 
	\eea
	by \eqref{sig_ij_and_gamma}.
	To prove \eqref{ii-1}, we only need to show that
	\bea
	\Big(1-\frac{1}{n} \Big) \tau_{0,ij}^2 + \frac{2 b_{0,ij}}{n} -1 - 4 \tau_{0,ij}^2 \Big\{{\frac{ C_1\log (n\vee p)}{n}} \Big\}^{1/2} &\ge& C_2 \Big\{{\frac{ \log (n\vee p)}{n}} \Big\}^{1/2},
	\eea
	when $\tau_{0,ij}^2 >1$, and 
	\bea
	\Big(1-\frac{1}{n} \Big) \tau_{0,ij}^2 + \frac{2 b_{0,ij}}{n} -1 + 4 \tau_{0,ij}^2 \Big\{{\frac{ C_1\log (n\vee p)}{n}} \Big\}^{1/2} &\le& -C_2 \Big\{{\frac{ \log (n\vee p)}{n}} \Big\}^{1/2}
	\eea
	when $\tau_{0,ij}^2 <1$.
	It is satisfied because we have condition (A1).
	Thus, we have proved that if a pair $(i,j)$ satisfies (A1),
	\bea
	\log B_{10}(\tilde{X}_i, \tilde{X}_j )  &\ge&  8^{-1}(C_2^2 - 4\alpha -8 ) \log (n\vee p)  + C
	\eea
	with probability at least $1- 4 (n\vee p)^{-C_1}$ for some constant $C$ and all sufficiently large $n$.

	\noindent{\bf{Case (iv).}}
	Suppose $\sigma_{0,ij}$ satisfies condition (A3).
	In this case, we have $n(\what{\tau}_i^2 - \what{\tau}_{ij}^2) / \tau_{0,ij}^2 \sim \chi_1^2(\lambda_{ij})$ given $\tilde{X}_j$, where $\lambda_{ij} = \| \tilde{X}_j \|_2^2 a_{0,ij}^2 / \tau_{0,ij}^2$.
	For a random variable $X\sim \chi_k^2(\lambda)$,
	\bea
	\bbP \left[ X \ge k + \lambda - 2 \{(k+2\lambda)x\}^{1/2}  \,\right] &\ge& 1- e^{-x}
	\eea
	for all $x>0$, by Lemma 8 in \cite{kolar2012marginal}.
	Then,
	\bean
	1 - (n\vee p)^{-C_1} 
	&\le& \bbP_0 \left( \frac{n}{2}\big( \what{\tau}_{i}^2 - \what{\tau}_{ij}^2 \big) \ge  \frac{\tau_{0,ij}^2}{2}\Big[ 1 +\lambda_{ij} - 2 \{(1+2\lambda_{ij}) C_1 \log (n\vee p) \}^{1/2} \,\Big] \,\,\Big|\,\, \tilde{X}_j  \right) \nonumber \\
	&\le& \bbP_0 \left( \frac{n}{2}\big( \what{\tau}_{i}^2 - \what{\tau}_{ij}^2 \big) \ge  \frac{\tau_{0,ij}^2}{2}\lambda_{ij} \Big[ 1 - 2 \Big\{\frac{(1+2\lambda_{ij}) C_1 \log (n\vee p)}{\lambda_{ij}^2} \Big\}^{1/2} \,\Big] \,\,\Big|\,\, \tilde{X}_j  \right)  . \nonumber
	\eean
	Also $\tau_{0,ij}^2 \lambda_{ij} /(a_{0,ij}^2 \sigma_{0,jj} ) = \|\tilde{X}_j\|_2^2 /\sigma_{0,jj} \sim \chi_n^2$, so we have
	\bean\label{iii-3}
	\bbP_0 \left( \frac{\tau_{0,ij}^2}{a_{0,ij}^2 \sigma_{0,jj} } \, \lambda_{ij}  \ge n \Big[1 -2 \Big\{\frac{C_1 \log (n\vee p)}{n} \Big\}^{1/2}\,\, \Big]  \right)  &\ge& 1- (n\vee p)^{-C_1}
	\eean
	by Lemma 1 in \cite{laurent2000adaptive}.
	On the event in \eqref{iii-3}, we have
	\bea
	\lambda_{ij} &\ge& \frac{a_{0,ij}^2 \sigma_{0,jj} }{\tau_{0,ij}^2} n \Big[1 -2 \Big\{\frac{C_1 \log (n\vee p)}{n} \Big\}^{1/2}\,\, \Big] \\
	&\ge& \frac{\sigma_{0,ij}^2 }{\sigma_{0,jj} \tau_{0,ij}^2 } n \Big[ 1 -2 \Big\{\frac{C_1 \log (n\vee p)}{n} \Big\}^{1/2}\,\, \Big] \\
	&\ge& \frac{9C_1}{(1-C_3)^2} \log (n\vee p)
	\eea
	with probability at least $1- (n\vee p)^{-C_1}$ by condition (A3), which implies 
	\bean\label{iii-1}
	1 - 2 \Big\{\frac{(1+2\lambda_{ij}) C_1 \log (n\vee p)}{\lambda_{ij}^2} \Big\}^{1/2} &\ge& C_3
	\eean
	for a given constant $0< C_3 <1$.
	Again by \eqref{iii-3} and condition (A3),
	\bea 
	\frac{\tau_{0,ij}^2}{2}\lambda_{ij} \Big[ 1 - 2 \Big\{\frac{(1+2\lambda_{ij}) C_1 \log (n\vee p)}{\lambda_{ij}^2} \Big\}^{1/2} \,\Big]
	&\ge& \frac{\tau_{0,ij}^2}{2}\lambda_{ij} C_3 \\
	&\ge& \frac{1}{2} \frac{\sigma_{0,ij}^2}{\sigma_{0,jj}} C_3 \, n  \Big[ 1 -2 \Big\{{\frac{ C_1\log (n\vee p)}{n}} \Big\}^{1/2}\,\, \Big]  \\
	&\ge& \frac{1}{2} C_4 (\alpha+2) \log (n\vee p)
	\eea
	with probability at least $1- (n\vee p)^{-C_1}$ for a given constant $C_4> 1$ and all large $n$.
	Note that \eqref{part3_1} is positive and \eqref{part3_3} is negligible compared to \eqref{part3_1}.
	Thus, by similar arguments used in Case (ii), 
	\bea
	\log B_{10}(\tilde{X}_i, \tilde{X}_j ) &\ge&  2^{-1} (C_4-1)(\alpha -2 ) \log (n\vee p) + C
	\eea
	with probability at least $1- 2 (n\vee p)^{-C'}$ for some constants $C, C'>0$ and all large $n$.

\end{proof}

\section{Proof of Theorem 2}

\begin{proof}
	
	For a given constant $C_\star>0$, define a parameter class
	\bea
	H_1^*(C_\star) &=&  \Big\{ \sg_\nu:  \sg_\nu = I_p +  \Big\{ C_\star \Big(\frac{\log p}{n}\Big)^{1/2} I(i=j=\nu) \Big\}_{1\le i,j \le p} , \, 1\le \nu \le p  \Big\}  ,
	\eea
	which trivially satisfies $H_1^*(C_\star) \subset H_1(C_\star)$.
	Let $\bbP_{mix} = p^{-1} \sum_\nu \bbP_{\sg_{\nu}} $ and $\bbE_{mix}$ be the corresponding expectation under $\bbP_{mix}$.
	For any $\sg_{\nu}\in H_1^*(C_\star)$ and test $\phi$, 
	\bea
	\sup_\nu \big\{ \bbE_{I_p}(\phi) + \bbE_{\sg_\nu} (1-\phi)  \big\}
	&\ge& \inf_{\phi} \sup_\nu \big\{ \bbE_{I_p}(\phi) + \bbE_{\sg_\nu}(1-\phi)   \big\} \\
	&\ge& \inf_{\phi} \frac{1}{p} \sum_{\nu} \big\{ \bbE_{I_p}(\phi) + \bbE_{\sg_\nu}(1-\phi)   \big\}  \\
	&=& \inf_{\phi}  \big\{ \bbE_{I_p}(\phi) + \bbE_{mix}(1-\phi)   \big\} \\
	&=& \int (f_{I_p} \wedge f_{mix})  \\
	&=& 1 - \frac{1}{2} \int |f_{mix} - f_{I_p}| ,
	\eea
	where $f_{mix}$ and $f_{\sg}$ are density functions of $\bbP_{mix}$ and $\bbP_{\sg}$, respectively.
	Also 
	\bean
	\Big( \int |f_{mix} - f_{I_p}|\,\Big)^2 &\le& \int |f_{mix} - f_{I_p}|^2 \nonumber \\
	&\le& \int | \frac{f_{mix}}{f_{I_p}} - 1|^2 f_{I_p} \nonumber \\
	&\le& \int \frac{f_{mix}^2}{f_{I_p}}   - 1 . \label{f_ratio}
	\eean
	Thus, for any test $\phi$,
	\bea
	\inf_{\sg \in H_1(C_\star) } \bbE_{\sg} (\phi) &\le& \inf_{\sg \in H_1^*(C_\star) } \bbE_{\sg}(\phi) \,\,\le\,\, \bbE_{I_p}(\phi) + \frac{1}{2} \Big( \int \frac{f_{mix}^2}{f_{I_p}}   - 1 \, \Big)^{1/2}
	\eea
	by the above arguments. 
	Now we only need to deal with the upper bound of \eqref{f_ratio}.
	An upper bound of $\int (f_{mix}^2/f_{I_p})$ can be derived as follows:
	\bea
	\int \frac{f_{mix}^2}{f_{I_p}} &=& \int \frac{1}{f_{I_p}} \Big( \frac{1}{p}\sum_{\nu} f_{\sg_{\nu}} \Big)^2 \\
	&=& \frac{1}{p^2} \sum_{\nu_1, \nu_2} \int \frac{f_{\sg_{\nu_1}}f_{\sg_{\nu_2}}}{f_{I_p}} \\
	&=& \frac{1}{p^2} \sum_{\nu_1, \nu_2} \left[ \det \Big\{ I_p - (\sg_{\nu_1}-I_p)(\sg_{\nu_2}-I_p) \Big\} \right]^{-n/2}\\
	&=& \frac{1}{p^2} \Big\{ p^2 -p + p \Big(1 - C_\star^2 \frac{\log p}{n} \Big)^{-n/2} \Big\} \\
	&\le& 1 - \frac{1}{p} + \frac{1}{p}p^{C_\star^2/2} \,\, \le \,\, 2 - \frac{1}{p}
	\eea
	because $C_\star^2 \le 2$.
	The third equality follows from Lemma B.3 of \cite{lee2018optimal}.
	It gives the upper bound
	\bea
	\inf_{\sg \in H_1(C_\star) } \bbE_{\sg} (\phi) &\le& \bbE_{I_p}(\phi) +  \frac{1}{2} \big( 2 -p^{-1} \big)^{1/2},
	\eea
	which completes the proof.
	
\end{proof}

\section{Proof of Theorem 3} 

\begin{proof}

	For a given pair $(i,j)$ such that $i\neq j$, suppose the null hypothesis is true, so that $\sigma_{0,ij} = 0$.
	We have
	\bea
	\log\tilde{B}_{10}(\tilde{X}_i, \tilde{X}_j) 
	&=& \frac{1}{2}\log \Big(\frac{\gamma}{1+\gamma} \Big) - \frac{n}{2}   \log \left(\frac{\what{\tau}_{ij,\gamma}^2  }{\what{\tau}_{i}^2 }  \right) \\
	&\le& \frac{1}{2}\log \Big(\frac{\gamma}{1+\gamma} \Big)
	+ \frac{n}{2} \frac{\what{\tau}_{i}^2 - \what{\tau}_{ij,\gamma}^2 }{\what{\tau}_{ij,\gamma}^2  } \\
	&\le& \frac{1}{2}\log \Big(\frac{\gamma}{1+\gamma} \Big)
	+ \frac{n}{2}\frac{\what{\tau}_{i}^2 - \what{\tau}_{ij}^2 }{\what{\tau}_{ij}^2 },
	\eea
	where the first inequality holds because $\log(1+x) \le x$ for all $x$.
	Note that $n\what{\tau}_{ij}^2/\sigma_{0,ii} \sim \chi_{n-1}^2$ and $n(\what{\tau}_i^2 - \what{\tau}_{ij}^2) /\sigma_{0,ii} \sim \chi_1^2$.
	By Lemma 1 in \cite{laurent2000adaptive},
	\bea
	\frac{n(\what{\tau}_{i}^2 - \what{\tau}_{ij}^2)}{\sigma_{0,ii}} &\le& 1 + 2 C \, C_4 \log (n\vee p)
	\eea
	and
	\bea
	\frac{n\what{\tau}_{ij}^2}{\sigma_{0,ii}} - (n-1)  &\ge& 2 \{C (n-1)\log (n\vee p)\}^{1/2}
	\eea
	with probability at least $1- 6 (n\vee p)^{-C}$ for any constant $C>2$ and $C_4>1$.
	This implies 
	\bea
	\log\tilde{B}_{10}(\tilde{X}_i, \tilde{X}_j) 
	&\le& \frac{1}{2}\log \Big(\frac{\gamma}{1+\gamma} \Big)
	+ \frac{n}{2} \frac{1 + 2 C \, C_4 \log (n\vee p)}{n-1 - 2 \{C (n-1)\log (n\vee p)\}^{1/2}  } \\
	&\le& \frac{1}{2}\log \Big(\frac{\gamma}{1+\gamma} \Big)
	+ \frac{1}{2} \frac{n + 2 C\, C_4 n\log (n\vee p)}{n-1 - 2 \{C (n-1)\log (n\vee p)\}^{1/2}  } 	 \\
	&\le& \frac{1}{2}\log \Big(\frac{\gamma}{1+\gamma} \Big)
	+ \frac{1}{2} C' \log (n\vee p) + C''
	\eea
	for any constants $C'> 2C\, C_4/(1- 3\epsilon_0{C}^{1/2})$, $C''>0$ with probability at least $1- 6 (n\vee p)^{-C}$ for some constant $C>2$ and large $n$.
	Since $\gamma =  (n\vee p)^{-\alpha}$ and $\alpha > 4/(1- 2^{1/2} 3\epsilon_0)$, by choosing $C>2$ and $C_4>1$ such that $\alpha>C'$, the log Bayes factor $\log\tilde{B}_{10}(\tilde{X}_i, \tilde{X}_j) $ tends to minus infinity as $n\to\infty$ on the above event.
	By similar arguments used in the proof of Theorem 1, this implies consistency of the maximum pairwise Bayes factor under $H_0$.

	Now suppose the alternative hypothesis is true. 
	Without loss of generality, assume that $\sigma_{0,ij}$ satisfies condition (15).
	Then,
	\bea
	-\log\tilde{B}_{10}(\tilde{X}_i, \tilde{X}_j) 
	&=& -\frac{1}{2}\log \Big(\frac{\gamma}{1+\gamma} \Big) - \frac{n}{2}  \log \left( 1 - \frac{\what{\tau}_{ij, \gamma}^2 - \what{\tau}_{i}^2  }{\what{\tau}_{ij, \gamma}^2 }  \right) \\
	&\le& -\frac{1}{2}\log \Big(\frac{\gamma}{1+\gamma} \Big) + \frac{n}{2}  \, \frac{\what{\tau}_{ij, \gamma}^2 - \what{\tau}_{i}^2 }{\what{\tau}_{i}^2 }
	\eea
	because $-\log(1-x) \le x/(1-x)$ for any $x<1$.
	Since $n\what{\tau}_i^2 / \sigma_{0,ii} \sim \chi_n^2$ and $n(\what{\tau}_i^2 - \what{\tau}_{ij}^2) /\tau_{0,ij}^2 \sim \chi_1^2(\lambda_{ij})$ given $\tilde{X}_j$, where $\lambda_{ij} = \|\tilde{X}_j\|_2^2 a_{0,ij}^2/ \tau_{0,ij}^2$, we have
	\bea
	\frac{1}{2\what{\tau}_{i}^2 } 
	&\ge&  \frac{1}{2} \Big( \sigma_{0,ii} \Big[1 + 4 \Big\{\frac{C_1\log (n\vee p)}{n} \Big\}^{1/2} \,\,\Big]  \Big)^{-1}  \\
	&\ge& \frac{1}{2} \Big\{ \sigma_{0,ii} (1+ 4\epsilon_0 {C_1}^{1/2})   \Big\}^{-1}
	\eea
	and
	\bea
	n(\what{\tau}_{ij, \gamma}^2 - \what{\tau}_{i}^2 ) 
	&=& - \frac{1}{1+\gamma} n(\what{\tau}_i^2 - \what{\tau}_{ij}^2) \\
	&\le& - \frac{C_3}{1+\gamma} \tau_{0,ij}^2 \lambda_{ij}  \\
	&=& - \frac{C_3}{1+\gamma} \sigma_{0,jj} a_{0,ij}^2  \frac{\|\tilde{X}_j\|_2^2}{\sigma_{0,jj}} \\
	&\le& - \frac{C_3}{1+\gamma} \frac{\sigma_{0,ij}^2 }{\sigma_{0,jj}}   n (1 - 2  {C_1}^{1/2} \epsilon_0)
	\eea
	with probability at least $1 - C (n\vee p)^{-c}$ for some constants $c>0$ and $C>0$ and all large $n$, by condition (15).
	The fourth inequality holds by condition (A4) and similar arguments used in \eqref{iii-1}.
	Thus, on this event,
	\bea
	\frac{n}{2}  \, \frac{\what{\tau}_{ij, \gamma}^2 - \what{\tau}_{i}^2 }{\what{\tau}_{i}^2 }
	&\le& - \frac{C_3}{1+\gamma} \frac{\sigma_{0,ij}^2 }{\sigma_{0,jj}} n (1 - 2\epsilon_0 {C_1}^{1/2}) \times \frac{1}{2} \Big\{  \sigma_{0,ii} (1+ 4\epsilon_0 {C_1}^{1/2}) \Big\}^{-1} \\
	&\le& - \frac{\alpha}{2}C_4  \log (n\vee p)
	\eea
	which implies
	\bea
	-\log\tilde{B}_{10}(\tilde{X}_i, \tilde{X}_j) 
	&\le&  \frac{\alpha}{2}\log (n\vee p) + \frac{1}{2} \log (1+\gamma) - \frac{\alpha}{2}C_4  \log (n\vee p)
	\eea
	for all sufficiently large $n$.
	Since $C_4>1$, this completes the proof.

\end{proof}

\section{Proof of Proposition 1} 

\begin{proof}
	
	Let $\what{\sigma}_{ij}^2 = n^{-1} ( \tilde{X}_i^T \tilde{X}_j  )^2$ and $\what{\rho}_{ij}^2  = \what{\sigma}_{ij}^2 / (\what{\tau}_{i}^2 \what{\tau}_{j}^2)$.
	Then 
	\bea
	\frac{\what{\tau}_{ij,\gamma}^2}{\what{\tau}_{i}^2} 
	&=& \frac{\what{\tau}_{ij}^2}{\what{\tau}_{i}^2}  + \frac{\what{\tau}_{ij,\gamma}^2 - \what{\tau}_{ij}^2}{\what{\tau}_{i}^2}  \\
	&=& 1 - \frac{\what{\sigma}_{ij}^2}{\what{\tau}_{i}^2 \what{\tau}_{j}^2} + \frac{\what{\tau}_{ij,\gamma}^2 - \what{\tau}_{ij}^2}{\what{\tau}_{i}^2} \\
	&\equiv& 1 - \what{\rho}_{ij}^2 + \what{\delta}_{ij}.
	\eea
	Then we have 
	\bea
	\log \tilde{B}_{\max, 10}(X) 
	&=& \frac{1}{2} \log \Big( \frac{\gamma}{1+\gamma}  \Big) - \frac{n}{2} \log \Big\{ 1 - \max_{i\neq j} ( \what{\rho}_{ij}^2 - \what{\delta}_{ij} )   \Big\}.
	\eea
	We will show that upper and lower bounds of $2 \log \tilde{B}_{\max, 10}(X) - C_{n,p}$ converges in distribution to an extreme distribution of type I with distribution function 
	\bea
	F(z) &=& \exp \big\{  - (8\pi)^{-1/2} e^{- z/2}  \big\} , \quad z \in \bbR .
	\eea
	
	By the definitions of $\what{\tau}_{ij}^2$ and $\what{\tau}_{ij,\gamma}^2$, we have $\what{\delta}_{ij} \ge 0$, thus 
	\bean
	&& 2 \log \tilde{B}_{\max, 10}(X) - C_{n,p}   \nonumber \\
	&\le& \log \Big( \frac{\gamma}{1+\gamma}  \Big) - n \log \big( 1 - \max_{i\neq j}  \what{\rho}_{ij}^2  \big)  - C_{n,p} \nonumber \\
	&=& \frac{\log \big( 1 - \max_{i\neq j}  \what{\rho}_{ij}^2  \big)}{ - \max_{i\neq j}  \what{\rho}_{ij}^2 }\, n  \, \max_{i\neq j}  \what{\rho}_{ij}^2 - 4\log p + \log (\log p) \nonumber \\ 
	&=&  \frac{\log \big( 1 - \max_{i\neq j}  \what{\rho}_{ij}^2  \big)}{ - \max_{i\neq j}  \what{\rho}_{ij}^2 } \big\{  n  \, \max_{i\neq j}  \what{\rho}_{ij}^2 - 4\log p + \log (\log p) \big\} \label{limit_null_eq1} \\ 
	&+& \Big\{  \frac{\log \big( 1 - \max_{i\neq j}  \what{\rho}_{ij}^2  \big)}{ - \max_{i\neq j}  \what{\rho}_{ij}^2 }  -1 \Big\} \big\{  4\log p - \log (\log p) \big\} . \label{limit_null_eq2}
	\eean
	By Theorem 4 in \cite{cai2011limiting}, $\log p \, \max_{i\neq j}  \what{\rho}_{ij}^2  \overset{p}{\lra} 0$ as $n\to\infty$ under $\bbP_0$. 
	Taylor expansion of $\log (1-x)$ gives  
	\bea
	0 \,\, \le \,\, \frac{\log \big( 1 - \max_{i\neq j}  \what{\rho}_{ij}^2  \big)}{ - \max_{i\neq j}  \what{\rho}_{ij}^2 }  -1  
	&=& \frac{1}{2} \max_{i\neq j}  \what{\rho}_{ij}^2  + \frac{1}{3} \max_{i\neq j}  \what{\rho}_{ij}^4 + \cdots  \\
	&\le& \frac{\max_{i\neq j}  \what{\rho}_{ij}^2 }{ 1- \max_{i\neq j}  \what{\rho}_{ij}^2 } ,
	\eea
	which implies that \eqref{limit_null_eq2} converges to 0 in probability and
	\bea
	\frac{\log \big( 1 - \max_{i\neq j}  \what{\rho}_{ij}^2  \big)}{ - \max_{i\neq j}  \what{\rho}_{ij}^2 } &\overset{p}{\lra}& 1 
	\eea
	as $n\to\infty$.
	Since $n  \, \max_{i\neq j}  \what{\rho}_{ij}^2 - 4\log p + \log (\log p)$ converges in distribution to an extreme distribution of type I by Theorem 4 in \cite{cai2011limiting}, Slutsky's theorem says that \eqref{limit_null_eq1} converges in distribution to an extreme distribution of type I.
	
	On the other hand, we have 
	\bea
	&& 2 \log \tilde{B}_{\max, 10}(X) - C_{n,p} \\
	&\ge& \log \Big( \frac{\gamma}{1+\gamma}  \Big) - n \log \big( 1 - \max_{i\neq j}  \what{\rho}_{ij}^2 +  \max_{i\neq j}  \what{\delta}_{ij}   \big)  - C_{n,p}  \\
	&=&  \frac{\log \big( 1 - \max_{i\neq j}  \what{\rho}_{ij}^2 +  \max_{i\neq j}  \what{\delta}_{ij}    \big)}{ - \max_{i\neq j}  \what{\rho}_{ij}^2 +  \max_{i\neq j}  \what{\delta}_{ij}   }\, n\big(   \max_{i\neq j}  \what{\rho}_{ij}^2 -  \max_{i\neq j}  \what{\delta}_{ij}  \big)  - 4\log p + \log (\log p)
	\eea
	Thus, if we show that $ n \, \max_{i\neq j}  \what{\delta}_{ij} \overset{p}{\lra} 0$ as $n\to\infty$ under $\bbP_0$, the same arguments used in the previous paragraph imply that the lower bound of $2 \log \tilde{B}_{\max, 10}(X) - C_{n,p}$ converges in distribution to an extreme distribution of type I, which completes the proof.
	By the proof of Theorem 1, for a given pair $(i,j)$,
	\bea
	\bbP_0 \Big[  1 - 2 \Big\{ \frac{C \log (n\vee p)}{n}  \Big\}^{1/2}  \le \frac{\what{\tau}_i^2}{\sigma_{0,ii}}  \le  1 + 4 \Big\{ \frac{C \log (n\vee p)}{n}  \Big\}^{1/2}  \Big] &\ge& 1 - (n\vee p)^{-2C}
	\eea
	and 
	\bea
	&&  \bbP_0 \Big(  0 < \frac{\what{\tau}_{ij,\gamma}^2 - \what{\tau}_{ij}^2}{\sigma_{0,ii}}  \le \frac{ \gamma }{1+\gamma} \Big[  1 + 2 \Big\{ \frac{C \log (n\vee p)}{n}  \Big\}^{1/2}  +   \frac{2C \log (n\vee p)}{n}   \Big]  \Big)  \\
	&\ge& 1 - 2(n\vee p)^{-C}
	\eea
	for some constant $C>2$.
	Since $\gamma = (n\vee p)^{-\alpha}$ and $\alpha>1$, it implies that $ n \, \max_{i\neq j}  \what{\delta}_{ij} \overset{p}{\lra} 0$ as $n\to\infty$ under $\bbP_0$.
	
\end{proof}

\section{Proof of Theorem 4} 

\begin{proof}
	
	For a given constant $C_{sel}>0$ and a pair $(i,j)$, define
	\bea
	\what{S}_{ij} &=& I \Big\{ \, \log \tilde{B}_{pair,10}(\tilde{X}_i, \tilde{X}_j) > C_{sel} \,   \Big\}
	\eea
	and $S_{ij}(\sg_0) = I(\sigma_{0,ij} \neq 0 )$.
	By the proof of Theorem 3, 
	\bea
	\bbP_0 \Big\{ \what{S}_{ij}=1 ,\, S_{ij}(\sg_0) =0  \Big\}
	&=&
	\bbP_0 \Big\{  \log \tilde{B}_{pair,10}(\tilde{X}_i, \tilde{X}_j) > C_{sel} ,\,  \sigma_{0,ij}  =0 \Big\}  \\
	&\le& (n\vee p)^{-c}
	\eea
	for some constant $c>2$ and all sufficiently large $n$.
	Now, assume that $\sigma_{0,ij}$ satisfies condition (A5).
	Condition (A5) is the same as condition (A4) except using $C_5$ instead of $C_1$.
	Thus, by similar arguments used in the proof of Theorem 3, it is easy to check that
	\bea
	\bbP_0 \Big\{ \what{S}_{ij}=0 ,\, S_{ij}(\sg_0) =1  \Big\}
	&=&
	\bbP_0 \Big\{  \log \tilde{B}_{pair,10}(\tilde{X}_i, \tilde{X}_j) \le C_{sel} ,\,  \sigma_{0,ij}  \neq 0 \Big\} \\
	&\le& (n\vee p)^{-c}
	\eea
	for some constant $c>2$ and all sufficiently large $n$.
	Therefore, we have
	\bea
	\bbP_0 \Big\{ \what{S} \neq S(\sg_0) \Big\}
	&\le& \sum_{i\neq j,\, i<j} \bbP_0 \Big\{ \what{S}_{ij} \neq S_{ij}(\sg_0)  \Big\} \,\,\lra\,\, 0
	\eea
	as $n\to\infty$.	
	
\end{proof}

\bibliographystyle{dcu}
\bibliography{bayes-cov-test}


\end{document}